\documentstyle[./epsf]{l-aa}

\def\ppmm{$\!\!\!\!\pm\!\!\!\!$}

\begin{document}

\thesaurus{06(02.01.2, 02.08.1, 02.09.1, 02.19.1, 03.13.4, 08.02.1)}

\title{Non-axisymmetric wind-accretion simulations}
\subtitle{II. Density gradients}

\author{M. Ruffert\thanks{e-mail: {\tt m.ruffert@ed.ac.uk}}}
\institute{Department of Mathematics \& Statistics, 
University of Edinburgh, Scotland EH9 3JZ, Great Britain}

%\date{Received; accepted}

\maketitle

\begin{abstract}
The hydrodynamics of a variant of classical Bondi-Hoyle-Lyttleton
accretion is investigated: a totally absorbing sphere
moves at various Mach numbers (3 and 10) relative to a
medium, which is taken to be an ideal gas having a {\it density}
gradient (of 3\%, 20\% or 100\% over one accretion radius)
perpendicular to the relative motion.
I examine the influence of the Mach number, the adiabatic index, and
the strength of the gradient upon the physical behaviour of the flow
and the accretion rates of the angular momentum in particular.
The hydrodynamics is modeled by the ``Piecewise Parabolic Method'' (PPM).
The resolution in the vicinity of the accretor is increased by
multiply nesting several grids around the sphere.

Similarly to the 3D models published previously,
both with {\it velocity} gradients and without,
the models with a {\it density} gradient presented here exhibit
non-stationary flow patterns, 
although the Mach cone remains fairly stable.
The accretion rates of mass, linear and angular momenta
do not fluctuate as strongly as published previously for 2D models.
No obvious trend of the dependency of mass accretion rate fluctuations
on the density gradient can be discerned.
The average specific angular momentum accreted is roughly between zero
and 70\% of the total angular momentum available in the accretion
cylinder in the cases where the average is prograde.
Due to the large fluctuations during accretion, the average angular
momentum of some models is retrograde by up to 25\%.
The magnitude is always smaller than the value of a vortex with
Kepler velocity around the surface of the accretor. 

The models with small density gradients {\it initially} display a
transient quasi-stable accretion phase in which the specific angular
momentum accreted is within 10\% of the total angular momentum
available in the accretion cylinder. Later, when the flow becomes
unstable, the average decreases.
I conclude that for accretion from a medium with both density
and/or velocity gradients, most of the angular momentum that is
available in the accretion cylinder is accreted together with mass.
Small gradients hardly influence the average accretion rates as
compared to accretion from a homogeneous medium, while very large ones
succeed to dominate and form an accretion flow in which the sense of
rotation is not inverted.

\end{abstract}

\begin{table*}
\caption[] {
Parameters and some computed quantities for all models.
${\cal M}_\infty \equiv {\cal M}_\infty (y=0)$ 
is the Mach number of the unperturbed flow,
$\varepsilon_\rho$ the parameter specifying the magnitude of the
density gradient,
$\gamma$ the ratio of specific heats,
$t_{\rm f}$ the total time of the run (units: $R_{\rm  A}/c_{\infty}$),
$\overline{\dot{M}}$ the integral average of the mass accretion rate,
$S$ one standard deviation around the mean $\overline{\dot{M}}$
of the mass accretion rate fluctuations,
$\widehat{\dot{M}}$ the maximum mass accretion rate,
$\dot{M}_{\rm BH}$ is defined in Eq.~(3) of Ruffert \& Arnett (1994),
$l_{\rm x}$, $l_{\rm y}$, $l_{\rm z}$,
are the averages of specific angular momentum components
together with their respective standard deviations
$\sigma_{\rm x}$, $\sigma_{\rm y}$, $\sigma_{\rm z}$,
$s$ is the entropy (Eq.~(4) in Ruffert \& Arnett~1994),
the radius of the accretor always is $R_\star=0.02R_{\rm A}$,
the number of grid nesting depth levels $g=9$,
the size of one zone on the finest grid $\delta=1/256R_{\rm A}$,
the softening parameter for the potential of the
accretor (see Ruffert~1994) $\epsilon=3$ zones,
the number $N$ of zones per grid dimension is 32,
and the size of the largest grid is $L=32R_{\rm A}$.
}
\label{tab:models}
\begin{flushleft}
\begin{tabular}{lccccrclcrclrclrclc}
\hline\\[-3mm]
   Model & ${\cal M}_\infty$ & $\varepsilon_\rho$ & $\gamma$  
   & $t_{\rm f}$ & $\overline{\dot{M}}$&\ppmm&$S$ & $\widehat{\dot{M}}$
   & $l_{\rm x}$&\ppmm&$\sigma_{\rm x}$ 
   & $l_{\rm y}$&\ppmm&$\sigma_{\rm y}$
   & $l_{\rm z}$&\ppmm&$\sigma_{\rm z}$ & $s$ \\
    & & & &  
      & \multicolumn{3}{c}{[$\dot{M}_{\rm BH}$]}  & [$\dot{M}_{\rm BH}$]
%      & \multicolumn{3}{c}{~[$R_{\rm A}c_\infty$]}
%      & \multicolumn{3}{c}{~[$R_{\rm A}c_\infty$]}
%      & \multicolumn{3}{c}{~[$R_{\rm A}c_\infty$]}
      & \multicolumn{3}{c}{[$\varepsilon_\rho R_{\rm A}v_0/4$]}
      & \multicolumn{3}{c}{[$\varepsilon_\rho R_{\rm A}v_0/4$]}
      & \multicolumn{3}{c}{[$\varepsilon_\rho R_{\rm A}v_0/4$]}
      & [${\cal R}$]
     \\[0.5mm] \hline\\[-3mm]
MV &1.4& 0.03 & 5/3 & 25.2 & 0.88&\ppmm&0.11 
   & 0.96 &  -0.01&\ppmm&0.01 &  -0.02&\ppmm&0.03 &  +0.76&\ppmm&0.17
      & 1.14 \\  % RAMV
MS &~3 & 0.03 & 5/3 & 10.1 & 0.62&\ppmm&0.09 
   & 0.77 &   0.01&\ppmm&0.50 &  -0.02&\ppmm&0.60 &  +0.38&\ppmm&1.22
      & 2.8  \\  % RAMS
MF &10 & 0.03 & 5/3 & 4.05 & 0.45&\ppmm&0.17 
   & 0.69 &  -0.20&\ppmm&1.22 &  +0.52&\ppmm&3.05 &  -0.22&\ppmm&2.36
      & 6.2  \\  % RAMF
NS &~3 & 0.20 & 5/3 & 7.96 & 0.38&\ppmm&0.09 
   & 0.64 &  -0.09&\ppmm&0.20 &  -0.07&\ppmm&0.26 &  +0.35&\ppmm&0.26
      & 2.9  \\  % RANS
NF &10 & 0.20 & 5/3 & 2.99 & 0.36&\ppmm&0.11 
   & 0.66 &  -0.04&\ppmm&0.20 &  -0.02&\ppmm&0.27 &  +0.40&\ppmm&0.39
      & 6.0  \\[0.7ex]  % RANF
PS &~3 & 0.03 & 4/3 & 9.99 & 1.03&\ppmm&0.14 
   & 1.43 &  +0.10&\ppmm&0.56 &  -0.23&\ppmm&0.75 &  +0.65&\ppmm&1.94
      & 6.3  \\  % RAPS
PF &10 & 0.03 & 4/3 & 3.99 & 0.81&\ppmm&0.15 
   & 1.36 &  -0.19&\ppmm&1.21 &  -0.81&\ppmm&2.51 &  -0.10&\ppmm&2.64
      & 13.1 \\  % RAPF
QS &~3 & 0.20 & 4/3 & 6.16 & 0.83&\ppmm&0.18 
   & 1.42 &   0.00&\ppmm&0.11 &   0.00&\ppmm&0.15 &  +0.55&\ppmm&0.26
      & 6.3  \\  % RAQS
QF &10 & 0.20 & 4/3 & 2.21 & 0.72&\ppmm&0.10 
   & 0.98 &  -0.01&\ppmm&0.12 &  -0.08&\ppmm&0.20 &  +0.37&\ppmm&0.29
      & 12.7 \\  % RAQF
VS &~3 & 1.00 & 4/3 & 6.24 & 0.35&\ppmm&0.08 
   & 0.71 &   0.00&\ppmm&0.02 &   0.02&\ppmm&0.04 &  +0.23&\ppmm&0.04
      & 5.5  \\[0.7ex]  % RAVS
TS &~3 & 0.03 & 1.01& 6.14 & 1.20&\ppmm&0.10 
   & 1.35 &  -0.02&\ppmm&0.08 &  +0.22&\ppmm&0.54 &  +1.11&\ppmm&1.50
      & 68.7 \\  % RATS
TF &10 & 0.03 & 1.01& 2.00 & 0.83&\ppmm&0.03 
   & 0.91 &  -1.12&\ppmm&0.08 &   0.75&\ppmm&1.11 &  -0.25&\ppmm&0.75
      & 201. \\  % RATF
US &~3 & 0.20 & 1.01& 6.09 & 1.19&\ppmm&0.11 
   & 1.43 &  +0.03&\ppmm&0.05 &  +0.08&\ppmm&0.11 &  +0.69&\ppmm&0.23
      & 76.4 \\  % RAUS
UF &10 & 0.20 & 1.01& 2.22 & 0.82&\ppmm&0.04 
   & 0.97 &  -0.14&\ppmm&0.02 &   0.16&\ppmm&0.05 &  -0.04&\ppmm&0.07
      & 210. \\  % RAUF
\hline
\end{tabular}
\end{flushleft}
\end{table*}

\section{Introduction\label{sec:intro}}

The simplicity of the classic Bondi-Hoyle-Lyttleton (BHL) accretion model
makes its use attractive in order to estimate roughly accretion rates
and drag forces applicable in many different astrophysical contexts.
Various aspects of the BHL flow have repeatedly been investigated
in the past by many authors.
In the BHL scenario a totally absorbing sphere of mass $M$ 
moves with velocity $v_\infty$ relative to a surrounding homogeneous
medium of density $\rho_\infty$ and sound speed  $c_\infty$.
In this second instalment, I extend the investigations started in 
Ruffert (1997, henceforth R1) to include {\it density} gradients of the 
surrounding medium.
Usually, the accretion rates of various quantities, like mass, angular 
momentum, etc., including drag forces, are of interest as well as the
bulk properties of the flow, (e.g.~distribution of matter and velocity,
stability, etc.). 
The results pertaining to total accretion rates tend to agree well
qualitatively (to within factors of two, and ignoring the
instabilities of the flow) with the original calculations of Bondi,
Hoyle and Lyttleton
(Hoyle \& Lyttleton~1939, 1940a, 1940b, 1940c; Bondi \& Hoyle~1944).
However, the question of whether and how much angular momentum is
accreted together with mass from an inhomogeneous medium has remained
largely unanswered, although R1 has attempted a first answer.
It has already been summarised in the introduction of R1, that the
strict application of the BHL recipe to inhomogeneous media, 
including some small constant gradient in the density or the velocity
distribution, yields that the accreted matter has zero angular momentum
by construction (Davies \& Pringle, 1980).

We recall that 
the largest radius $R_{\rm a}$ from which matter is still accreted by the
BHL-procedure turns out to be the so-called Hoyle-Lyttleton accretion 
radius
\begin{equation}
   R_{\rm a} = \frac{2GM}{v_\infty^2}  \quad,
\label{eq:accrad1}
\end{equation}
where $G$ is the gravitational constant.
The mass accretion rate follows to be
\begin{equation}
   \dot{M}_{\rm HL} = \pi R^2_{\rm a} \rho_\infty v_\infty  \quad.
  \label{eq:accmass1}
\end{equation}
I will refer to the volume upstream of the accretor from which matter
is accreted as accretion cylinder.
One can additionally calculate 
(Dodd \& McCrea~1952; Illarionov \& Sunyaev~1975; 
Shapiro \& Lightman~1976; Wang~1981)
how much angular momentum is present in the accretion
cylinder for a non-axisymmetric flow which has a gradient in its
density or velocity distribution
perpendicular to the mean velocity direction.
Then, assuming no redistribution of angular momentum,
the amount accreted is equal to (or at least is a large fraction
of) the angular momentum present in the accretion cylinder.

The uncertainty about how much angular momentum can actually be
accreted in a BHL flow stems from these two opposing views involving
either a large  
or a very small fraction of what is present in the accretion cylinder.
In the first paper~R1, I showed that the answer is not clear cut, but
depends on the initial and boundary conditions.
Roughly 7\% to 70\% of the total angular momentum available in the
accretion cylinder is accreted.

\begin{figure}
 \epsfxsize=8.8cm \epsfclipon \epsffile{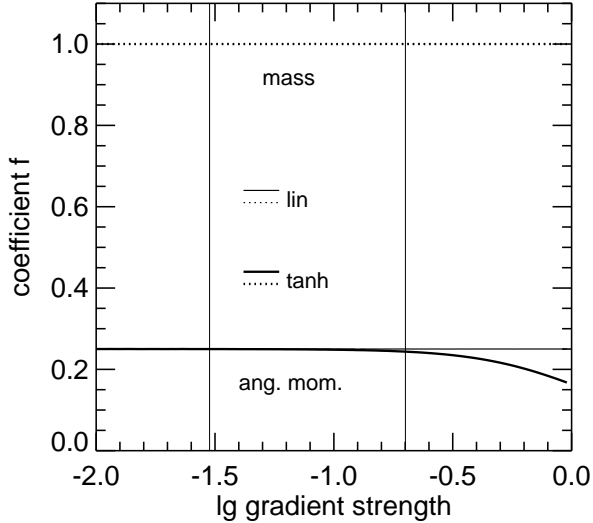}
\caption[]{The coefficient for the mass accretion rate $f_{\rm m}$
(dotted), defined by Eq.~(\ref{eq:coeffmass}), and for the specific
angular momentum $f_{\rm j}$ (solid), defined by
Eq.~(\ref{eq:coeffspec}), as a function of $\varepsilon_\rho$.
The thin curves show the values of $f$ for
a simple linear relation between $v_{{\rm x}\infty}$ 
and $\varepsilon_\rho$, while the bold curves apply to a relation
including the ``tanh''-term as given in Eq.~(\ref{eq:vgrad}).
The two vertical straight lines indicate the gradients that were used
in the numerical models (cf.~Table~\ref{tab:models}), 
$\varepsilon_\rho=0.03$ and $\varepsilon_\rho=0.2$.
Compare this figure with Fig.~1 in paper~R1 and note the differences.
}
\label{fig:shali}
\end{figure}

In this second paper I would like to compare the accretion rates of
the angular momentum of numerically modeled accretion flows 
with {\it density} gradients to the previous results of accretion
with {\it velocity} gradients (R1).
Although several investigations of {\it two}-dimensional flows with
gradients exist (Anzer et al.~1987; Fryxell \& Taam~1988; 
Taam \& Fryxell~1989; Ho et al.~1989; Benensohn et al~1997,
Shima et al, 1998), {\it three}-dimensional simulations are scarce
due to their inherently high computational load.
Livio et al.~(1986) first attempted a three-dimensional model
including gradients, but due to their 
low numerical resolution the results were only tentative.
In the models of Ishii et al.~(1993) the accretor was only
coarsely resolved, while the results of Boffin (1991) and 
Sawada et al.~(1989) are only indicative, because due to the numerical
procedure the flows remained stable (too few SPH particles in Boffin~1991 
and local time stepping in Sawada et al.~1989 which was
described to be appropriate only for stationary flows).
Also Sawada et al.~(1989) only investigated velocity gradients.

In order to be able to compare the new results to previous models
of~R1, I will use the same values for the gradients as in R1. 
However, as will be mentioned in connection with Eq.~\ref{eq:accmass} 
(which had already been presented in R1),
the expected angular momentum accretion is reduced by a factor of six,
when changing from velocity to density gradients.
Thus the clear separation between angular momentum from the bulk flow 
one from the unstable, fluctuating flow blurs.
On the other hand arbitrarily large specific angular momenta cannot be
accreted because of the ang.~mom.~barrier felt by matter spiralling
into the accretor.
A simulation with a very large density gradient was done, too,
mainly to be able to compare to a previous 2D model (Fryxell \& Taam, 1988).

In Sect.~\ref{sec:numer} I give an only short summary of the
numerical procedure used. 
Sections~\ref{sec:descr1} and~\ref{sec:descr2} present the results,
which I summarise in Sect.~\ref{sec:conc}.

\section{Numerical Procedure and Initial Conditions
     \label{sec:numer}}

Since the numerical procedures and initial conditions are mostly
identical to what has already been described and used in previous
papers (cf.~Ruffert~1997, R1, and references therein) I will refrain
from repeating every detail, but only give a brief summary.

\subsection{Numerical Procedure\label{sec:numproc}}

The distribution of matter is discretised on multiply nested
equidistant Cartesian grids (e.g.~Berger \& Colella, 1989) 
with zone size $\delta$ and is evolved using the ``Piecewise
Parabolic Method'' (PPM) of Colella \& Woodward (1984).
The equation of state is that of a perfect gas with a specific
heat ratio $\gamma$ (see Table~\ref{tab:models}).
The model of the maximally accreting, vacuum sphere in a softened
gravitational potential is summarised in Ruffert \& Arnett (1994) and
Ruffert \& Anzer (1995).

\begin{figure*}
 \begin{tabular}{cc}
 \epsfxsize=8.6cm  \epsfclipon \epsffile{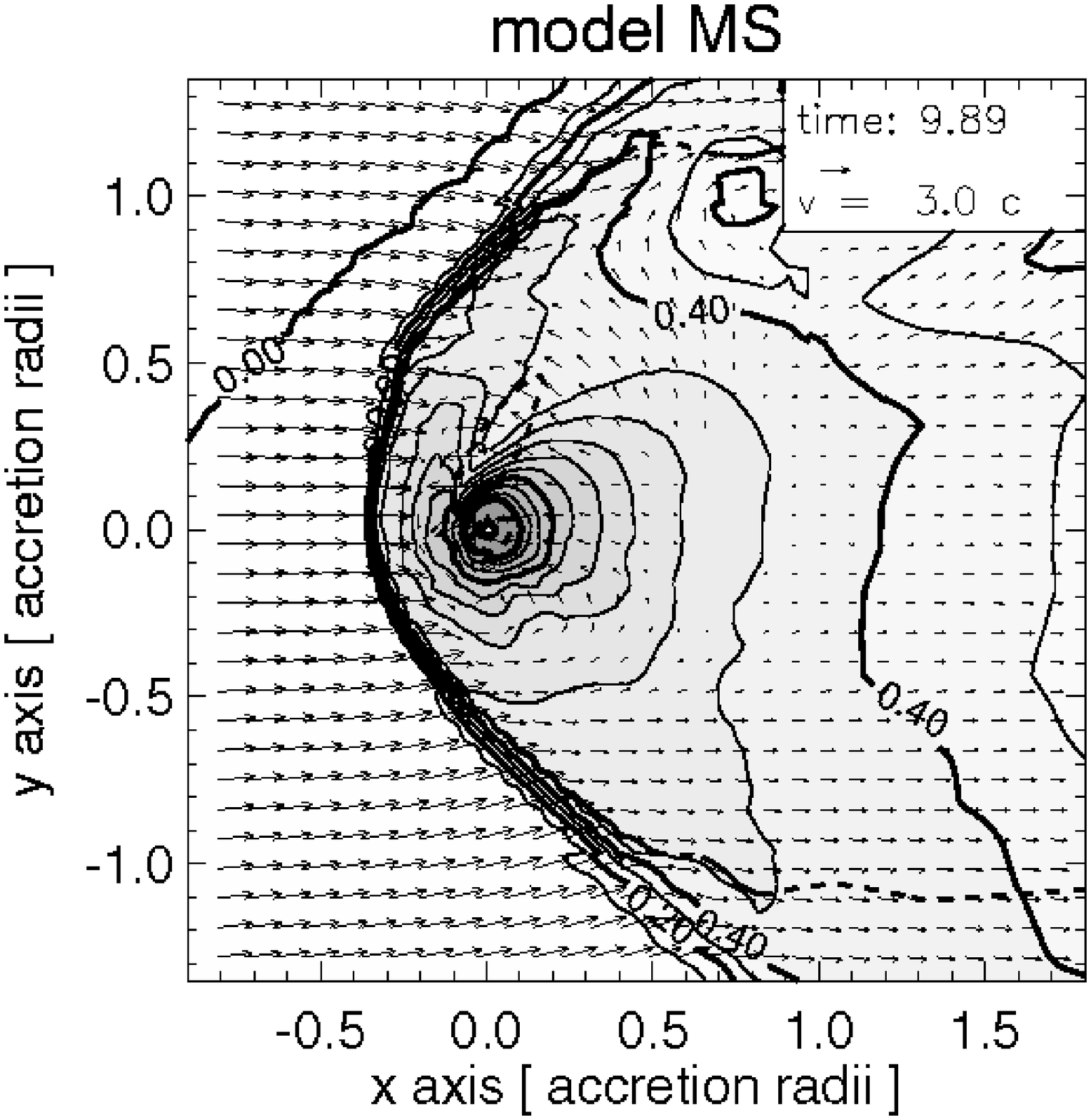} &
 \epsfxsize=8.6cm  \epsfclipon \epsffile{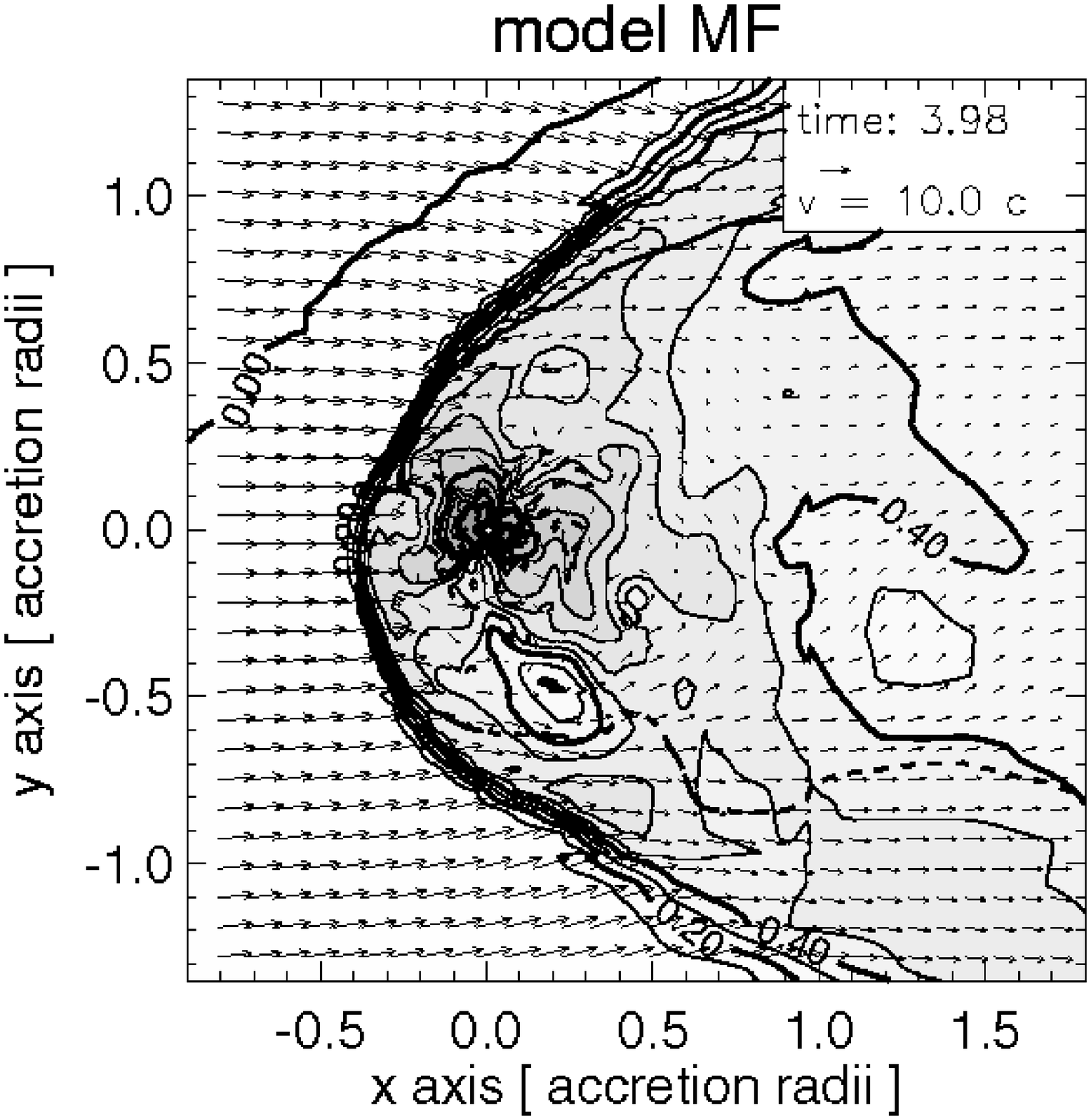}\\[-2.5ex]
 \epsfxsize=8.6cm  \epsfclipon \epsffile{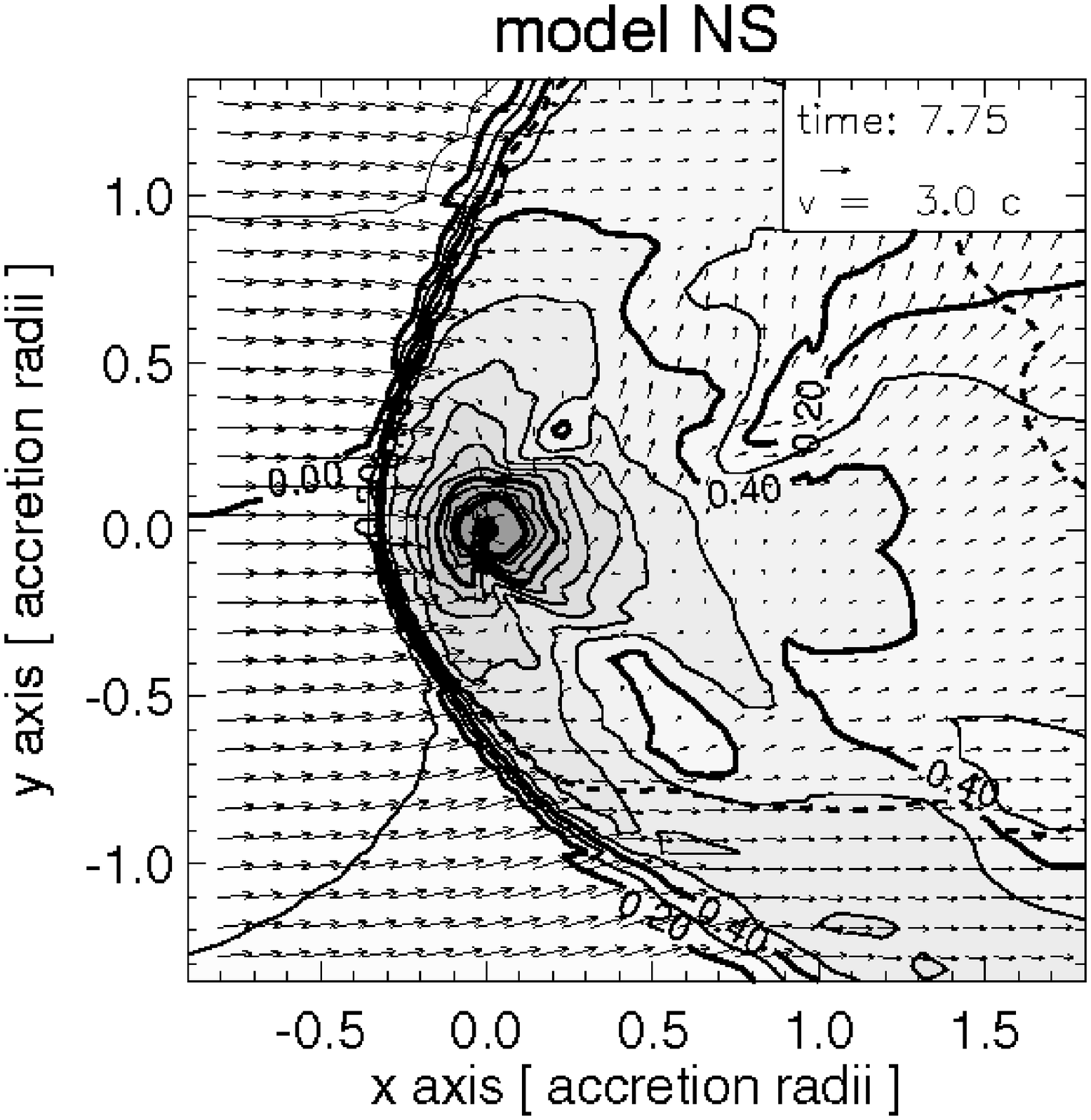} &
 \epsfxsize=8.6cm  \epsfclipon \epsffile{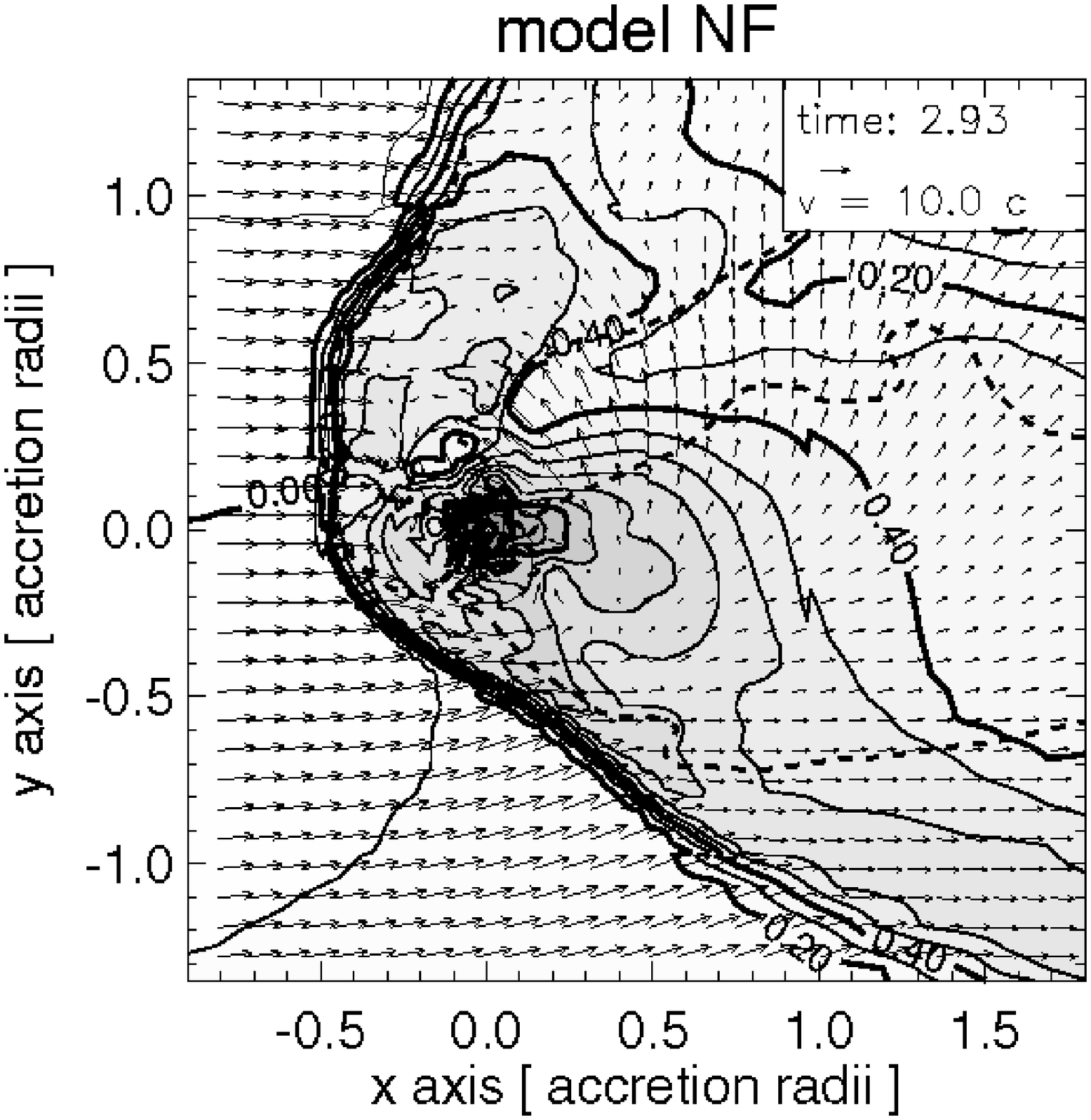}\\[-2.5ex]
 \epsfxsize=8.6cm  \epsfclipon \epsffile{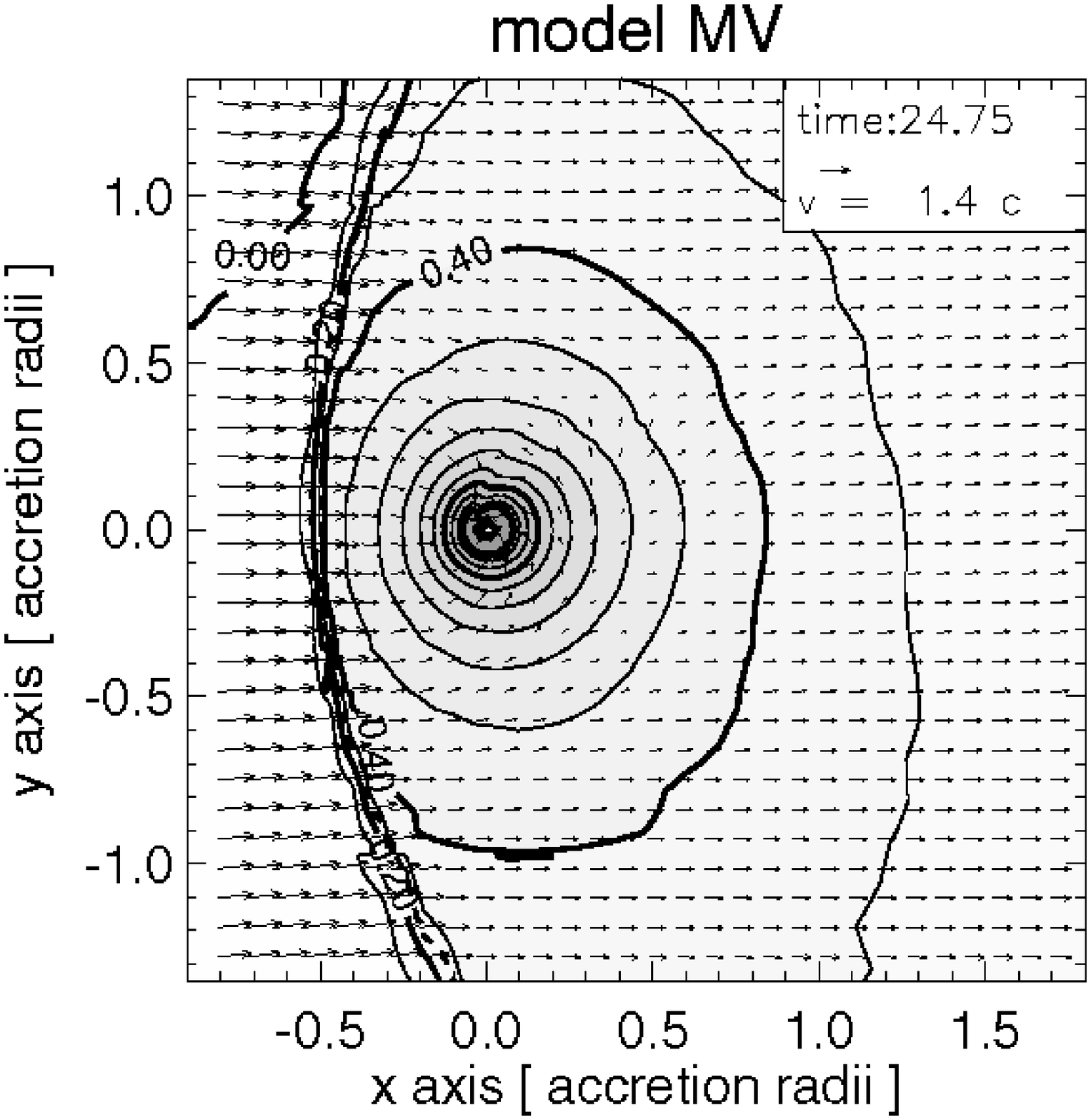} &
\raisebox{5cm}{\parbox[t]{8.6cm}{
\caption[]{\label{fig:Mdens}
Contour plots showing snapshots of the density together
with the flow pattern for all models with an adiabatic index of 5/3.
The contour lines are spaced logarithmically in intervals of 0.1~dex.
The bold contour levels are sometimes labeled with their respective
values (0.0, 0.2, and 0.4).
Darker shades of gray indicate higher densities.
The dashed contour delimits supersonic from subsonic regions.
The time of the snapshot together with the velocity scale is given in
the legend in the upper right hand corner of each panel.
}}}
 \end{tabular}
\end{figure*}

\begin{figure*}
 \tabcolsep = 0mm
 \begin{tabular}{cc}
  \epsfxsize=8.5cm  \epsfclipon \epsffile{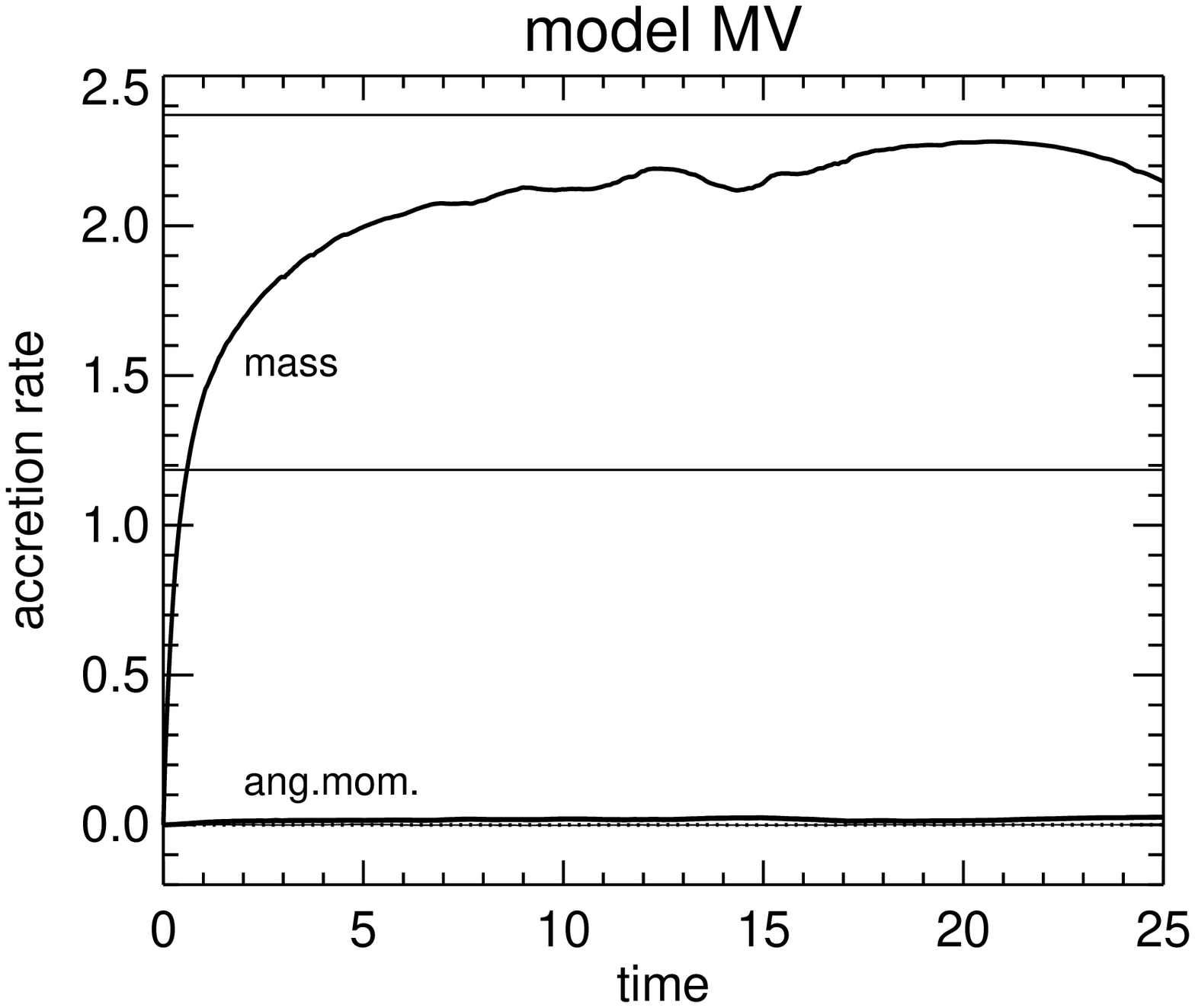} &
  \epsfxsize=8.5cm  \epsfclipon \epsffile{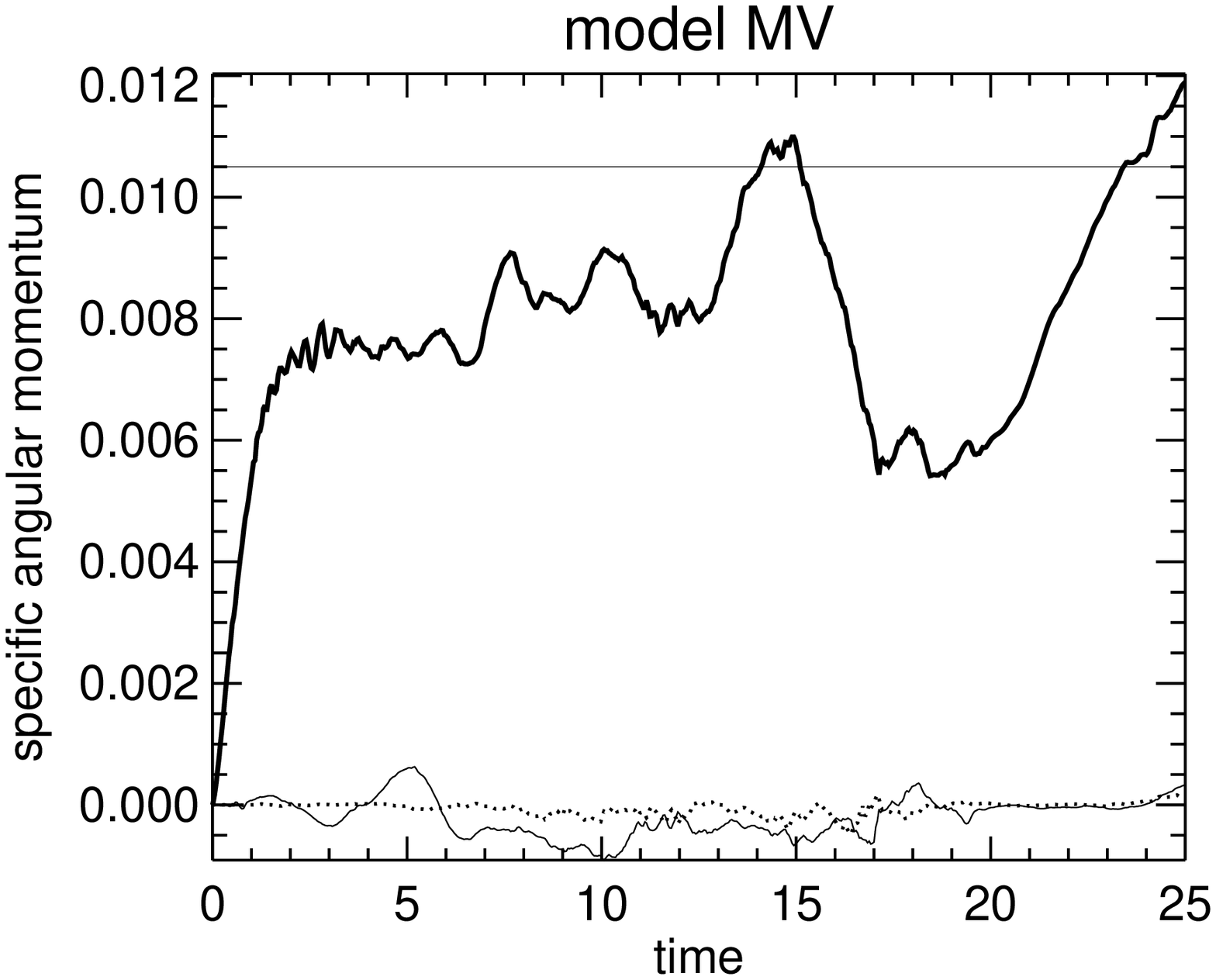}
 \end{tabular}
\caption[]{
The accretion rates of several quantities are plotted as a
function of time for model~MV (${\cal M}_\infty$=1.4, gradient of~3\%,
adiabatic index of $\gamma=5/3$).
The left panel contains the mass and angular momentum accretion rates,
the right panel the specific angular momentum of the matter
that is accreted.
In the left panel, the straight horizontal lines show the analytical
mass accretion rates: the top
solid line is the Bondi-Hoyle approximation formula (Eq.~(3) in
Ruffert~1994; Bondi~1952) and the lower one half that value.
The upper solid bold curve represents the
numerically calculated mass accretion rate.
The lower three curves of the left panel trace the x~(dotted),
y~(thin solid) and z~(bold solid) component of the angular momentum
accretion rate.
The same components apply to the right panel.
The horizontal line in the right panel shows the
specific angular momentum value as given by Eq.~(\ref{eq:specmomang}).
}
\label{fig:valueMV}
\end{figure*}

\begin{figure*}
 \tabcolsep = 0mm
 \begin{tabular}{cc}
%\framebox[8.8cm]{\raisebox{8cm}{~}} &
%\framebox[8.8cm]{\raisebox{8cm}{~}} \\
%\framebox[8.8cm]{\raisebox{8cm}{~}} &
%\framebox[8.8cm]{\raisebox{8cm}{~}} \\
%\framebox[8.8cm]{\raisebox{8cm}{~}} &
%\framebox[8.8cm]{\raisebox{8cm}{~}} 
  \epsfxsize=8.8cm  \epsfclipon \epsffile{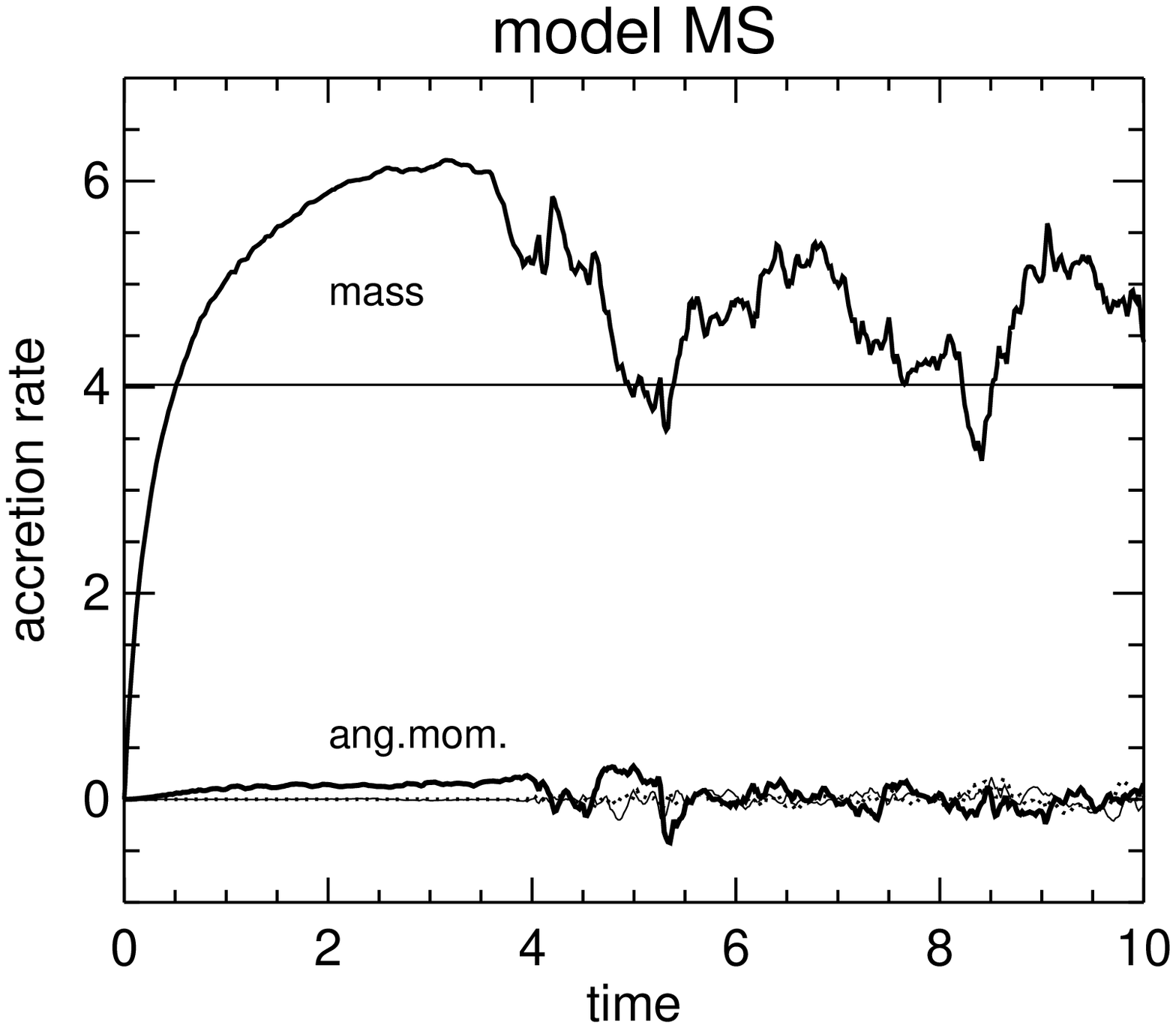} &
  \epsfxsize=8.8cm  \epsfclipon \epsffile{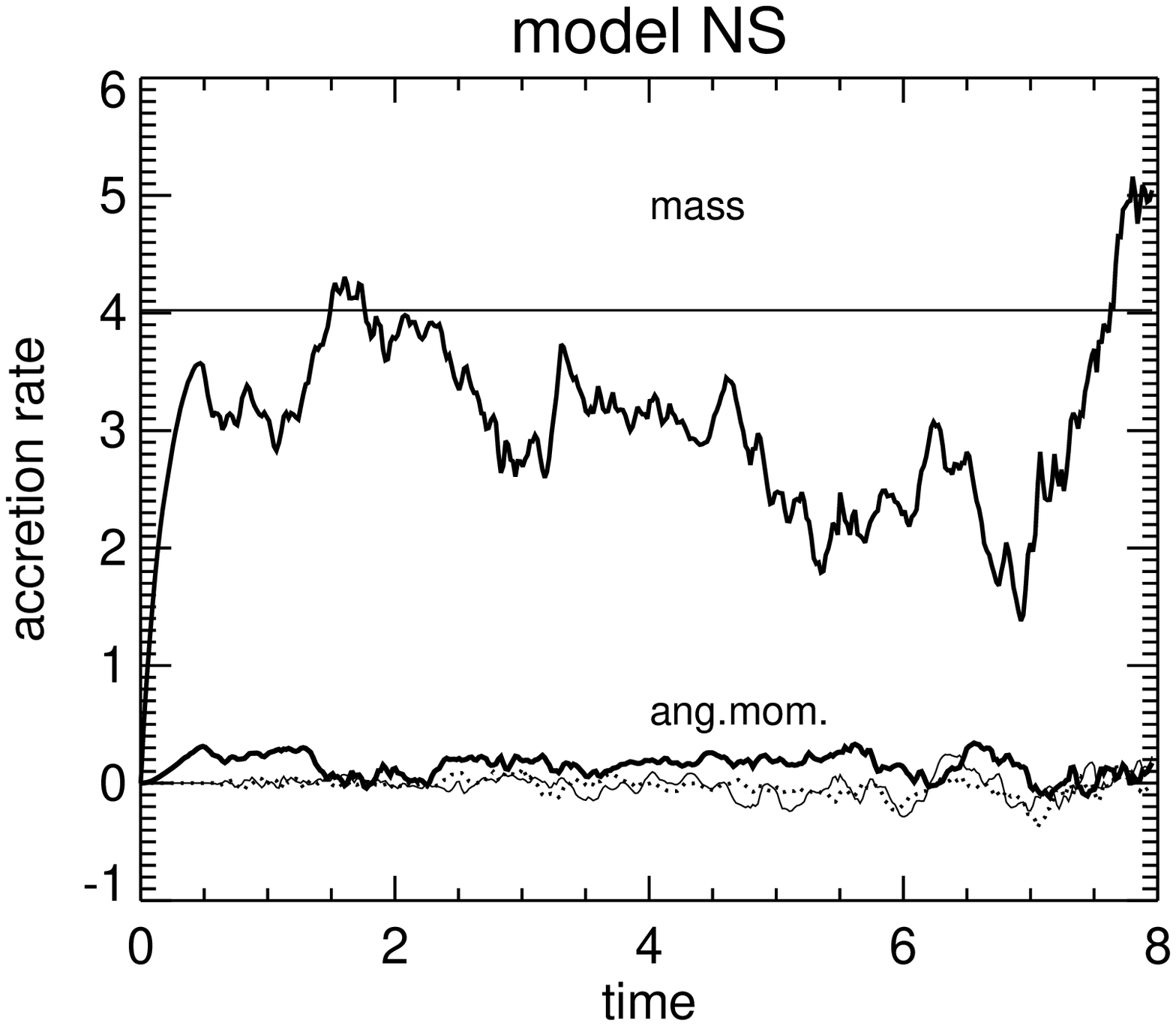} \\
  \epsfxsize=8.8cm  \epsfclipon \epsffile{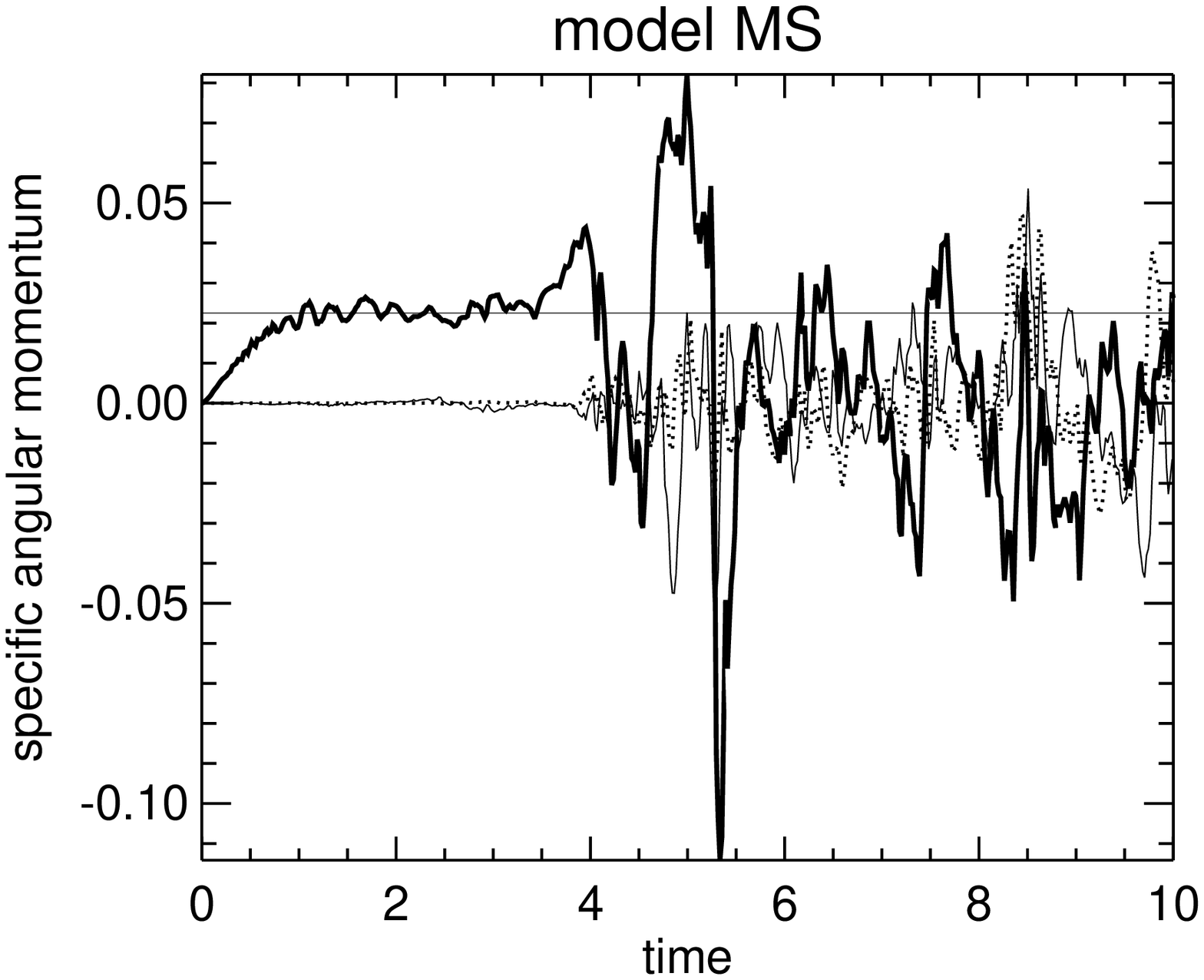} &
  \epsfxsize=8.8cm  \epsfclipon \epsffile{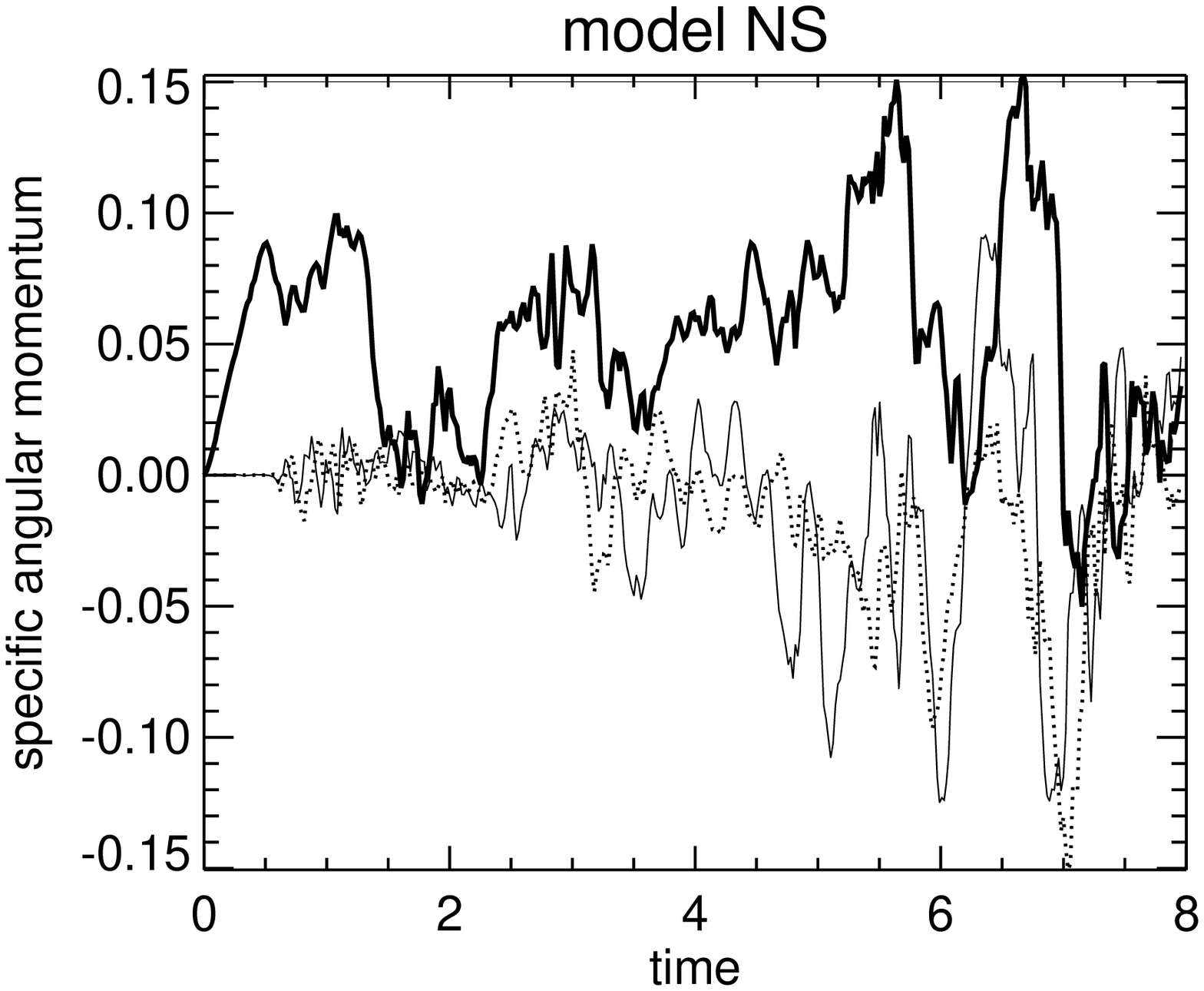}
 \end{tabular}
\caption[]{
The accretion rates of several quantities are plotted as a
function of time for the moderately supersonic (${\cal M}_\infty$=3)
models~MS and~NS with an index $\gamma=5/3$.
The top panels contain the mass and angular momentum accretion rates,
the bottom panels the specific angular momentum of the matter
that is accreted.
In the top panels, the straight horizontal lines show the analytical
mass accretion rates: half the value of 
the Bondi-Hoyle approximation formula (Eq.~(3) in
Ruffert~1994; Bondi~1952).
The upper solid bold curve represents the
numerically calculated mass accretion rate.
The lower three curves of the top panels trace the x~(dotted),
y~(thin solid) and z~(bold solid) component of the angular momentum
accretion rate.
The same components apply to the bottom panels.
The horizontal line in the bottom panels show the
specific angular momentum value as given by Eq.~(\ref{eq:specmomang}).
}
\label{fig:valueMS}
\end{figure*}

A gravitating, totally absorbing ``sphere'' moves relative to a medium
that has a distribution of density and velocity far upstream
(at $x\to-\infty$) given by
\begin{equation}
 \rho_\infty = \rho_{\rm 0} 
        \left( 1 - \frac{1}{2} 
       \tanh \left[ 2 \varepsilon_\rho \frac{y}{R_{\rm A}} \right] \right)
      \quad,
  \label{eq:rhograd}
\end{equation}
\begin{equation}
 v_{{\rm x}\infty} 
     = v_{\rm 0} \left( 1 + \frac{1}{2} 
       \tanh \left[ 2 \varepsilon_{\rm v} \frac{y}{R_{\rm A}} \right] \right),
  \,\, v_{{\rm y}\infty} = 0, \,\, v_{{\rm z}\infty} = 0 \,\,,
  \label{eq:vgrad}
\end{equation}
with the redefined accretion radius (note slight difference to
Eq.~\ref{eq:accrad1})
\begin{equation}
    R_{\rm A}= \frac{2GM}{v_0^2} \quad.
  \label{eq:accrad}
\end{equation}

In this paper I only investigate models with gradients of the
{\it density} distribution, thus for all models I set
$\varepsilon_{\rm v}\equiv0$. 
The values of $\varepsilon_\rho$ can be found in Table~\ref{tab:models}.
In order to keep the pressure constant throughout the upstream boundary,
I changed the internal energy or equivalently the sound speed $c$
accordingly ($c^2=\gamma p/\rho$).
Thus the Mach number ${\cal M}_\infty (y) \equiv v/c_\infty (y)$ 
of the incoming flow varies
too, since the velocity is kept constant.
The Mach numbers given in the Table~\ref{tab:models} refer to the
values along the axis ($y$=0) upstream of the accretor.

The function ``$\tanh$'' is introduced in Eq.~(\ref{eq:rhograd})
and~(\ref{eq:vgrad}) to serve 
as a cutoff at large distances $y$ for large gradients
$\varepsilon$.
The units I use in this paper
are (1) the on-axis sound speed $c_{\infty}(y=0)$ as velocity unit;
(2) the accretion radius (Eq.~(\ref{eq:accrad}))
as unit of length, and
(3) the on-axis density $\rho_0$ .
Thus the unit of time is $R_{\rm A}/c_{\infty}$.

I'll assume both $\varepsilon_{\rm v}\neq0$ and 
$\varepsilon_\rho\neq0$ for the following estimates.
Accreting all mass within the accretion cylinder,
and taking the matter to have the density and velocity distributions
as given by Eqs.~(\ref{eq:rhograd}) and~(\ref{eq:vgrad})
one obtains the mass accretion rate to lowest order in
$\varepsilon_{\rm v}$ and $\varepsilon_\rho$ to be
\begin{equation}
   \dot{M} = \pi R^2_{\rm A} \rho_0 v_0  \quad,
  \label{eq:accmass}
\end{equation}
an equation very similar to Eq.~(\ref{eq:accmass1}).
Further assuming that all angular momentum within the deformed
accretion cylinder is accreted too, the specific angular momentum of
the accreted matter follows to be
(Ruffert \& Anzer 1994; Shapiro \& Lightman 1976; again to lowest
order in $\varepsilon_{\rm v}$)
\begin{equation}
  j_{\rm z} = \frac{1}{4}
     \left( 6 \varepsilon_{\rm v} + \varepsilon_\rho \right) v_0 R_{\rm A}  
  \label{eq:specmomang}\quad.
\end{equation}
For positive $\varepsilon_\rho$ 
the density is lower on the positive side of the
$y$-axis, then the vortex formed around the accretor is in the
anticlockwise direction, i.e.~the angular momentum component in
$z$-direction is positive.

The values of the specific angular momentum obtained from the
numerical simulations  will be compared to the values that follow from
this Eq.~(\ref{eq:specmomang}) to conclude which of the above
mentioned views --- low or high specific angular momentum of the
accreted material --- is most more appropriate.
As was stated further above, R1 found that a sizable amount (between
7\% and 70\%) is accreted, when velocity gradients are present.
I will implicitly assume the component $j_{\rm z}$ when discussing
properties like fluctuations, magnitudes, etc.
From the symmetry of the boundary conditions the average of the $x$
and $y$ components of the angular momentum should be zero, although
their fluctuations can be quite large.
Taking the numerically obtained accretion rates of mass $\dot{M}(t)$
and angular momentum $\dot{J}(t)$, I calculate the instantaneous
specific angular momentum $j(t)=\dot{J}(t)/\dot{M}(t)$, the 
temporal mean $l$ of which is listed in Table~\ref{tab:models}, too.

Apart from serving as cut-off, the tanh-dependencies in
Eqs.~(\ref{eq:rhograd}) and~(\ref{eq:vgrad}) have a gradient that is
slightly less steep than simply linear. Thus less specific angular
momentum is 
present at a given radius from the accretor and smaller values in the
magnitude of the accreted quantity result.

One can numerically approximate the integrals
of the mass flux and angular momentum over the
cross section of the accretion cylinder, to obtain the
coefficients $f$ in the relations Eq.~(\ref{eq:accmass}) 
and~Eq.~(\ref{eq:specmomang}):
\begin{equation}
   \dot{M} = f_{\rm m}(\varepsilon_\rho)
             \pi R^2_{\rm A} \rho_0 v_0  \quad,
  \label{eq:coeffmass}
\end{equation}
\begin{equation}
  \jmath_{\rm z} = f_{\rm j} (\varepsilon_\rho)
      \varepsilon_\rho  v_0 R_{\rm A}  
  \label{eq:coeffspec}\quad.
\end{equation}
Here, I will only consider the effect of a density gradient.
The unitless functions $f$ are a function of $\varepsilon_\rho$ and 
the functional relation of $\rho_\infty[\varepsilon_\rho]$,
i.e.~whether $\rho_\infty$ depends purely linearly on
$\varepsilon_\rho$ or as in Eq.~(\ref{eq:rhograd}) via the ``tanh''-term.
Figure~\ref{fig:shali} shows the values of the functions $f$ for the
mass and specific angular momentum and for both the linear and
``tanh'' case.
The only minimal deviation of the  ``tanh'' curves from the linear
ones indicates that the tanh-cutoff hardly acts within the
accretion cylinder.
Since $f_{\rm m}\approx1$ and is practically constant for 
$\varepsilon_\rho\la1$, Eq.~(\ref{eq:accmass}) is a good
approximation in this range.
If the prescription is correct that everything in the accretion
cylinder is accreted, no difference in the accretion rate 
should be observable in the models with differing magnitude of
gradient $\varepsilon_\rho$ (cf.~Table~\ref{tab:models}).
The same constancy applies to the accretion of specific angular momentum: 
its coefficient $f_{\rm j}$ remains relatively constant 
$f_{\rm j}\approx 0.25$ in the range $\varepsilon_\rho\la0.4$,
which includes both $\varepsilon_\rho$ for which models were simulated.
So although I used the same magnitudes for the gradients, the effects
I expect in the accretion rates are markedly different from what was
presented in paper~R1.
Only for model~VS with $\varepsilon_\rho=1.0$, does $f_{\rm j}$
deviate appreciably from 0.25: it is $f_{\rm j} \approx 0.16$.

\begin{figure*}
 \tabcolsep = 0mm
 \begin{tabular}{cc}
  \epsfxsize=8.8cm  \epsfclipon \epsffile{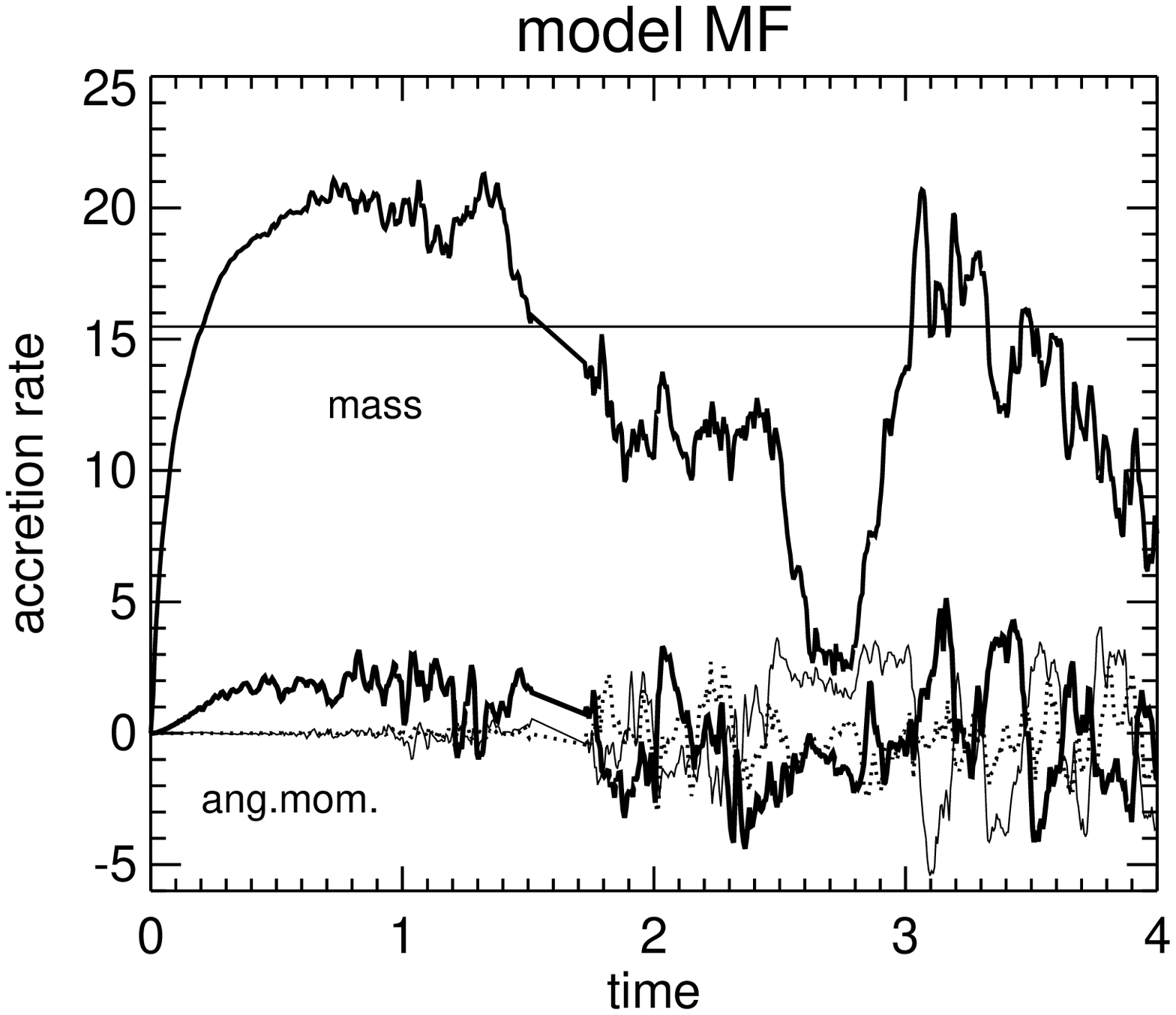} &
  \epsfxsize=8.8cm  \epsfclipon \epsffile{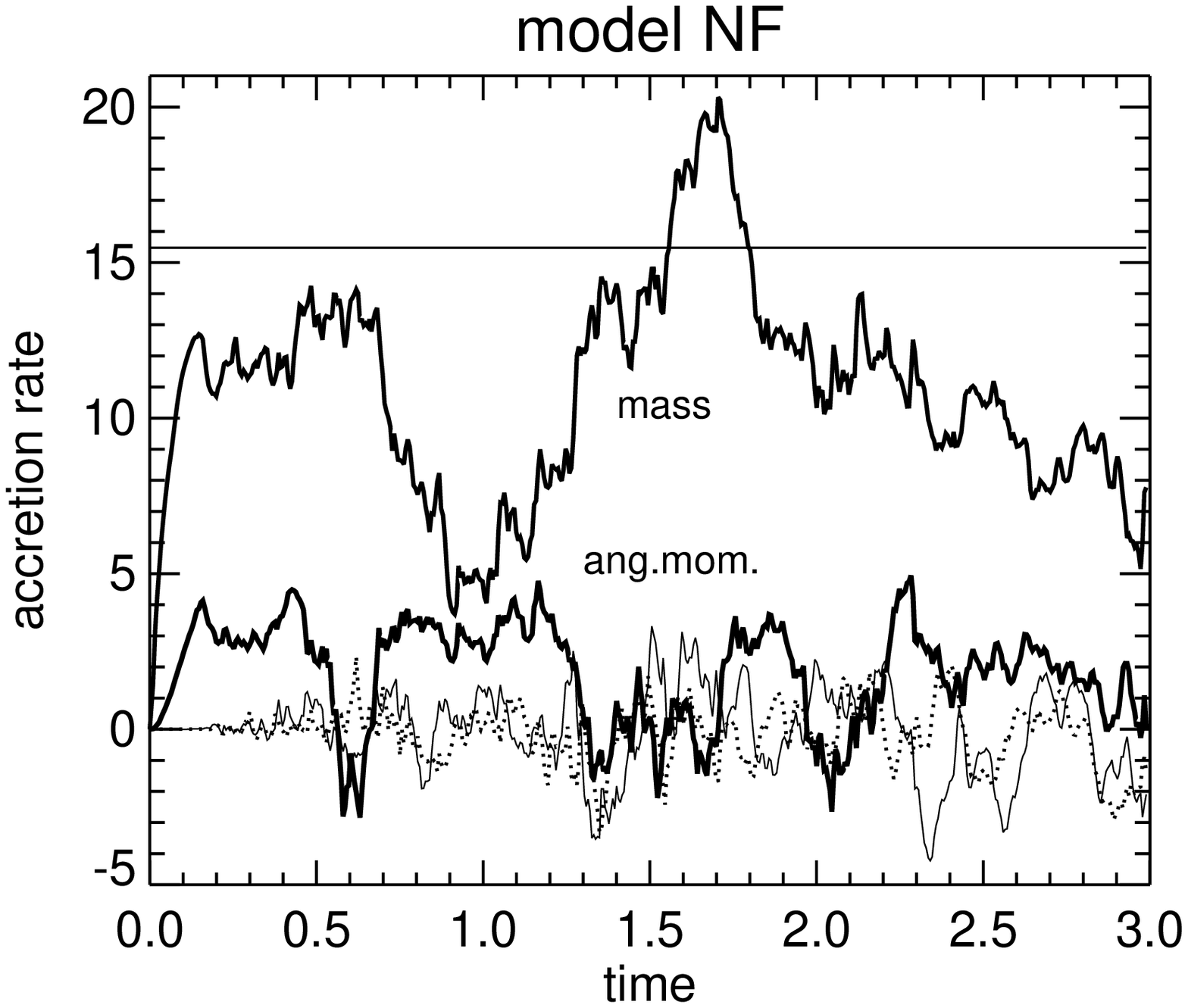} \\
  \epsfxsize=8.8cm  \epsfclipon \epsffile{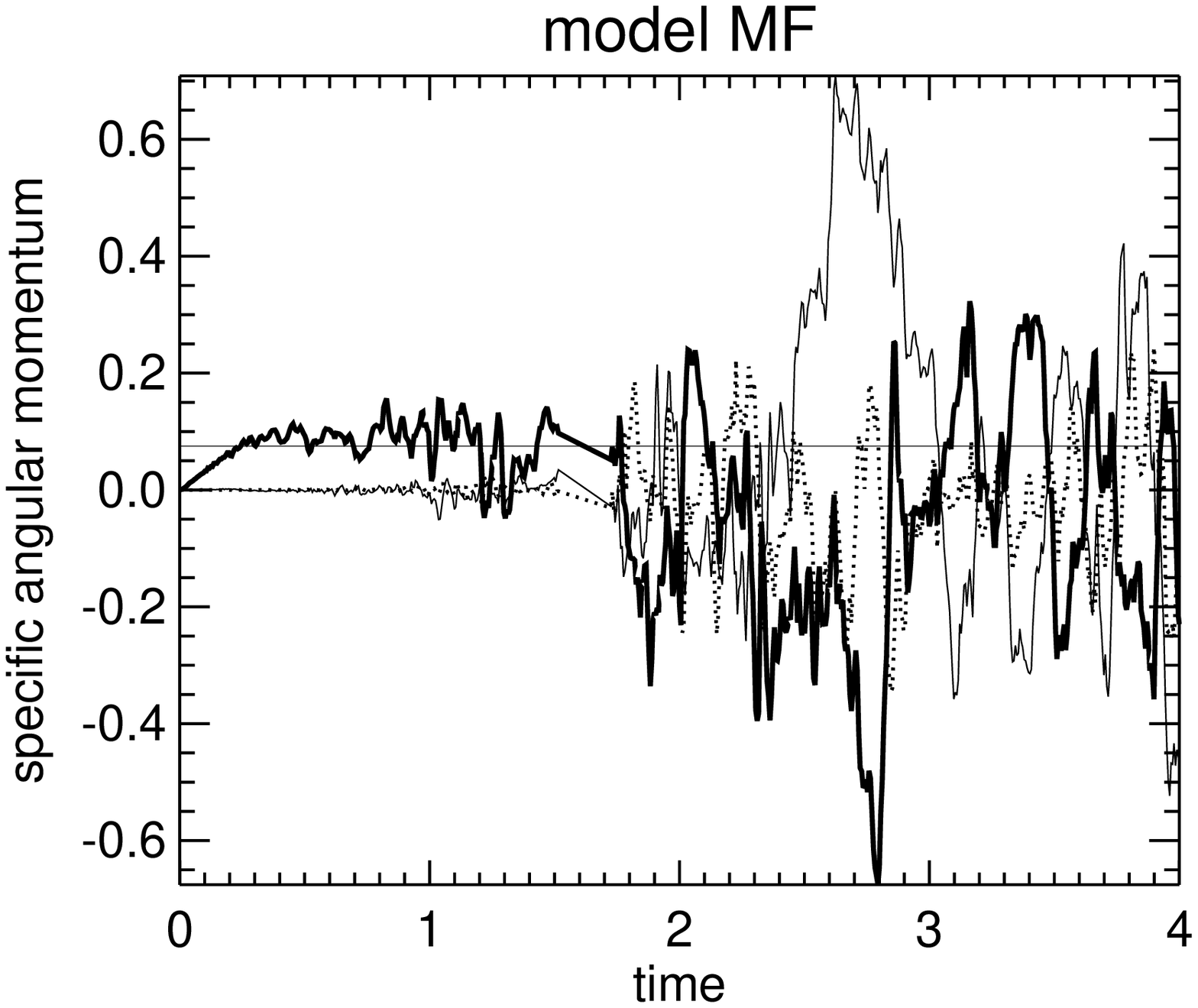} &
  \epsfxsize=8.8cm  \epsfclipon \epsffile{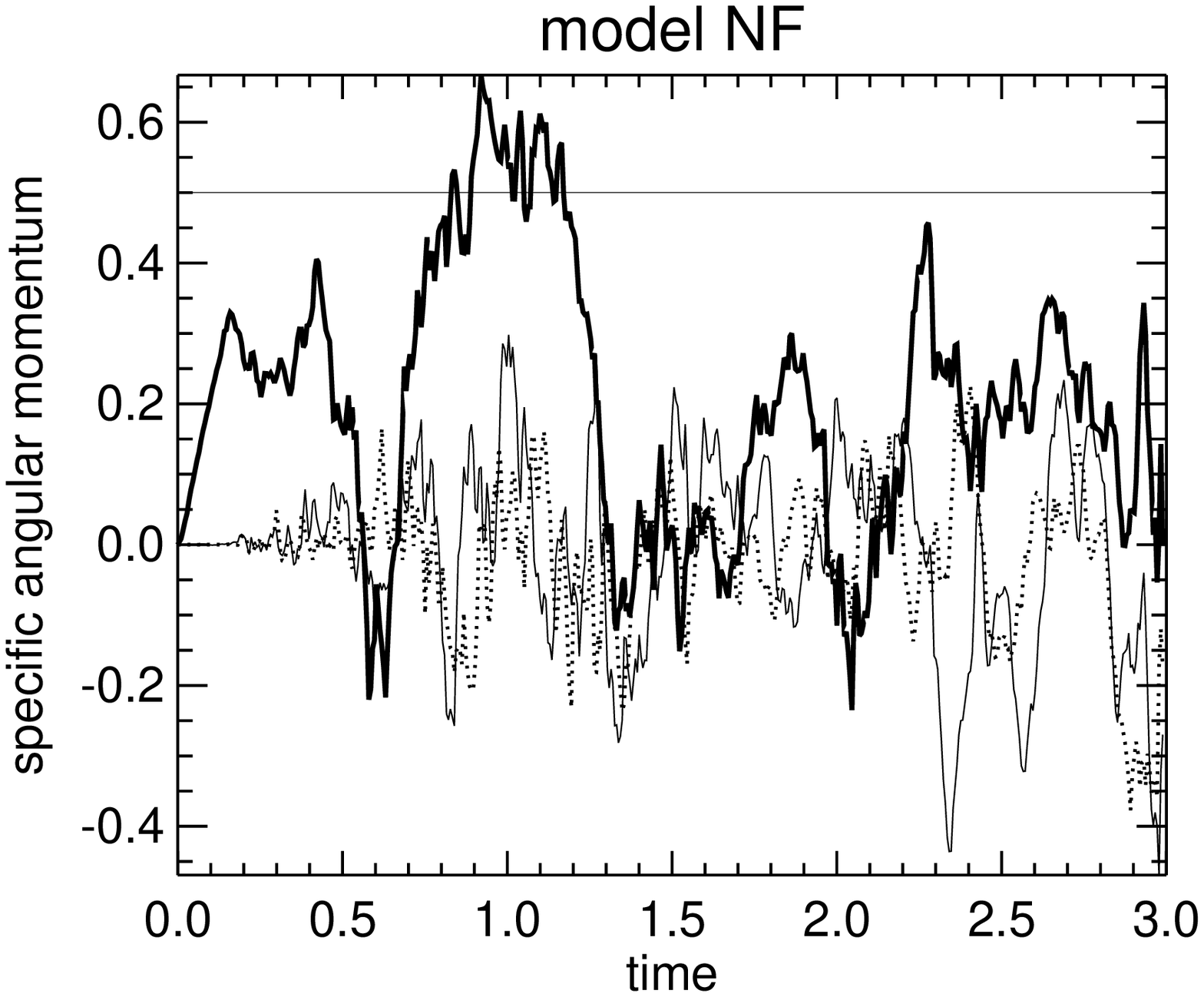}
 \end{tabular}
\caption[]{
The accretion rates of several quantities are plotted as a
function of time for the highly supersonic (${\cal M}_\infty$=10)
models~MF and~NF with an adiabatic index of $\gamma=5/3$.
The top panels contain the mass and angular momentum accretion rates,
the bottom panels the specific angular momentum of the matter
that is accreted.
In the top panels, the straight horizontal lines show the analytical
mass accretion rates: half the value of
the Bondi-Hoyle approximation formula (Eq.~(3) in
Ruffert~1994; Bondi~1952).
The upper solid bold curve represents the
numerically calculated mass accretion rate.
The lower three curves of the top panels trace the x~(dotted),
y~(thin solid) and z~(bold solid) component of the angular momentum
accretion rate.
The same components apply to the bottom panels;
the horizontal line shows the
specific angular momentum value as given by Eq.~(\ref{eq:specmomang}).
}
\label{fig:valueMF}
\end{figure*}

\begin{figure*}
 \tabcolsep = 0mm
 \begin{tabular}{cc}
  \epsfxsize=8.8cm  \epsfclipon \epsffile{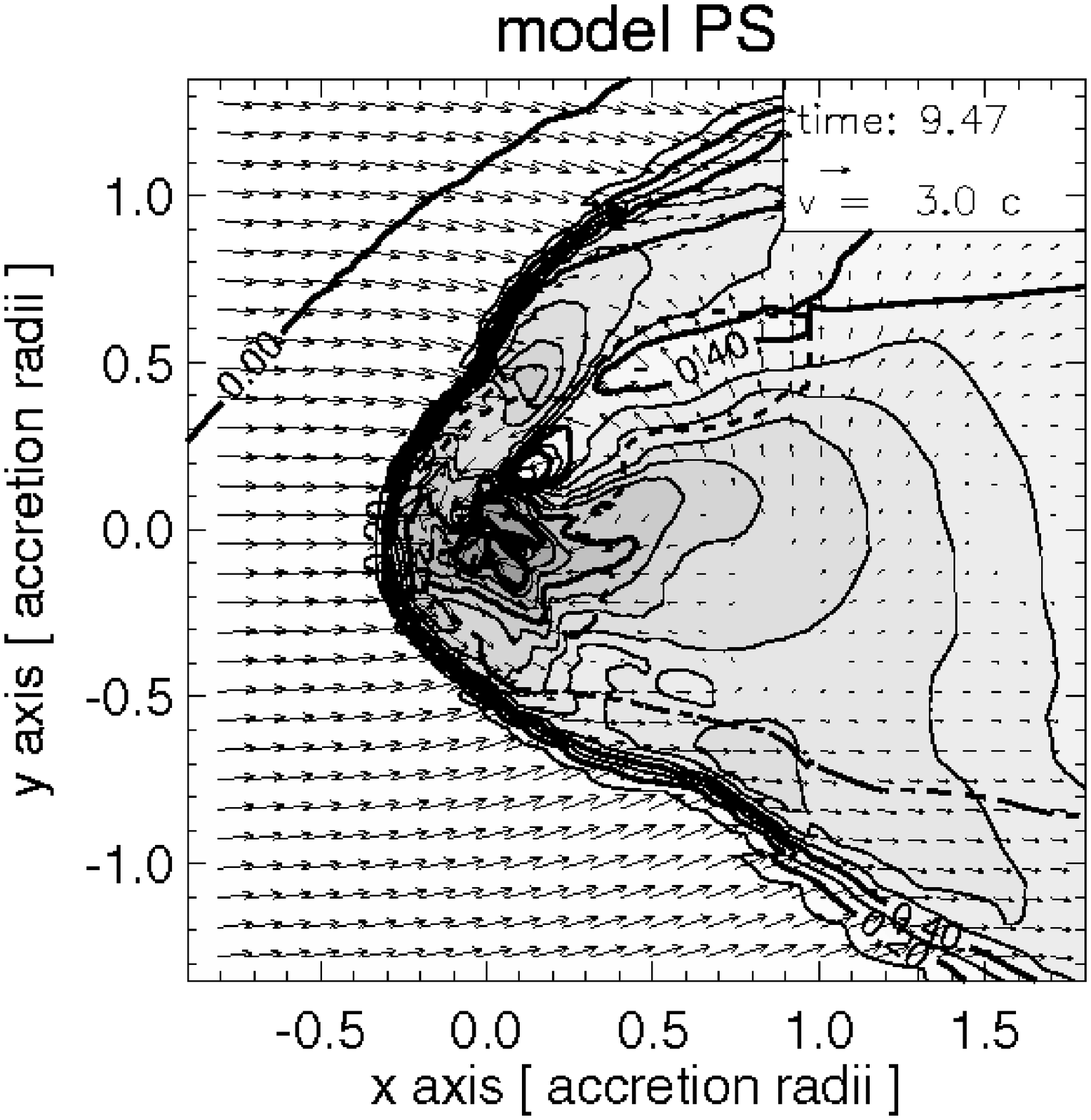} &
  \epsfxsize=8.8cm  \epsfclipon \epsffile{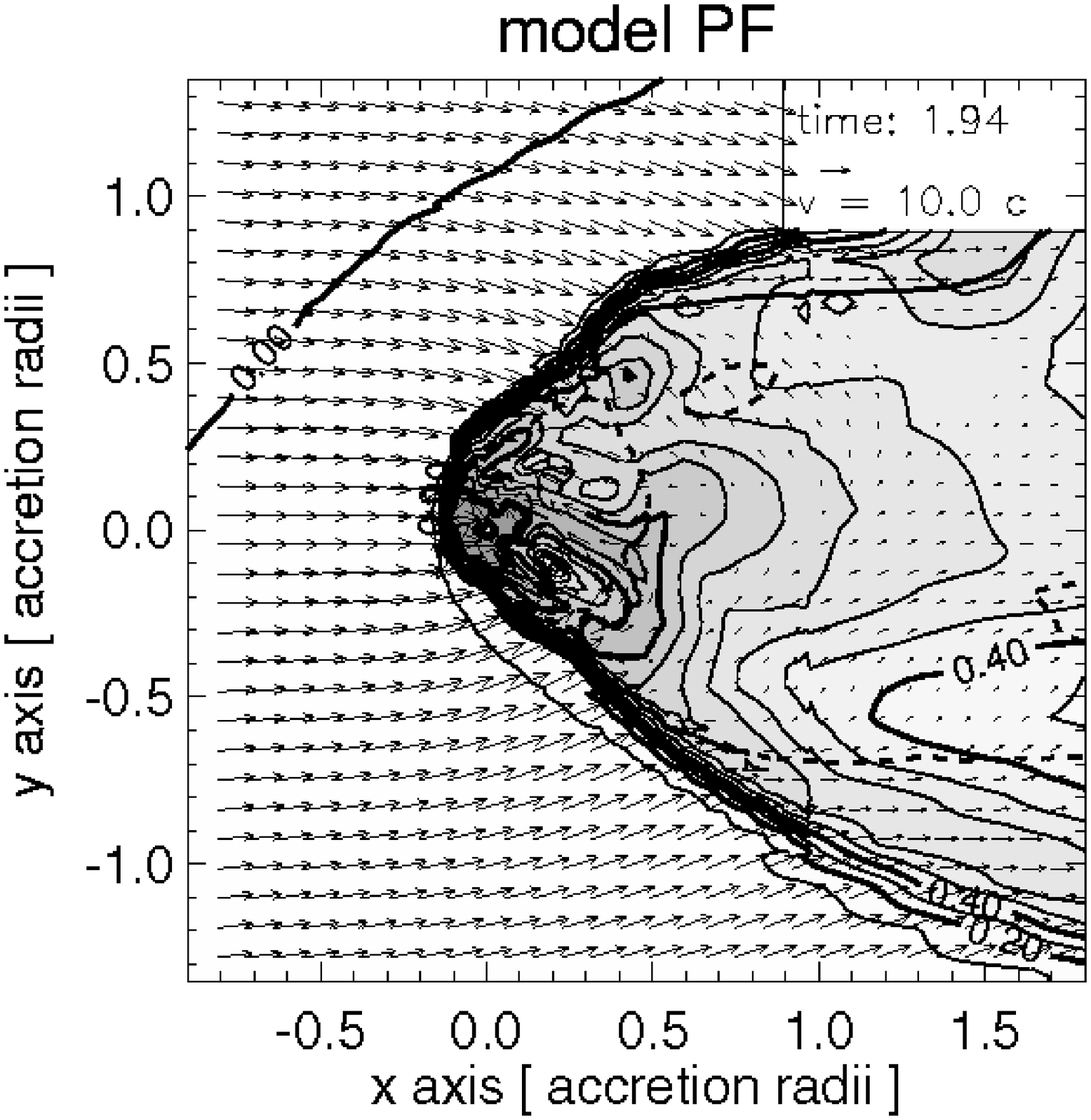} \\
  \epsfxsize=8.8cm  \epsfclipon \epsffile{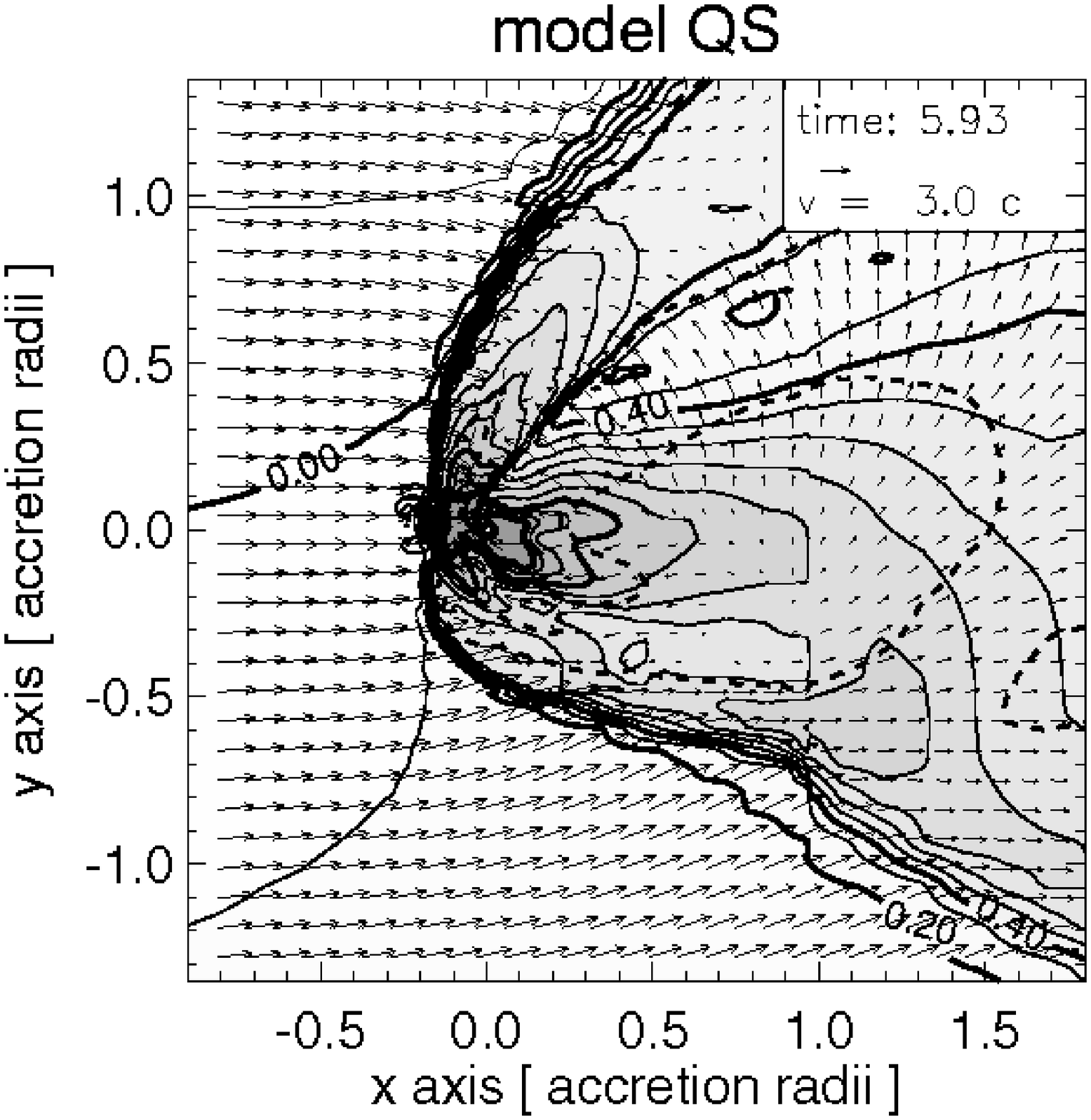} &
  \epsfxsize=8.8cm  \epsfclipon \epsffile{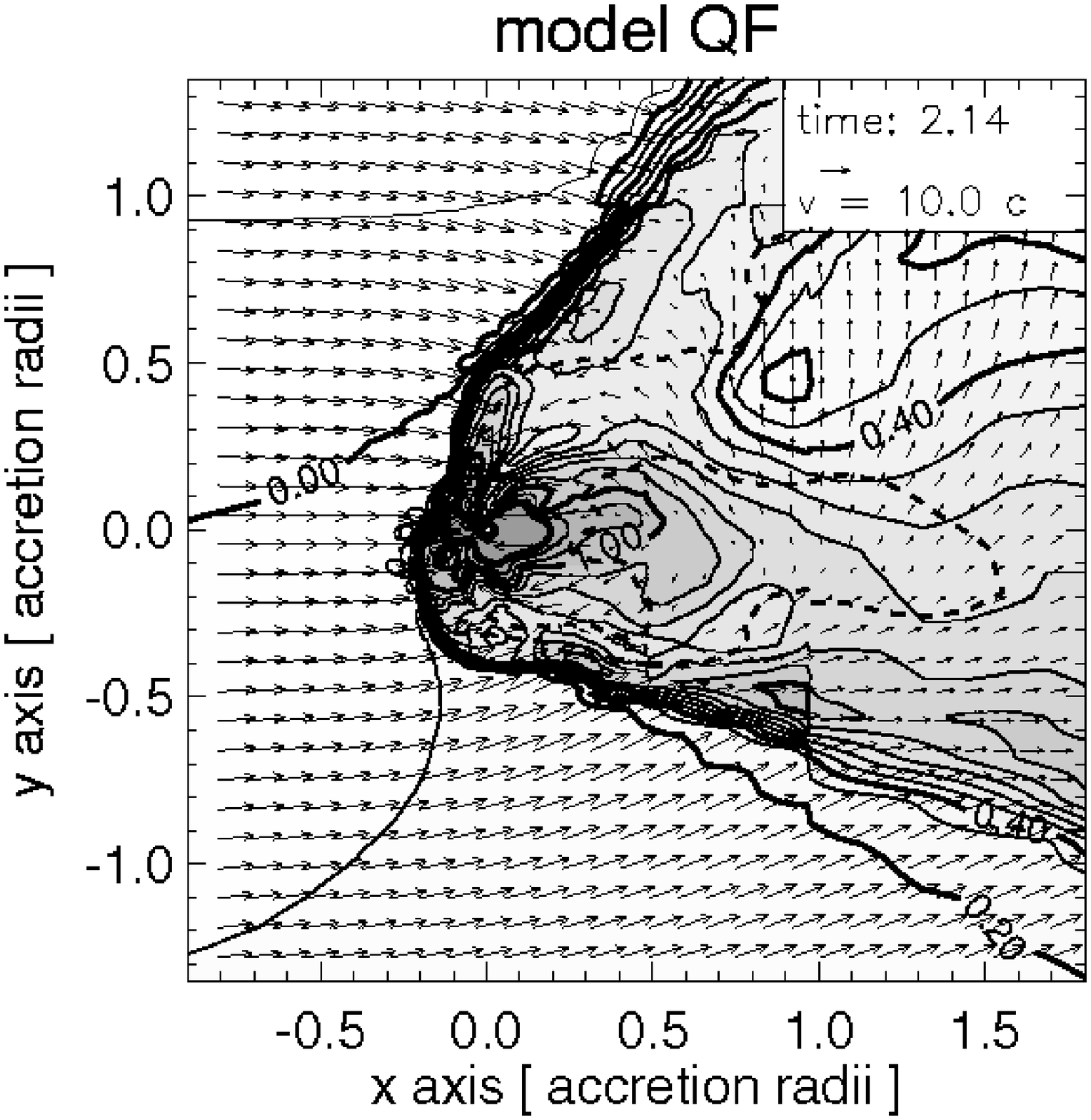}
 \end{tabular}
\caption[]{Contour plots showing snapshots of the density together
with the flow pattern in a plane containing the centre of the accretor
for all models with an adiabatic index of 4/3.
The contour lines are spaced logarithmically in intervals of 0.1~dex.
The bold contour levels are labeled with their respective values 
(0.0 or~1.0).
Darker shades of gray indicate higher densities.
The dashed contour delimits supersonic from subsonic regions.
The time of the snapshot together with the velocity scale is given in
the legend in the upper right hand corner of each panel.
}
\label{fig:Pdens}
\end{figure*}

\begin{figure*}
 \tabcolsep = 0mm
 \begin{tabular}{cc}
  \epsfxsize=8.8cm  \epsfclipon \epsffile{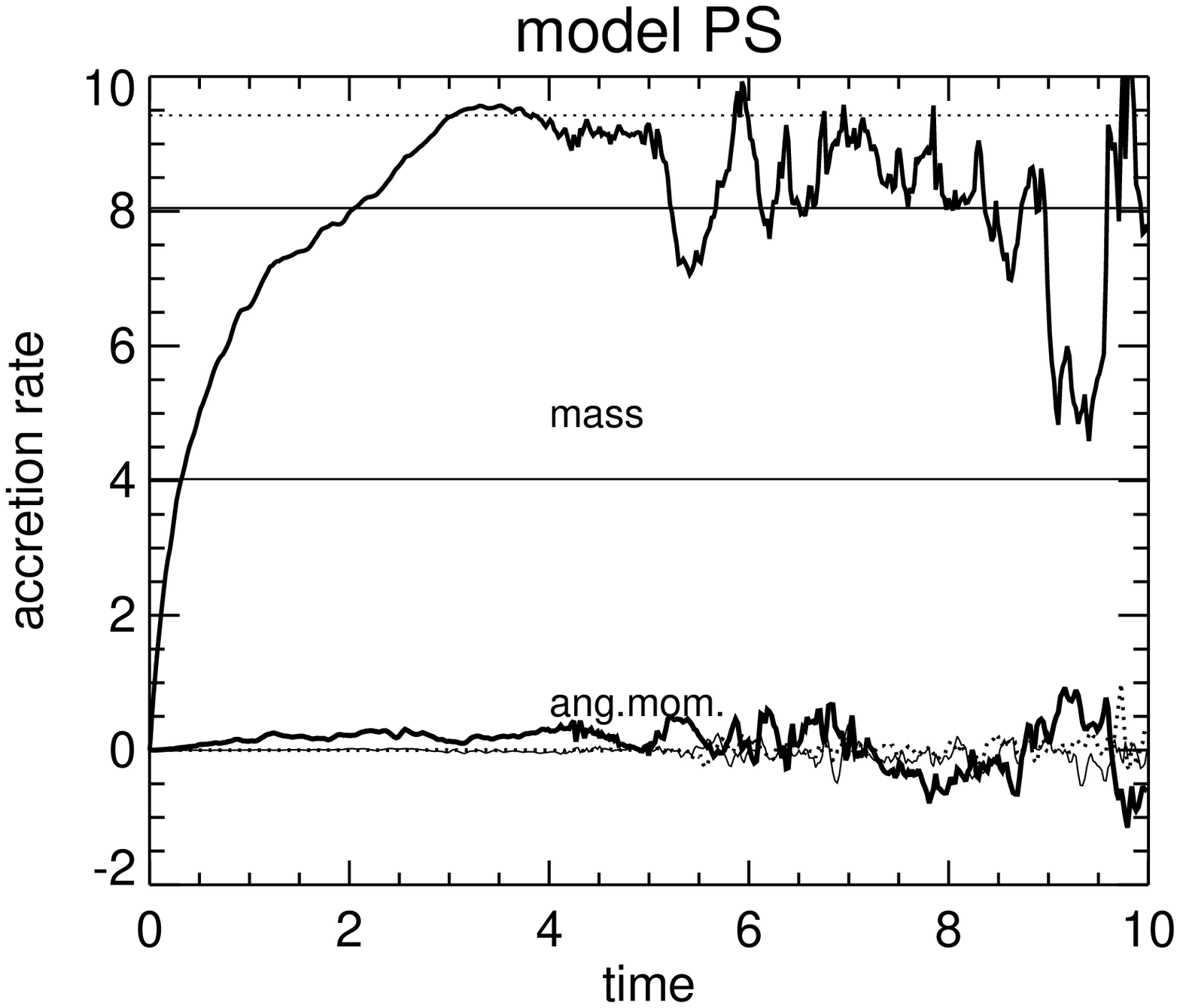} &
  \epsfxsize=8.8cm  \epsfclipon \epsffile{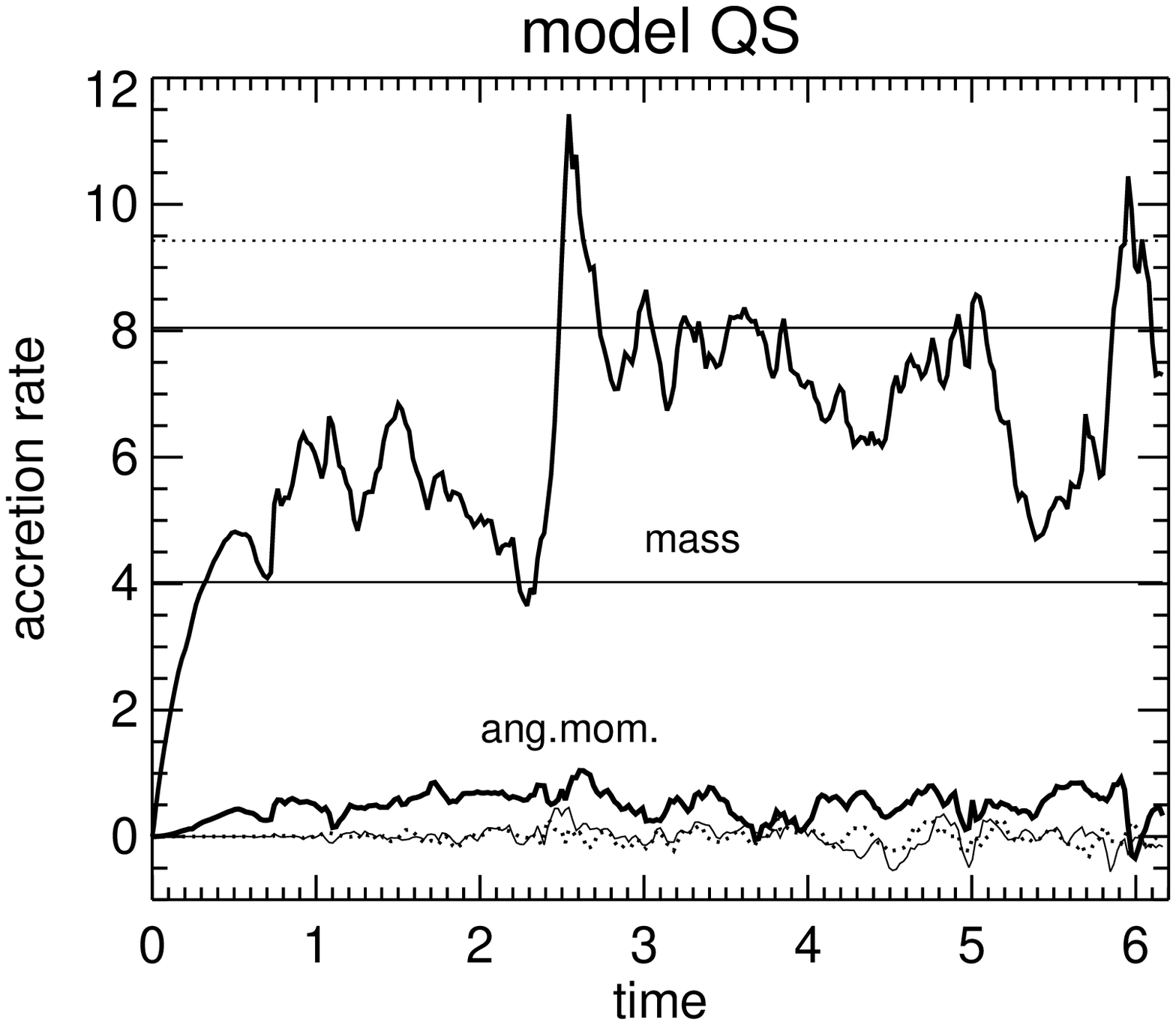} \\
  \epsfxsize=8.8cm  \epsfclipon \epsffile{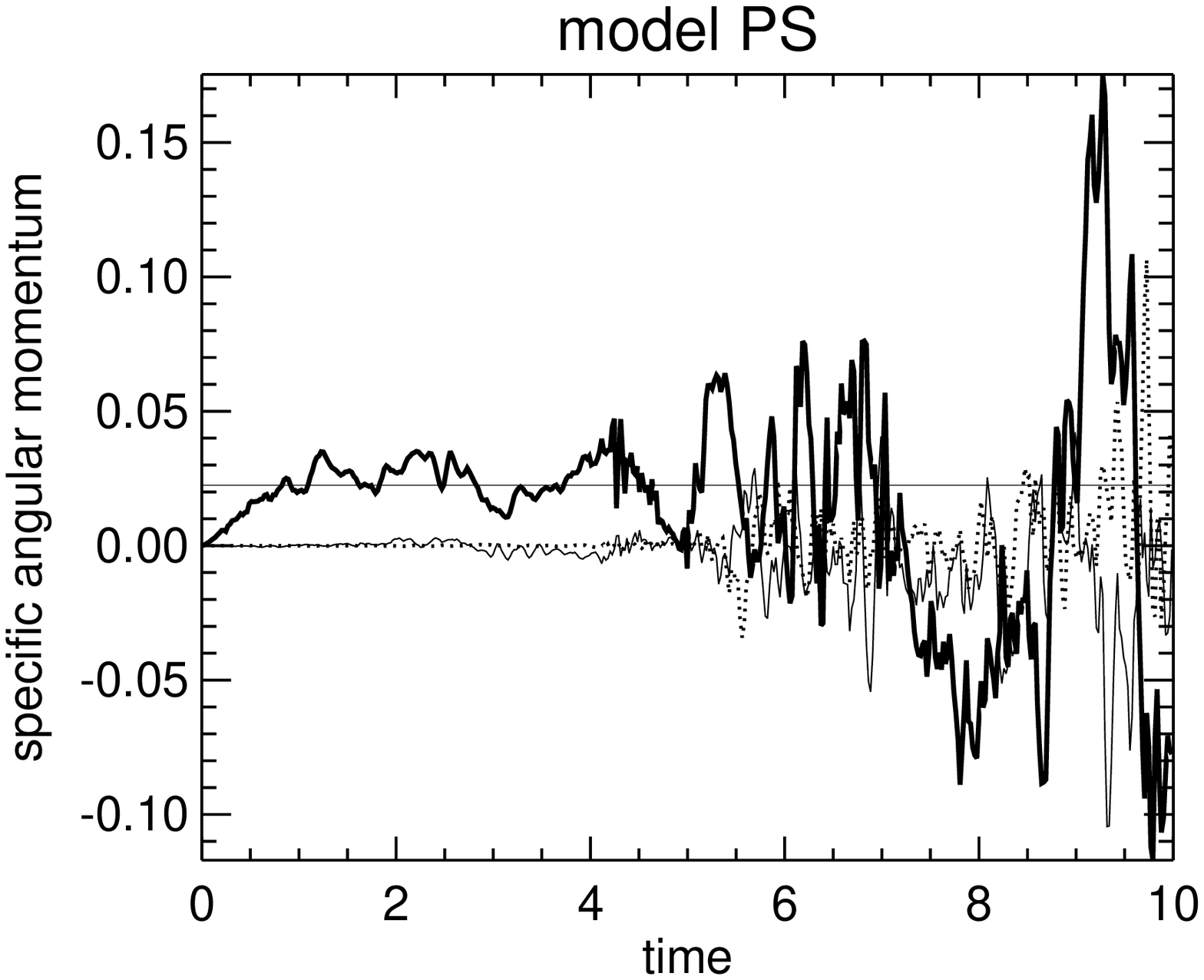} &
  \epsfxsize=8.8cm  \epsfclipon \epsffile{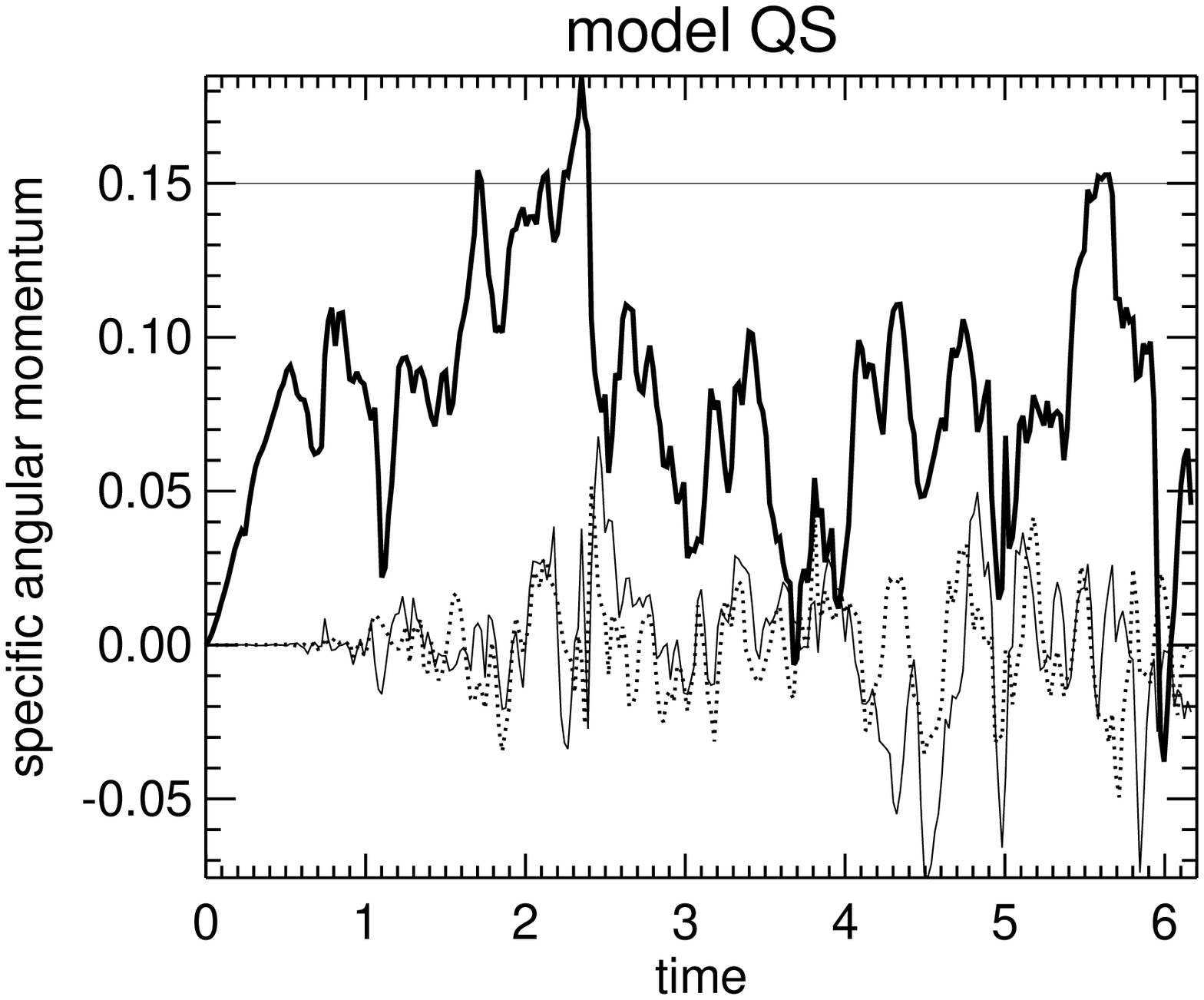}
 \end{tabular}
\caption[]{
The accretion rates of several quantities are plotted as a
function of time for the moderately supersonic (${\cal M}_\infty$=3)
models~PS and~QS with an adiabatic index of $\gamma=4/3$.
The top panels contain the mass and angular momentum accretion rates,
the bottom panels the specific angular momentum of the matter
that is accreted.
In the top panels, the straight horizontal lines show the analytical
mass accretion rates: dotted is the Hoyle-Lyttleton rate
(Eq.~(1) in Ruffert~1994), 
solid is the Bondi-Hoyle approximation formula (Eq.~(3) in
Ruffert~1994; Bondi~1952) and half that value.
The upper solid bold curve represents the
numerically calculated mass accretion rate.
The lower three curves of the top panels trace the x~(dotted),
y~(thin solid) and z~(bold solid) component of the angular momentum
accretion rate.
The same components apply to the bottom panels;
the horizontal line shows the
specific angular momentum value as given by Eq.~(\ref{eq:specmomang}).
}
\label{fig:valuePS}
\end{figure*}

\begin{figure*}
 \tabcolsep = 0mm
 \begin{tabular}{cc}
  \epsfxsize=8.8cm  \epsfclipon \epsffile{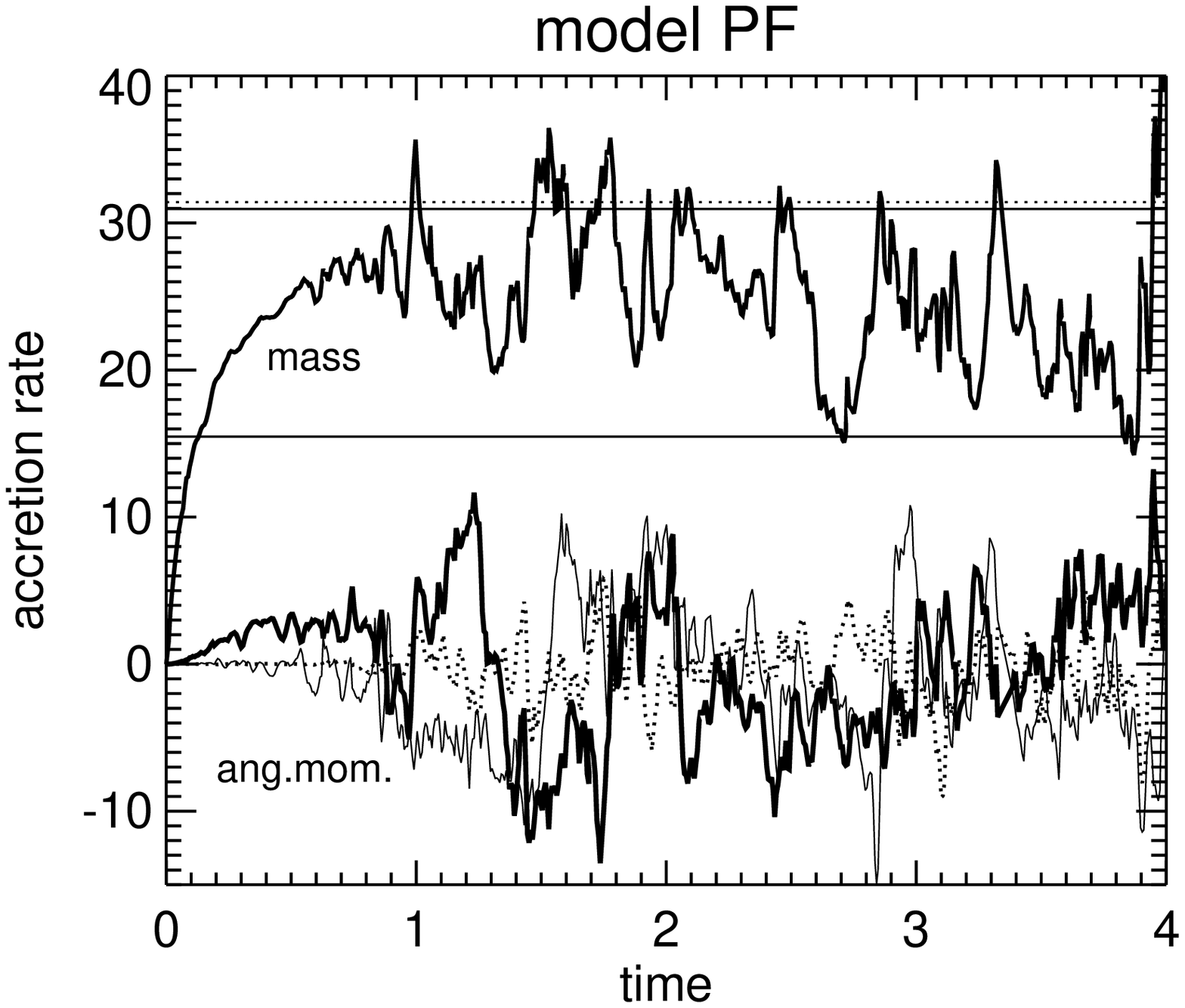} &
  \epsfxsize=8.8cm  \epsfclipon \epsffile{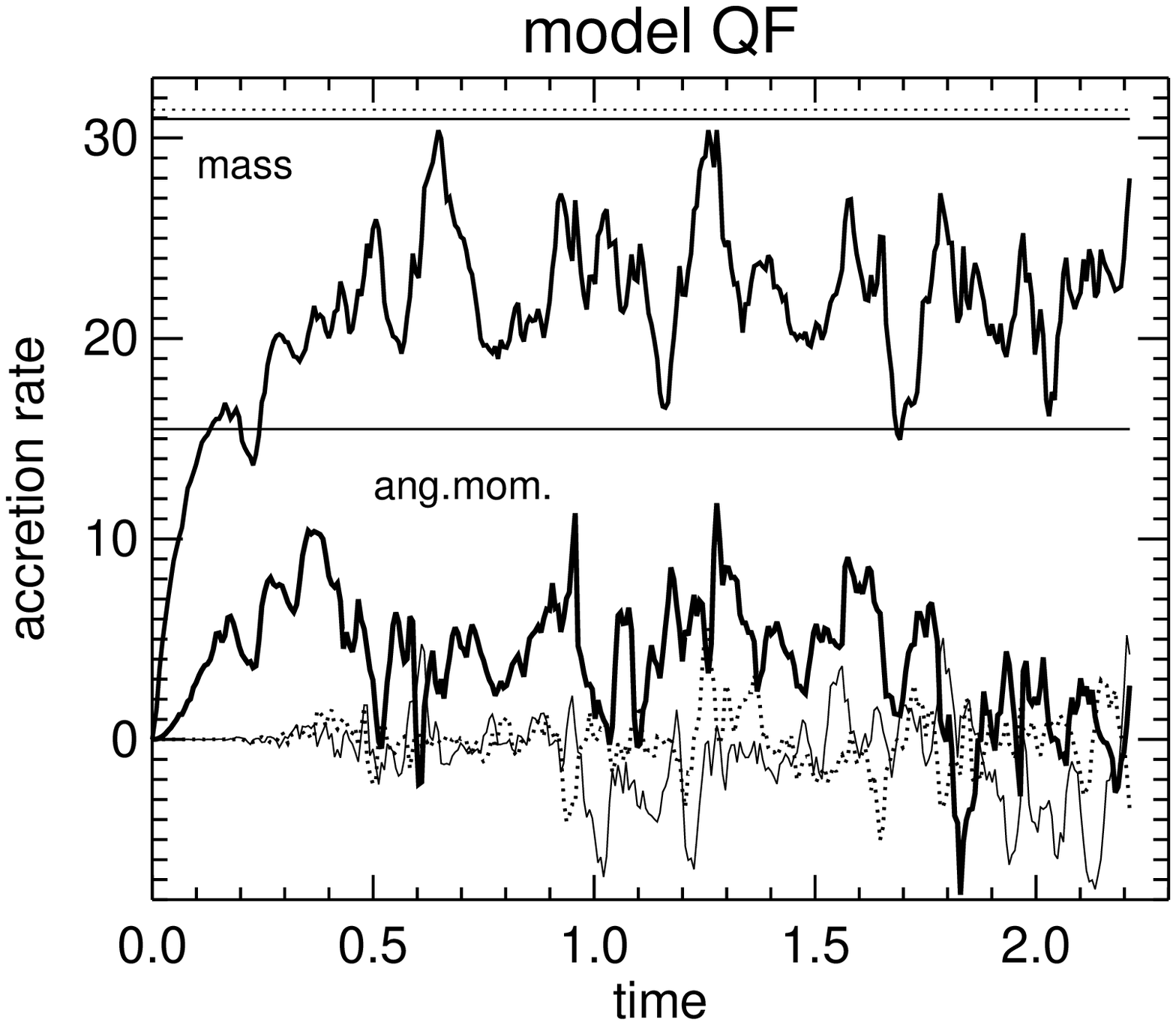} \\
  \epsfxsize=8.8cm  \epsfclipon \epsffile{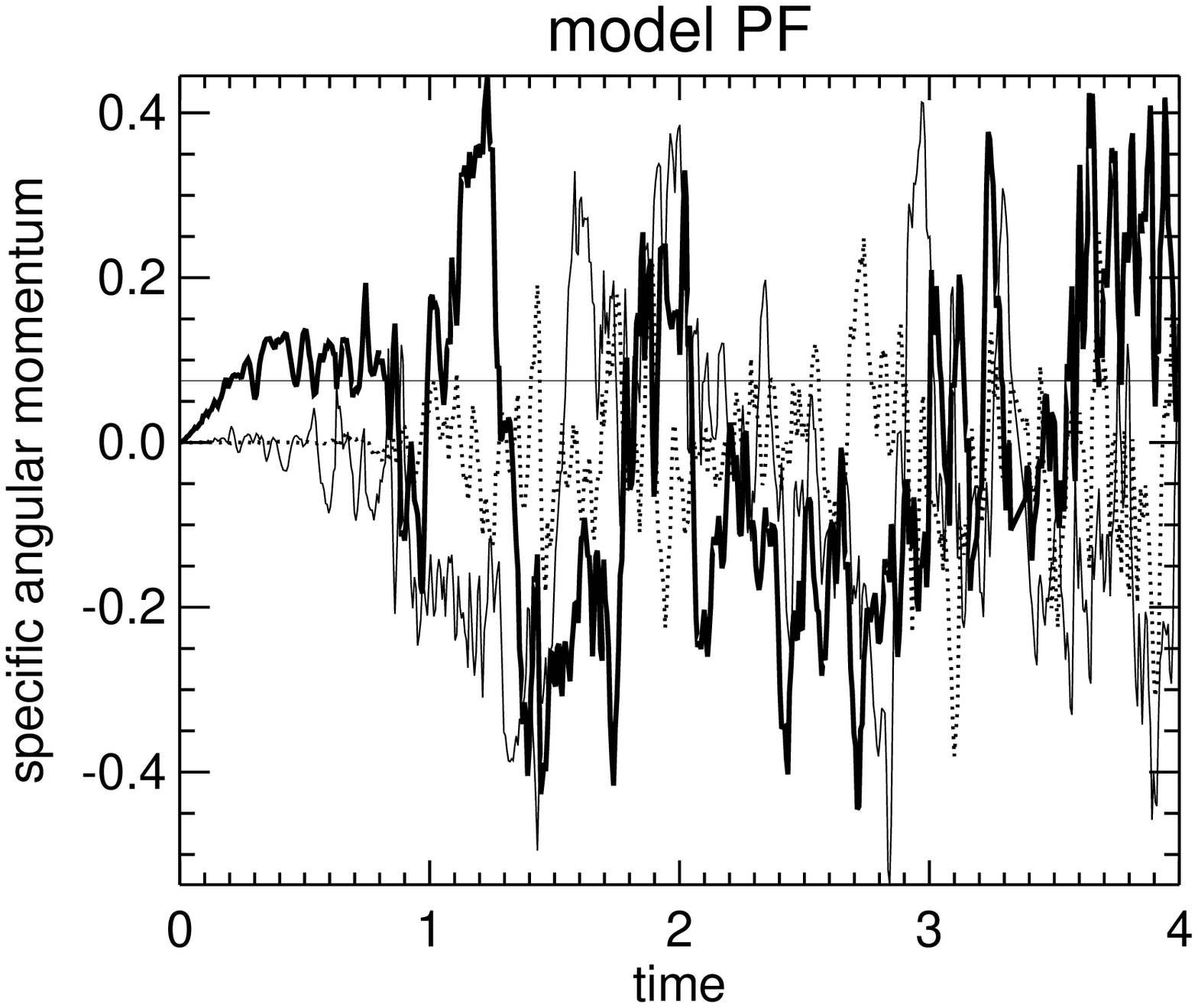} &
  \epsfxsize=8.8cm  \epsfclipon \epsffile{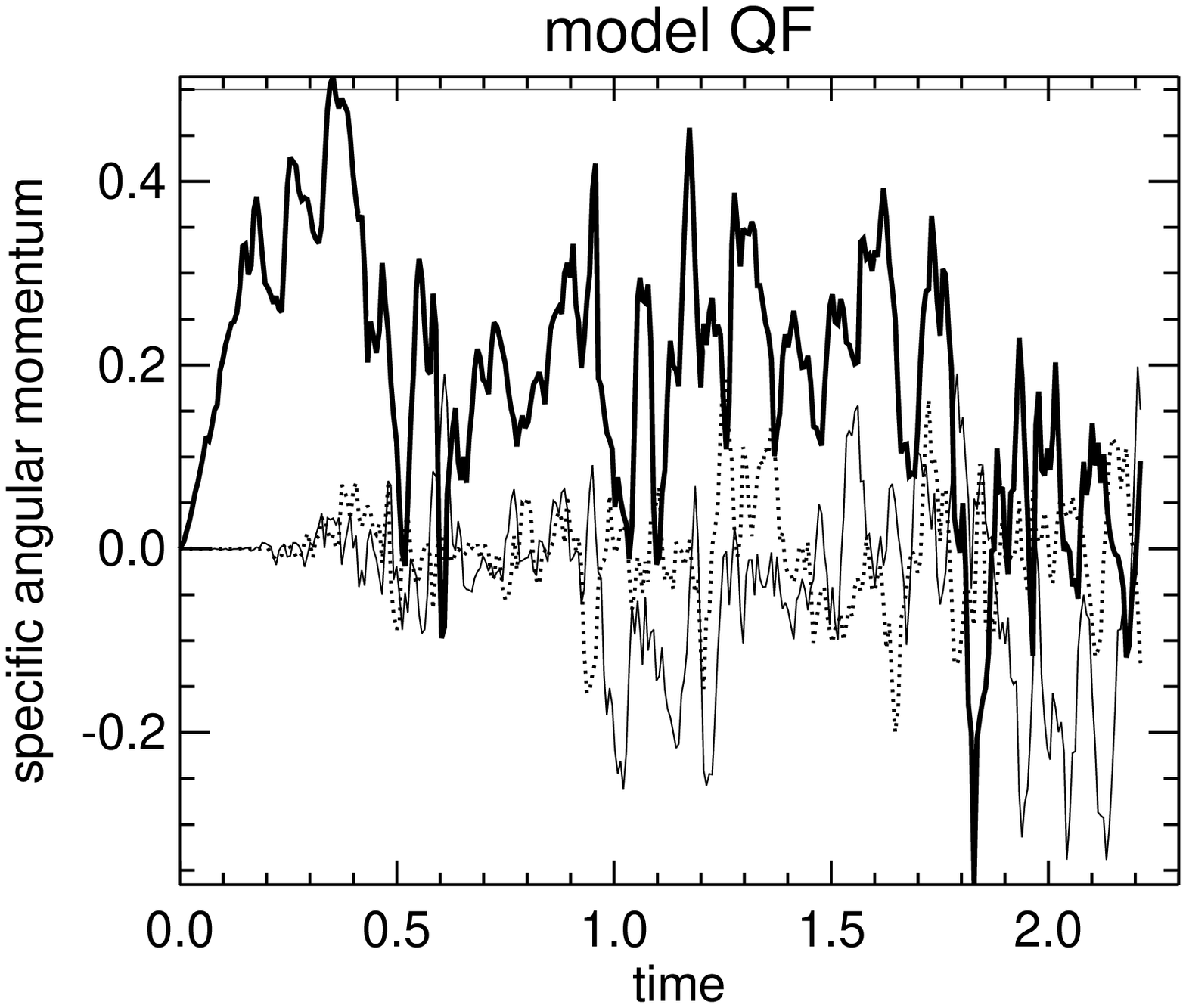}
 \end{tabular}
\caption[]{
The accretion rates of several quantities are plotted as a
function of time for the highly supersonic (${\cal M}_\infty$=10)
models~PF and~QF with an adiabatic index of $\gamma=4/3$.
The top panels contain the mass and angular momentum accretion rates,
the bottom panels the specific angular momentum of the matter
that is accreted.
In the top panels, the straight horizontal lines show the analytical
mass accretion rates: dotted is the Hoyle-Lyttleton rate
(Eq.~(1) in Ruffert~1994), 
solid is the Bondi-Hoyle approximation formula (Eq.~(3) in
Ruffert~1994; Bondi~1952) and half that value.
The upper solid bold curve represents the
numerically calculated mass accretion rate.
The lower three curves of the top panels trace the x~(dotted),
y~(thin solid) and z~(bold solid) component of the angular momentum
accretion rate.
The same components apply to the bottom panels;
the horizontal line shows the
specific angular momentum value as given by Eq.~(\ref{eq:specmomang}).
}
\label{fig:valuePF}
\end{figure*}

\begin{figure*}
 \tabcolsep = 0mm
 \begin{tabular}{cc}
  \epsfxsize=8.8cm  \epsfclipon \epsffile{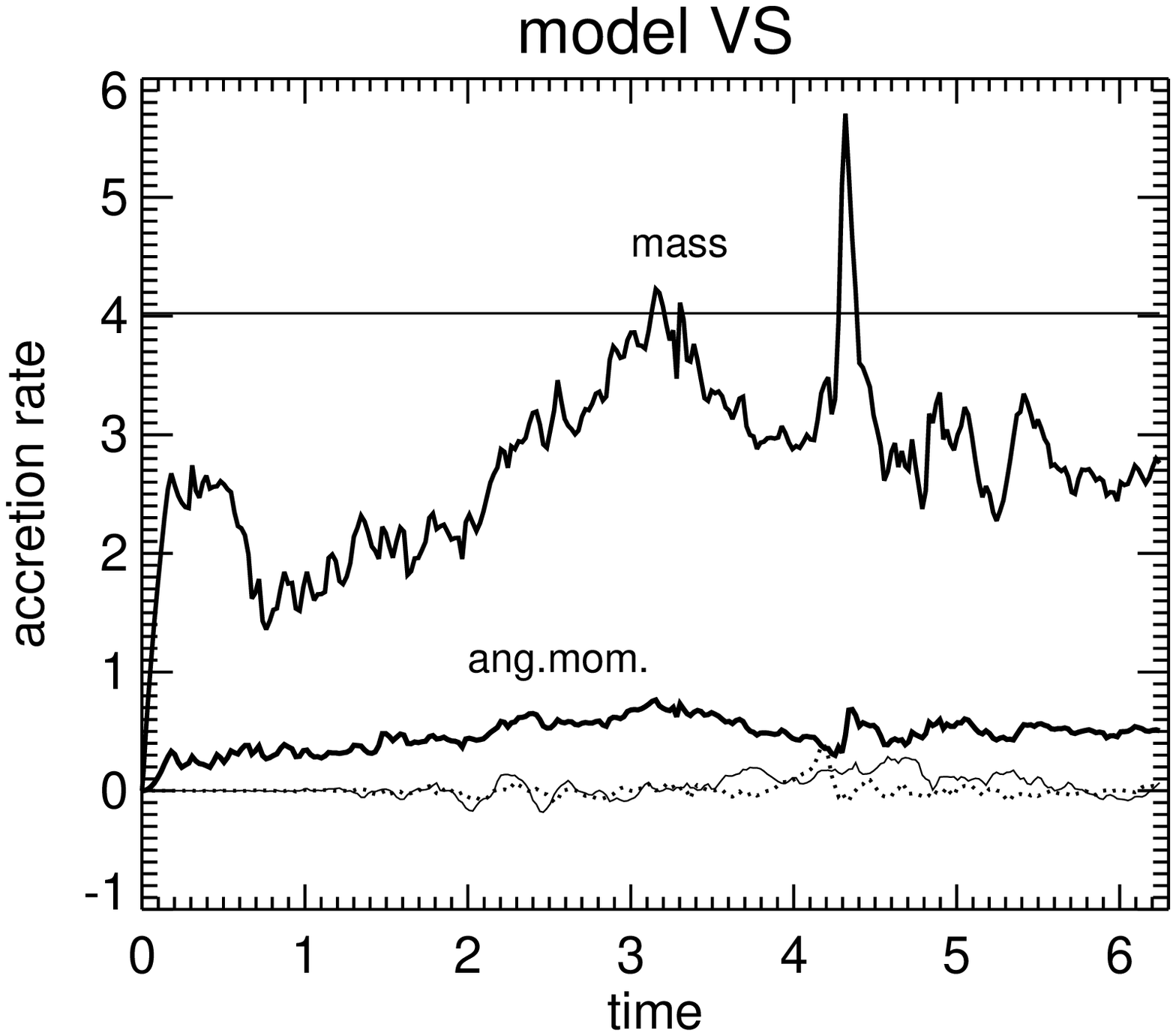} &
  \epsfxsize=8.8cm  \epsfclipon \epsffile{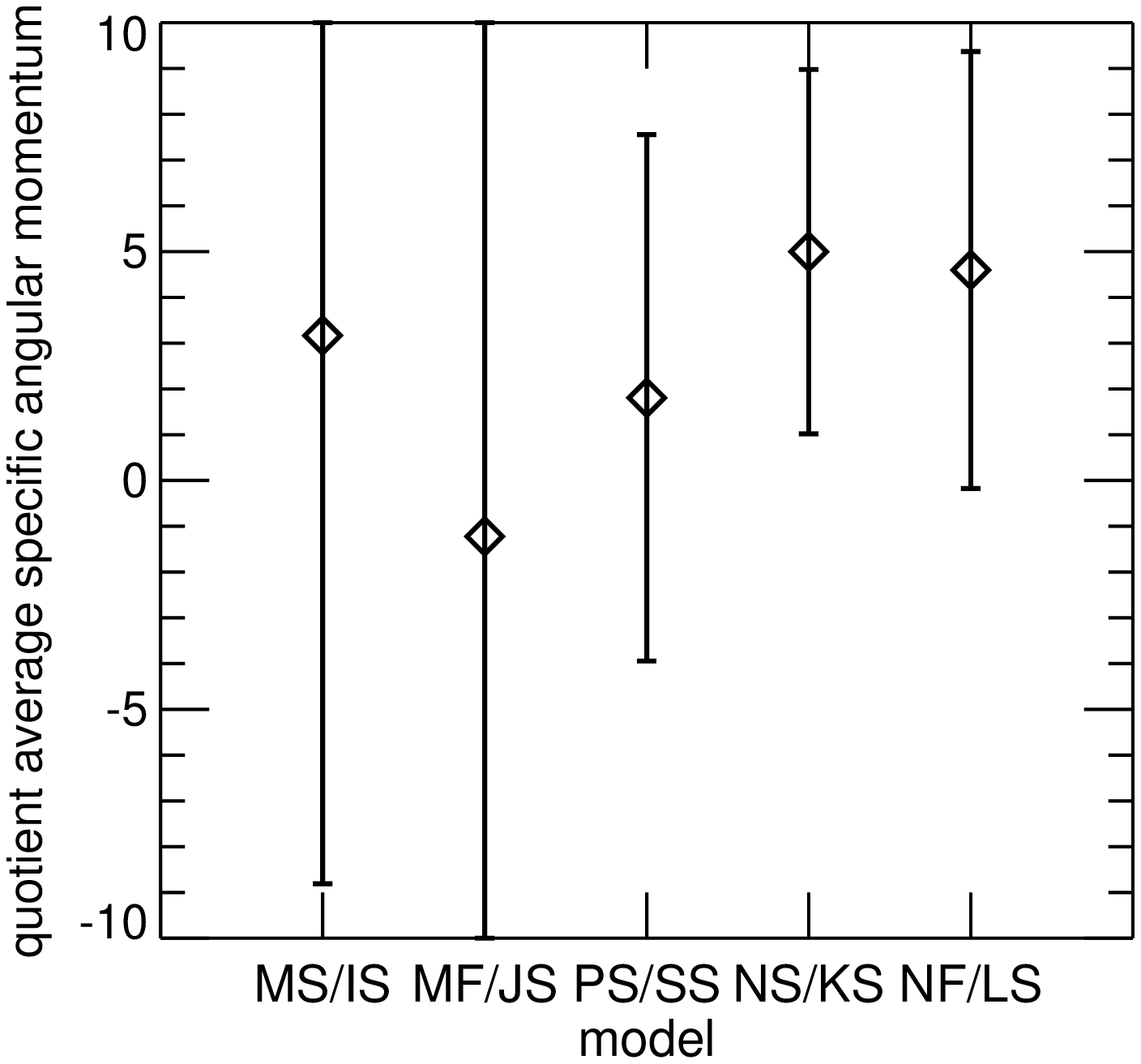} \\
  \epsfxsize=8.8cm  \epsfclipon \epsffile{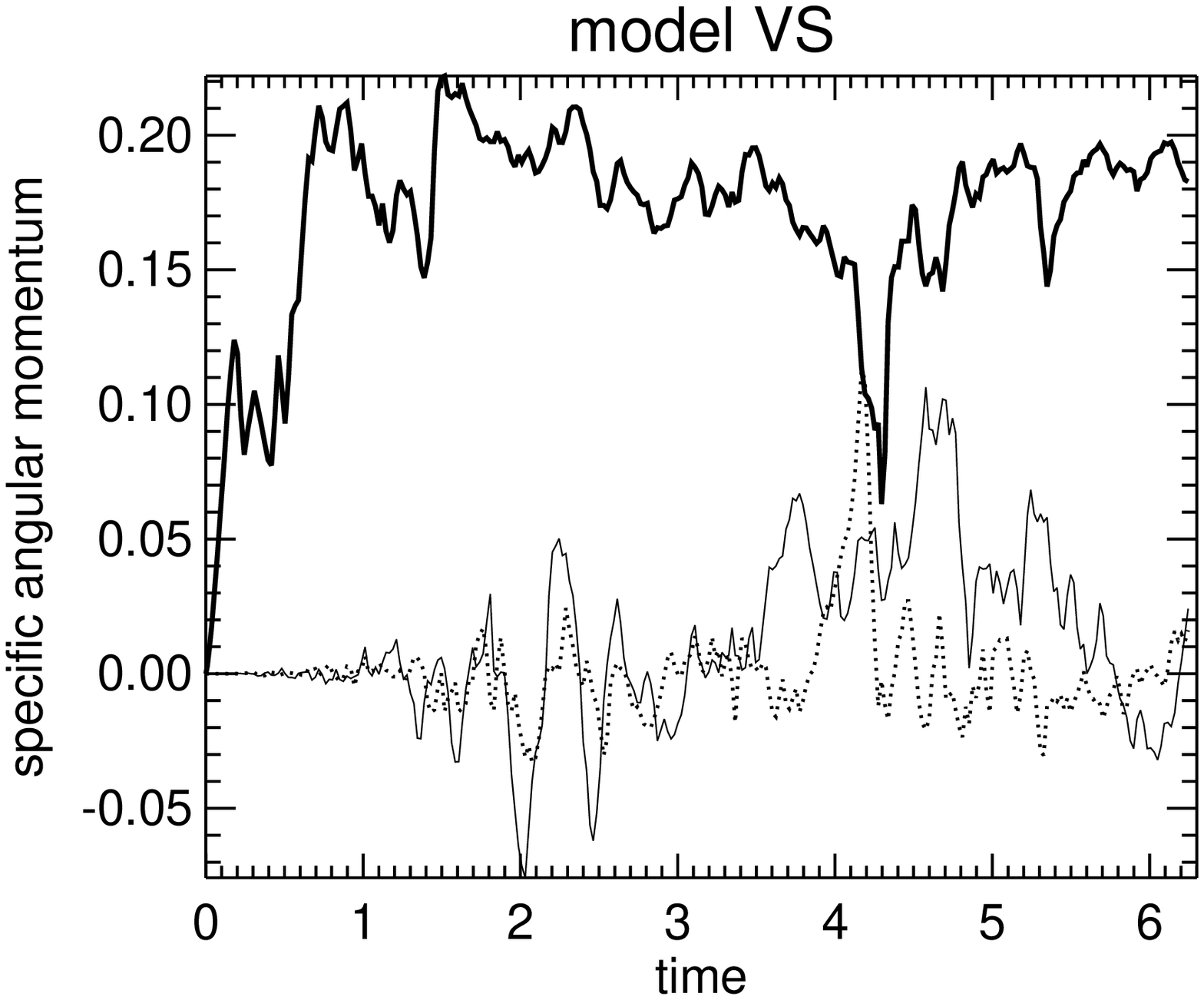} &
\raisebox{7cm}{\parbox[t]{8.6cm}{
\caption[]{\label{fig:valueVS}
{\bf Left panels:}
The accretion rates of several quantities are plotted as a
function of time for the model with largest gradient 
($\varepsilon_\rho=1.0$).
The top panel contains the mass and angular momentum accretion rates,
the bottom panels the specific angular momentum of the matter
that is accreted.
In the top panel, the straight horizontal line shows the analytical
half the Bondi-Hoyle value from the approximation formula (Eq.~(3) in
Ruffert~1994; Bondi~1952).
The upper solid bold curve represents the
numerically calculated mass accretion rate.
The lower three curves of the top panel trace the x~(dotted),
y~(thin solid) and z~(bold solid) component of the angular momentum
accretion rate.
The same components apply to the bottom panel.
{\bf Right panel:}
Comparison of the ratio of specific angular momenta accreted for
models with density gradients (MS, MF, PS, NS, NF) to
models with velocity gradients (IS, JS, SS, KS, LS taken from paper
R1), keeping all other parameters equal. }}}
\end{tabular}
\end{figure*}

\begin{figure*}
 \tabcolsep = 0mm
 \begin{tabular}{cc}
  \epsfxsize=8.8cm  \epsfclipon \epsffile{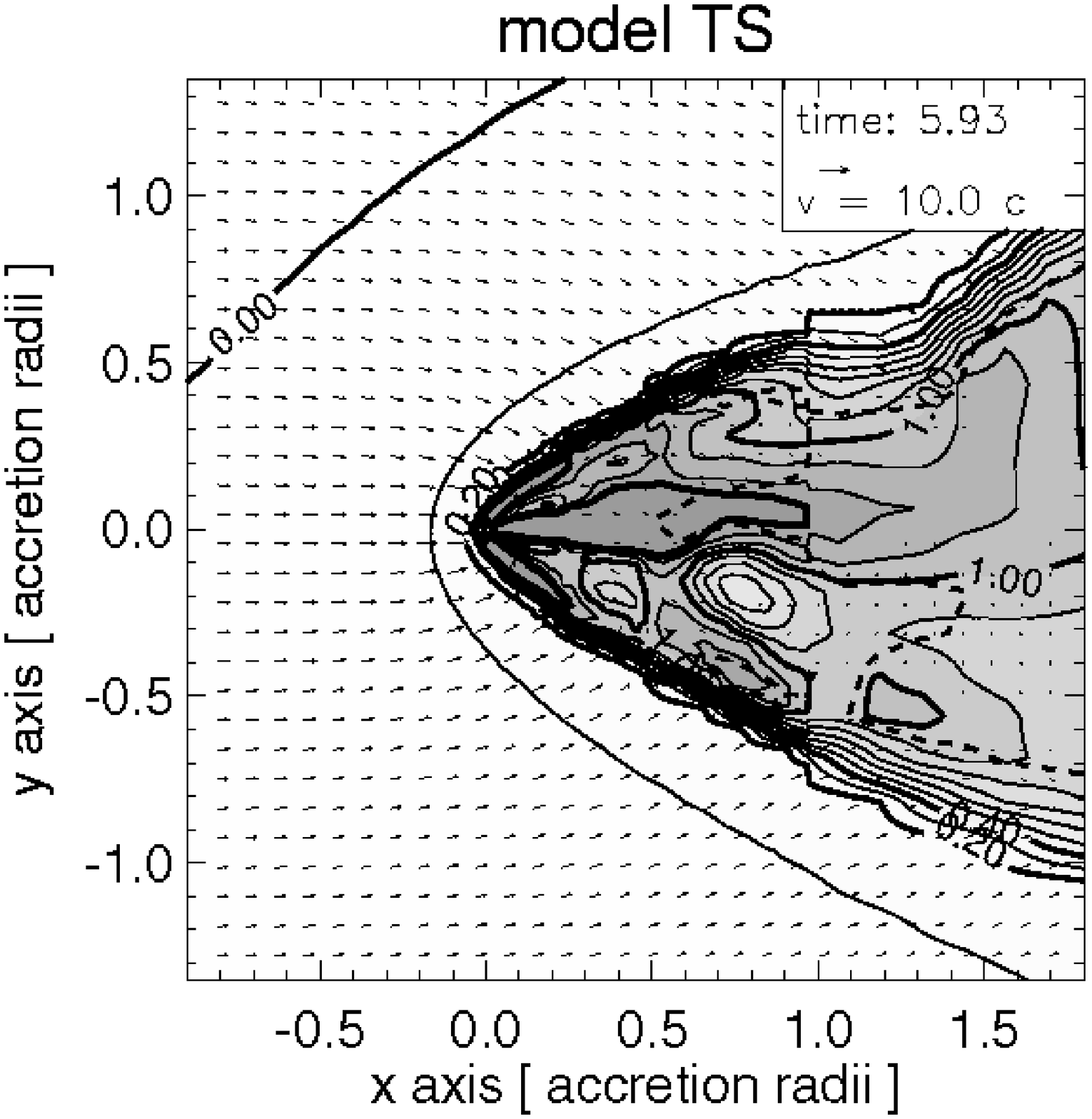} &
  \epsfxsize=8.8cm  \epsfclipon \epsffile{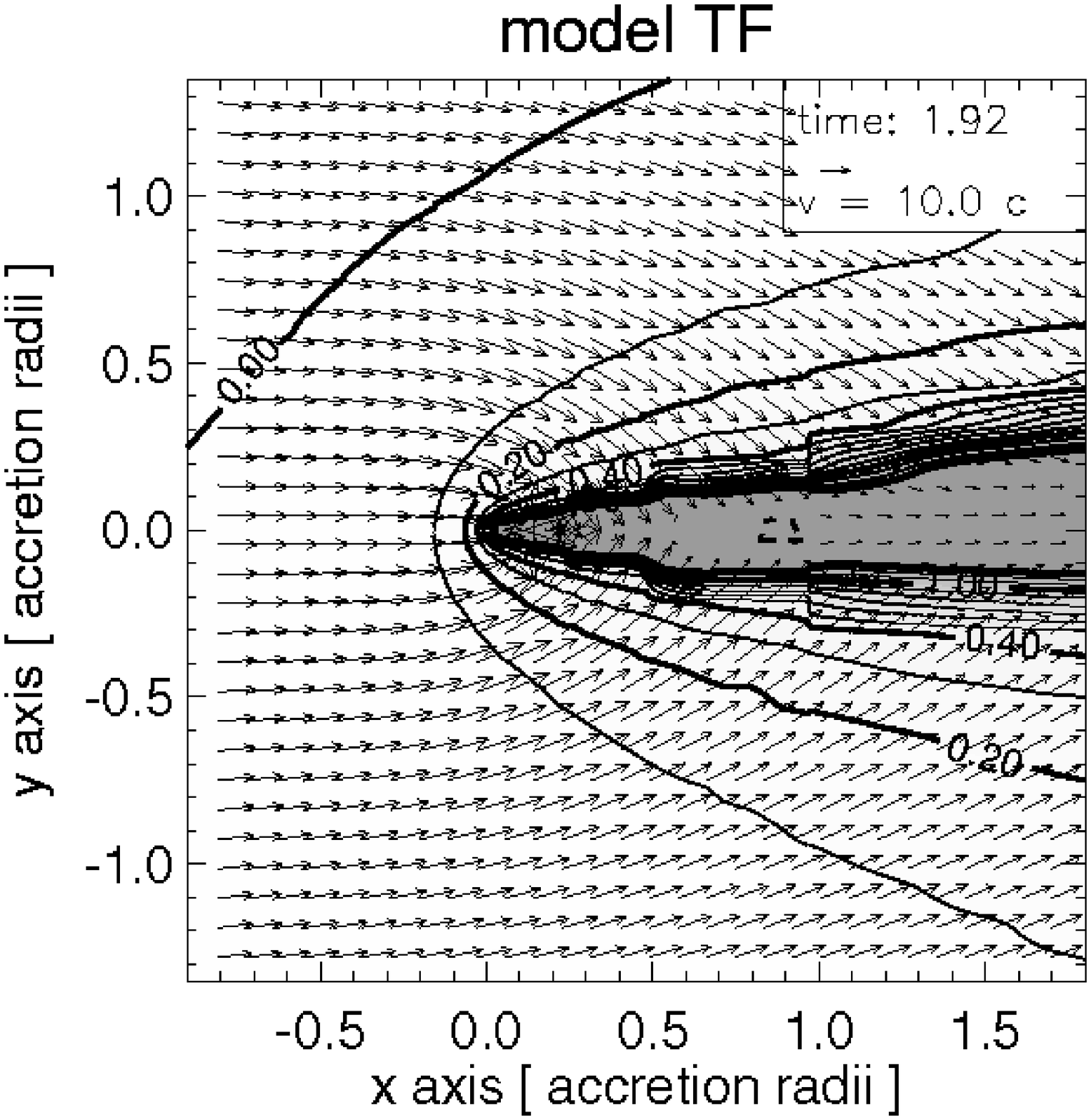} \\
  \epsfxsize=8.8cm  \epsfclipon \epsffile{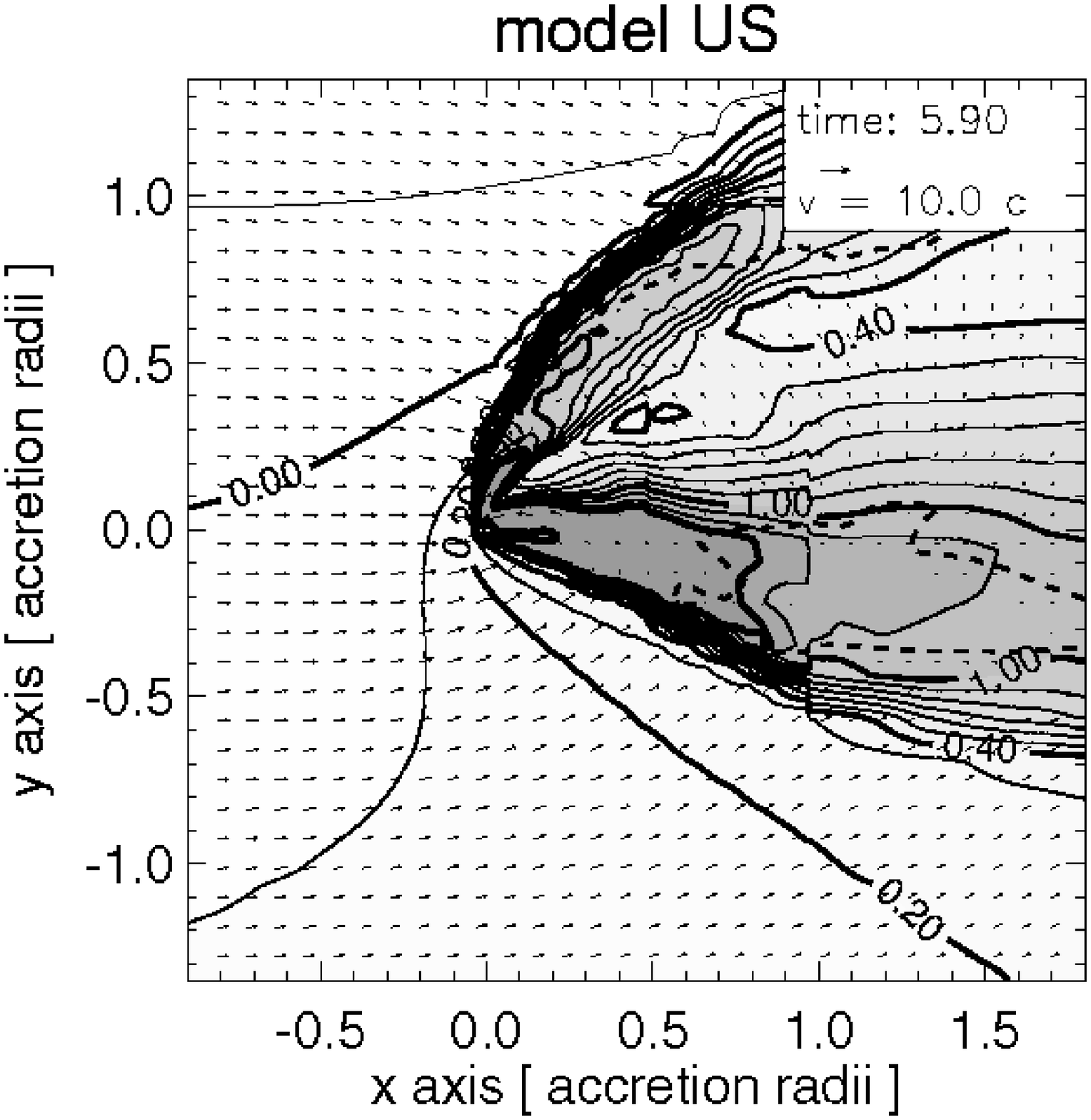} &
  \epsfxsize=8.8cm  \epsfclipon \epsffile{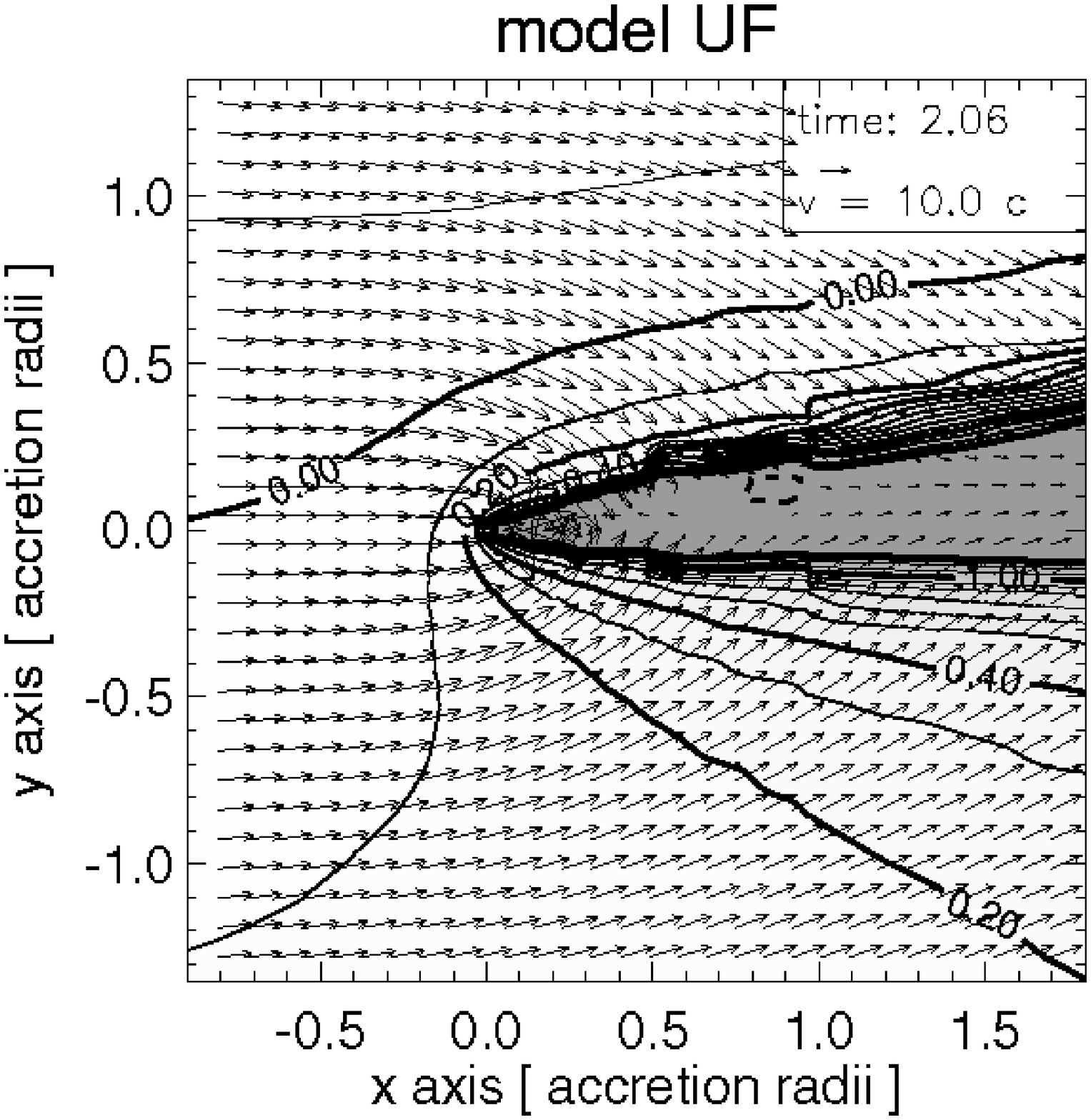}
 \end{tabular}
\caption[]{Contour plots showing snapshots of the density together
with the flow pattern in a plane containing the centre of the accretor
for all models with an adiabatic index of 1.01
The contour lines are spaced logarithmically in intervals of 0.1~dex.
The bold contour levels are labeled with their respective values 
(0.0 or~1.0).
Darker shades of gray indicate higher densities.
The dashed contour delimits supersonic from subsonic regions.
The time of the snapshot together with the velocity scale is given in
the legend in the upper right hand corner of each panel.
}
\label{fig:Tdens}
\end{figure*}

\begin{figure*}
 \tabcolsep = 0mm
 \begin{tabular}{cc}
  \epsfxsize=8.8cm  \epsfclipon \epsffile{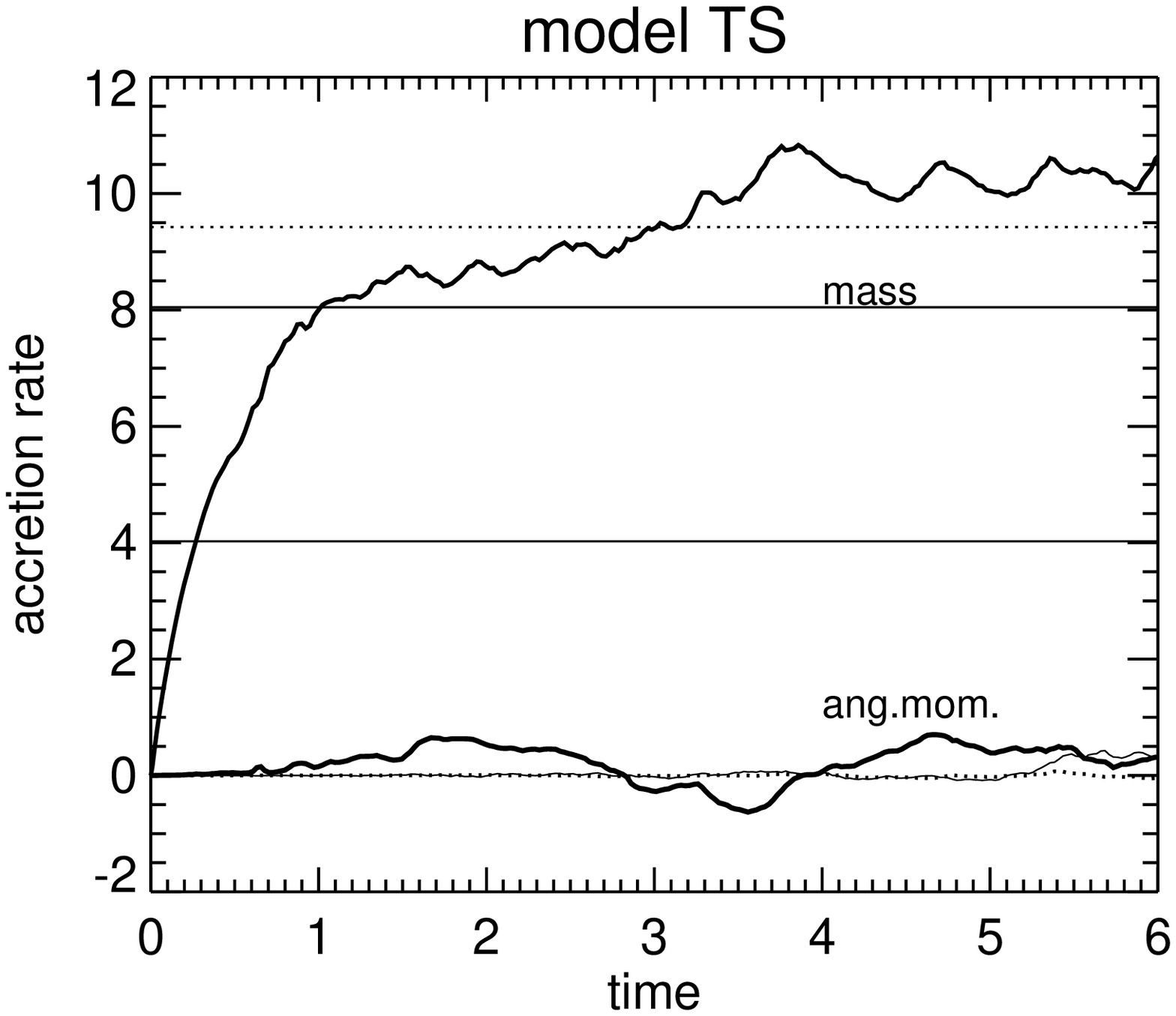} &
  \epsfxsize=8.8cm  \epsfclipon \epsffile{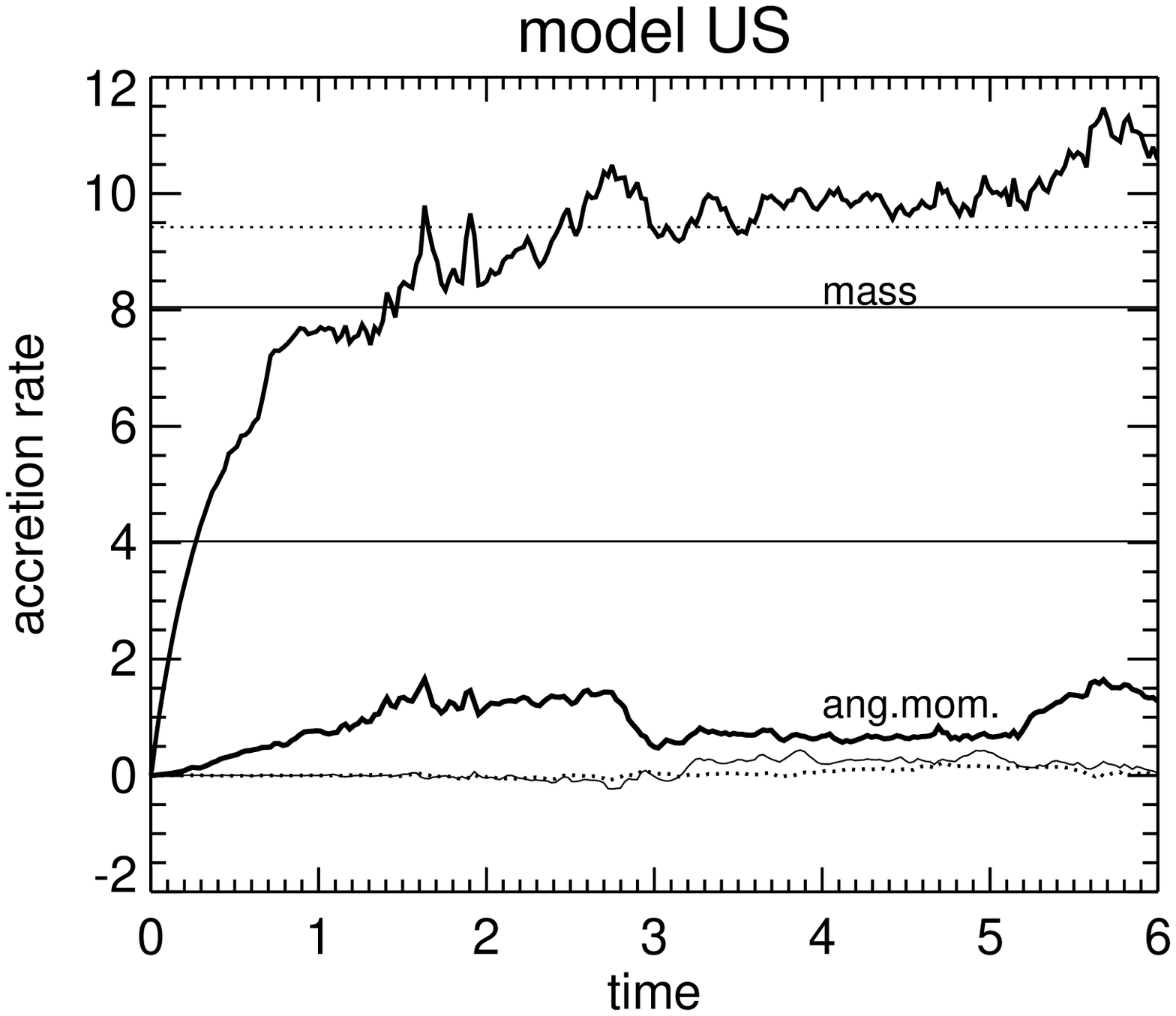} \\
  \epsfxsize=8.8cm  \epsfclipon \epsffile{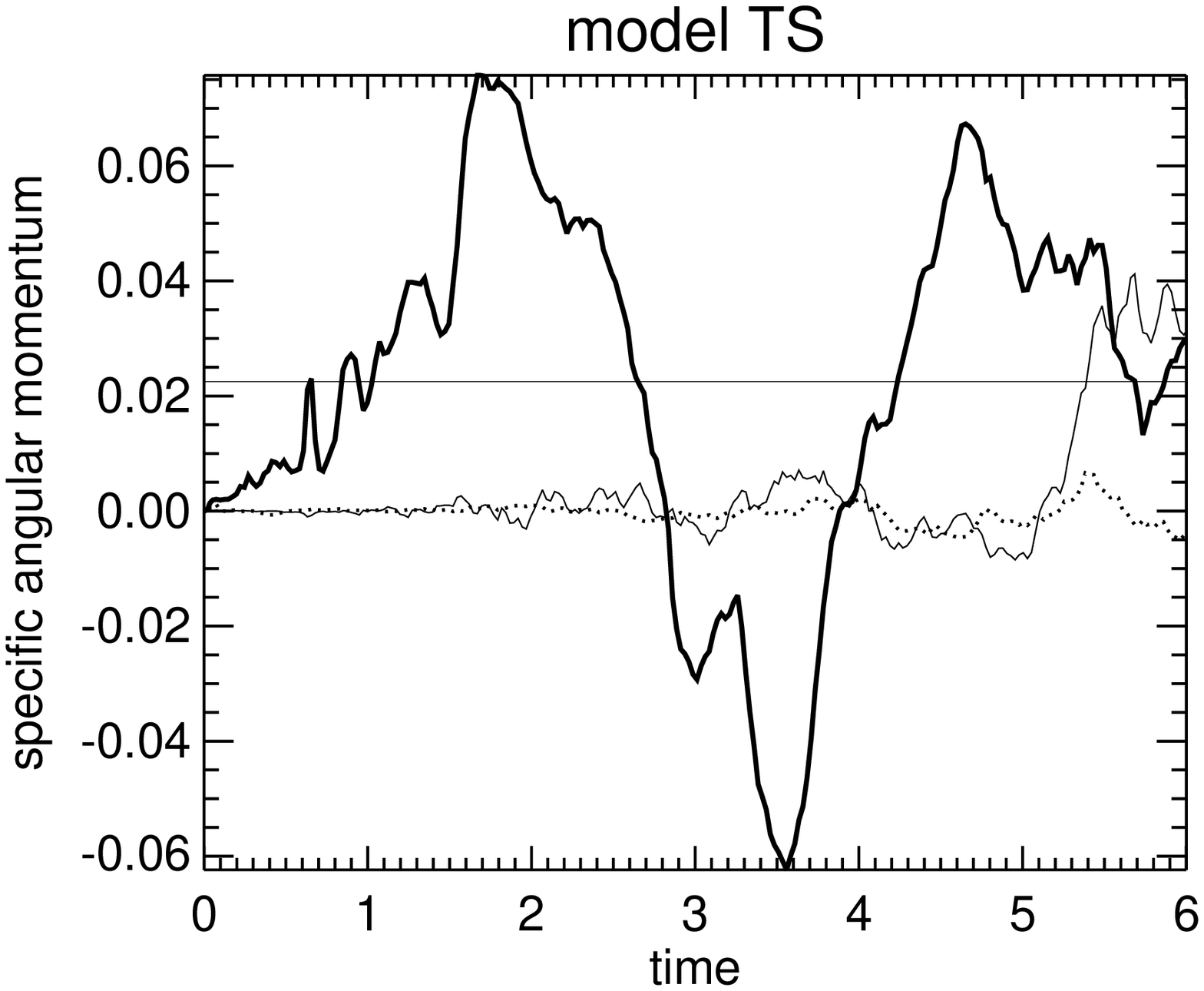} &
  \epsfxsize=8.8cm  \epsfclipon \epsffile{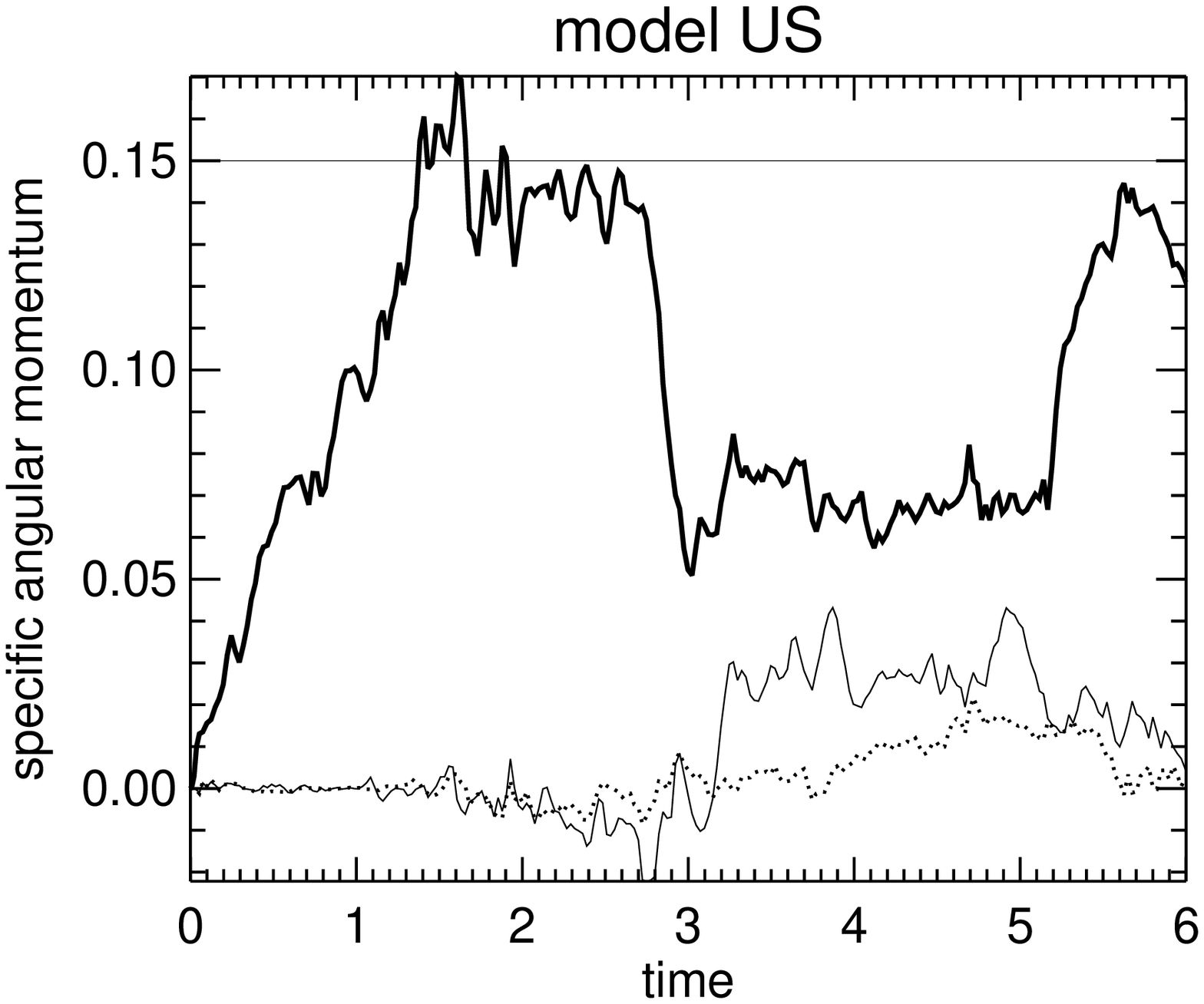}
 \end{tabular}
\caption[]{
The accretion rates of several quantities are plotted as a
function of time for the moderately supersonic (${\cal M}_\infty$=3)
models~TS and~US with an adiabatic index of $\gamma=1.01$.
The top panels contain the mass and angular momentum accretion rates,
the bottom panels the specific angular momentum of the matter
that is accreted.
In the top panels, the straight horizontal lines show the analytical
mass accretion rates: dotted is the Hoyle-Lyttleton rate
(Eq.~(1) in Ruffert~1994), 
solid is the Bondi-Hoyle approximation formula (Eq.~(3) in
Ruffert~1994; Bondi~1952) and half that value.
The upper solid bold curve represents the
numerically calculated mass accretion rate.
The lower three curves of the top panels trace the x~(dotted),
y~(thin solid) and z~(bold solid) component of the angular momentum
accretion rate.
The same components apply to the bottom panels;
the horizontal line shows the
specific angular momentum value as given by Eq.~(\ref{eq:specmomang}).
}
\label{fig:valueTS}
\end{figure*}

\begin{figure*}
 \tabcolsep = 0mm
 \begin{tabular}{cc}
  \epsfxsize=8.8cm  \epsfclipon \epsffile{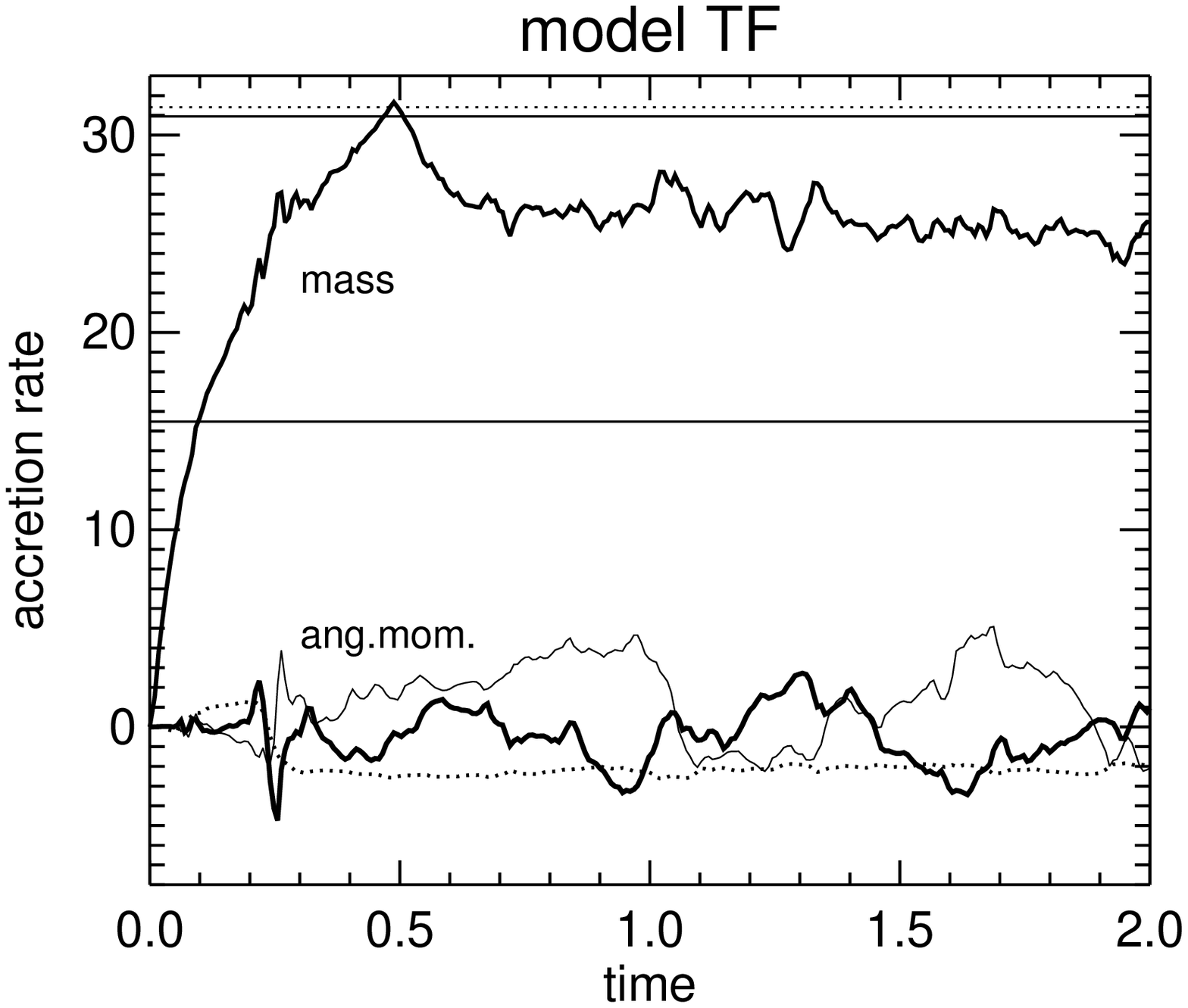} &
  \epsfxsize=8.8cm  \epsfclipon \epsffile{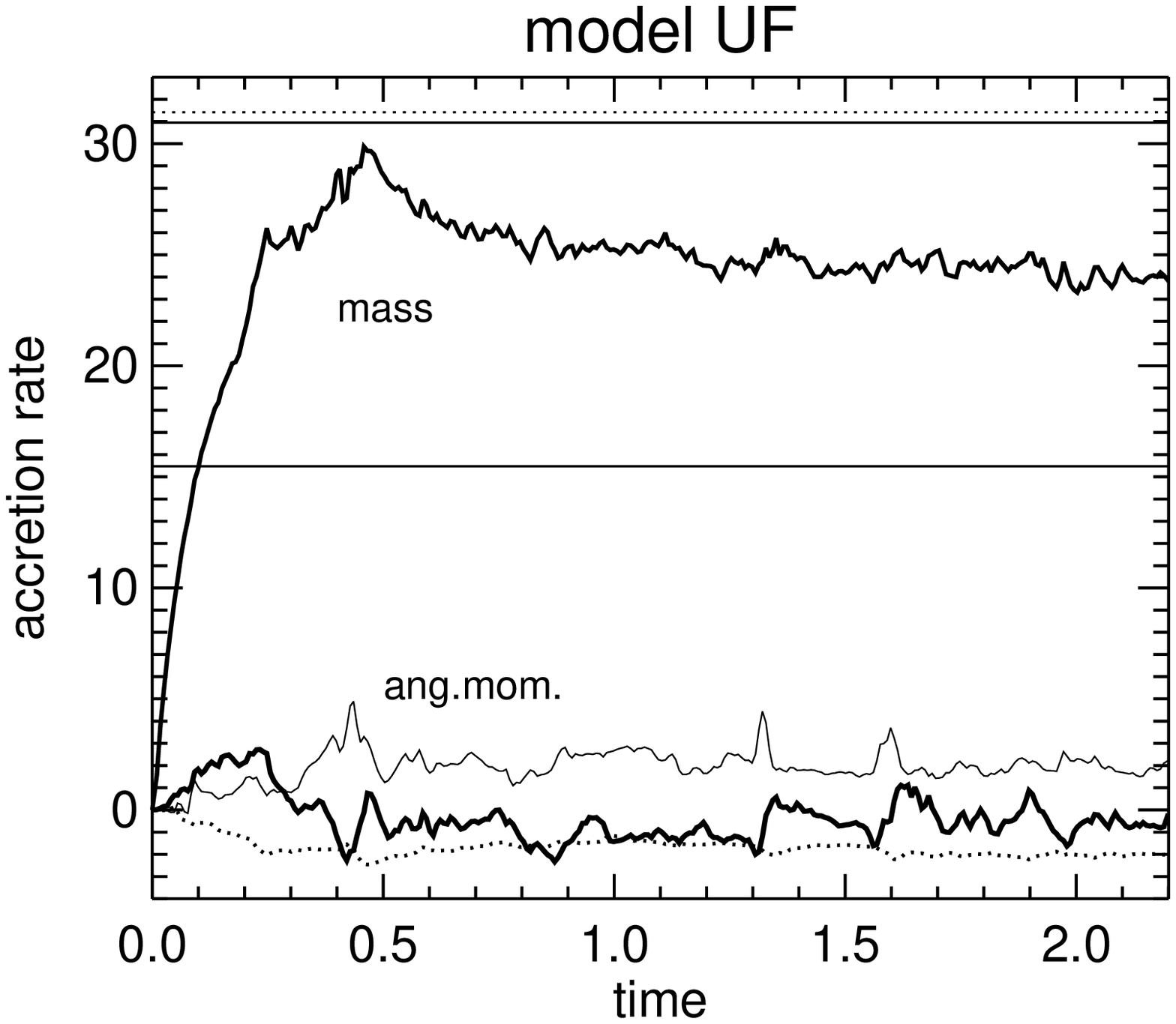} \\
  \epsfxsize=8.8cm  \epsfclipon \epsffile{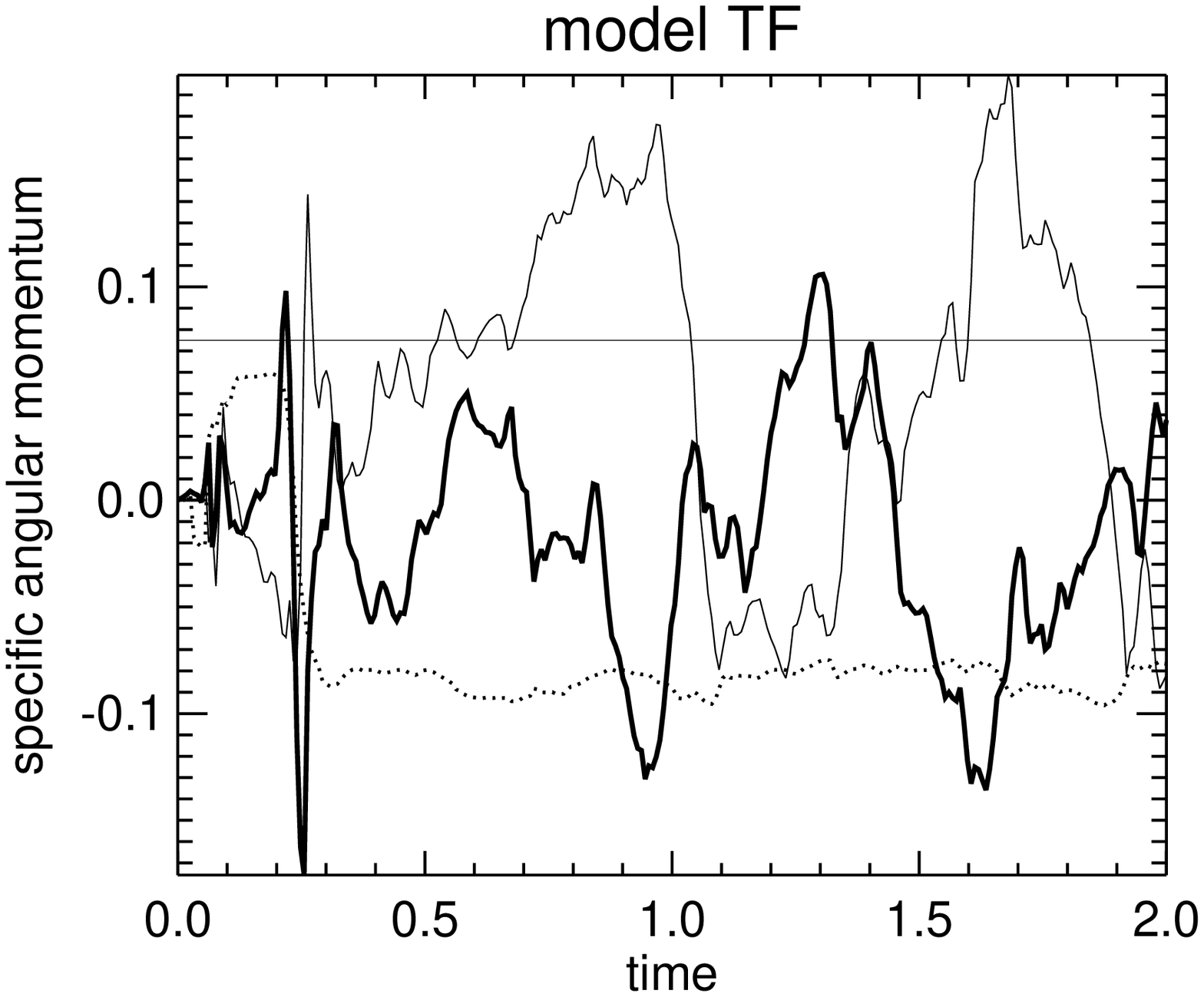} &
  \epsfxsize=8.8cm  \epsfclipon \epsffile{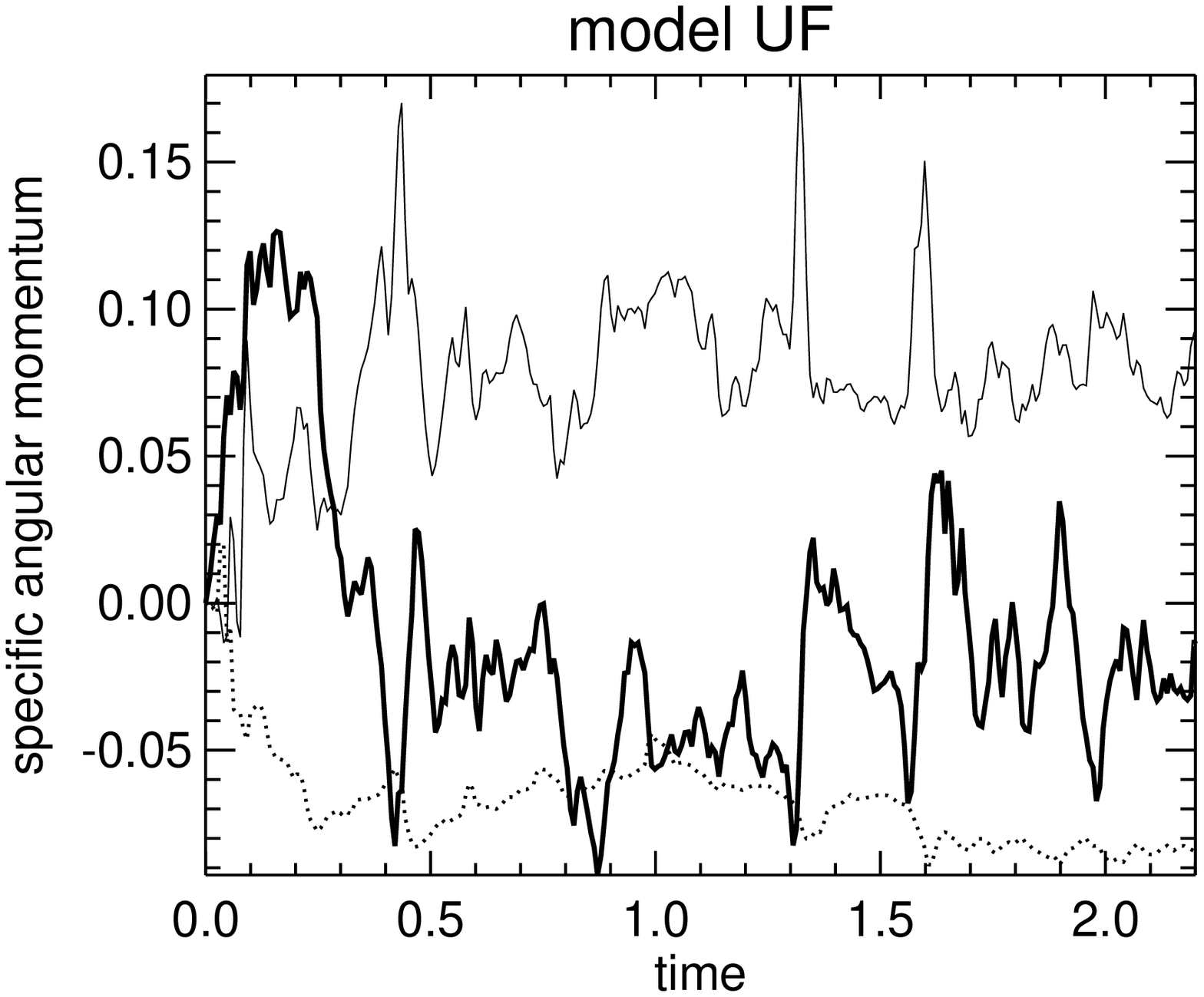}
 \end{tabular}
\caption[]{
The accretion rates of several quantities are plotted as a
function of time for the highly supersonic (${\cal M}_\infty$=10)
models~TF and~UF with an adiabatic index of $\gamma=1.01$.
The top panels contain the mass and angular momentum accretion rates,
the bottom panels the specific angular momentum of the matter
that is accreted.
In the top panels, the straight horizontal lines show the analytical
mass accretion rates: dotted is the Hoyle-Lyttleton rate
(Eq.~(1) in Ruffert~1994), 
solid is the Bondi-Hoyle approximation formula (Eq.~(3) in
Ruffert~1994; Bondi~1952) and half that value.
The upper solid bold curve represents the
numerically calculated mass accretion rate.
The lower three curves of the top panels trace the x~(dotted),
y~(thin solid) and z~(bold solid) component of the angular momentum
accretion rate.
The same components apply to the bottom panels.
The horizontal line in the bottom panels show the
specific angular momentum value as given by Eq.~(\ref{eq:specmomang}).
}
\label{fig:valueTF}
\end{figure*}

\subsection{Models\label{sec:models}}

The combination of parameters that I varied, together with some results are
summarised in Table~\ref{tab:models}.
The first letter in the model designation indicates the strength of
the gradient:  
M, P, and T have $\varepsilon_\rho=0.03$, while 
N, Q, and U have $\varepsilon_\rho=0.2$ and V is $\varepsilon_\rho=1.0$.
The second letter specifies the relative wind flow speeds,
F (fast), S (slow) and V stand for 
Mach numbers of~10, 3 and~1.4, respectively.
I basically simulated models with all possible combinations of the
two higher flow speeds (Mach numbers of~3 and~10), the two gradients 
(3\% and~20\%) and varying the adiabatic index between 5/3,
4/3 and 1.01.
Model~VS with the largest gradient of 100\%
facilitates a comparison to a previous
two-dimensional simulation by Fryxell \& Taam (1988).
The grids are nested to a depth $g=9$ such that the radius of the
accretor $R_\star=0.02R_{\rm A}$ spans several zones on the finest grid.

As far as computer resources permitted, I aimed at evolving the models
for at least as long as it takes the flow to move
from the boundary to the position of the accretor which is at the centre
(crossing time scale).
This time is given by $L/2{\cal M}_\infty$ and ranges from about 1
to about 10 time units.
The actual time $t_{\rm f}$ that the model is run can be found in
Table~\ref{tab:models}, as well as the parameters of the grids
($L$, $g$, etc.).

When modeling a BHL flow with a density gradient, one has to pay
attention to the fact that matter parcels with possibly very different
densities (which initially are separated upstream) will be focussed to
find themselves close to each other along the accretion axis
(U.~Anzer, personal communication).
This might render the flow additionally unstable.
Although true in principle, this problem does not affect, in practice,
the models I will present: a closer inspection of 
Eq.~(\ref{eq:rhograd}) reveals that the density jump from one end of
the accretion radius to the other is only a factor 1.5 for the case
with large gradient ($\varepsilon=0.2$).
This does not seem to have an additional influence as compared to
models with constant density presented in paper~R1.

The calculations are performed on a Cray-YMP~4/64 and a Cray J90 8/512.
They need about 12--16 MWords of main memory
and take approximately 160 CPU-hours per simulated time unit.

\begin{figure*}
 \begin{tabular}{cc}

  \epsfxsize=8.5cm \epsfclipon \epsffile{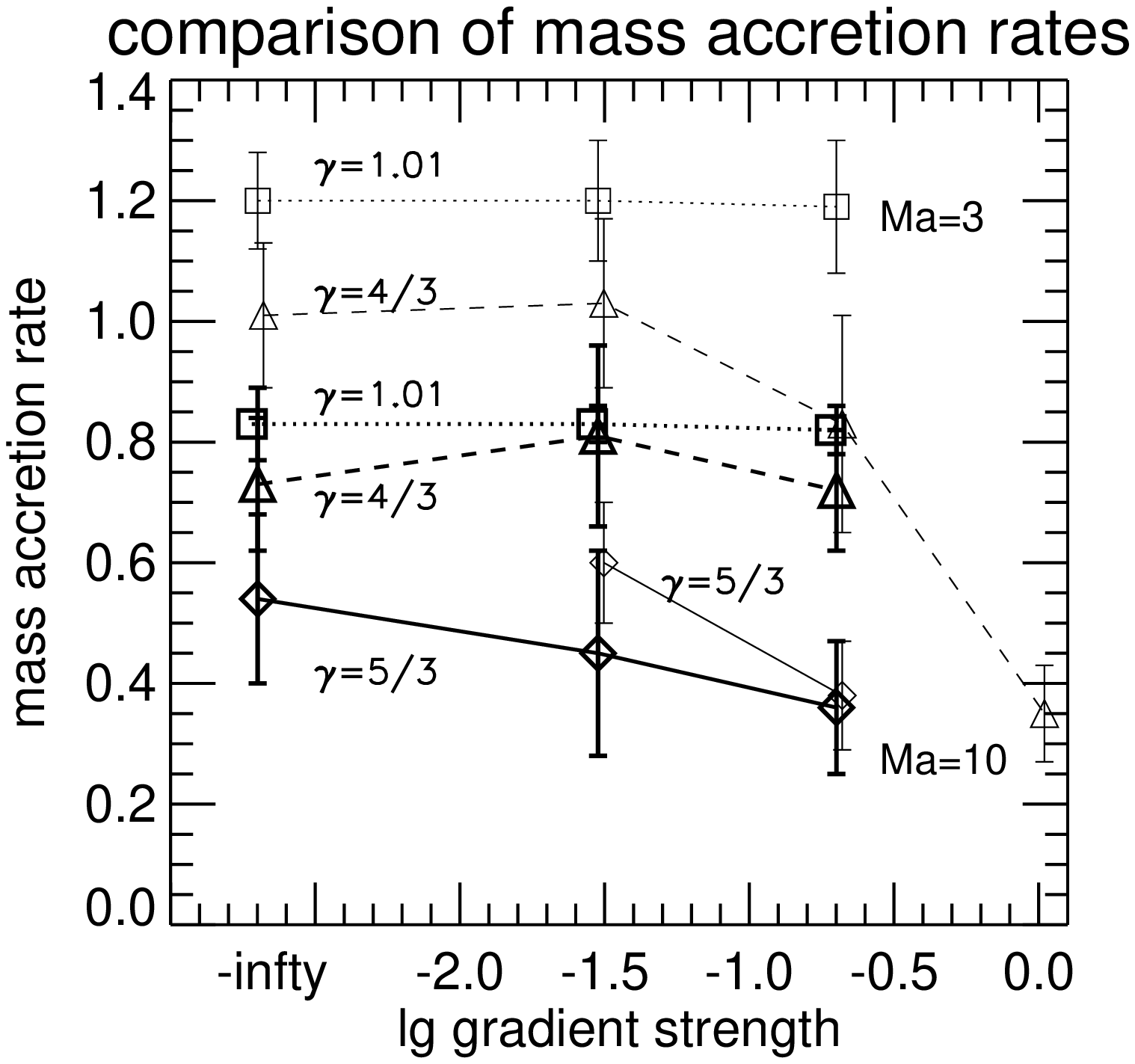} &
  \epsfxsize=8.5cm \epsfclipon \epsffile{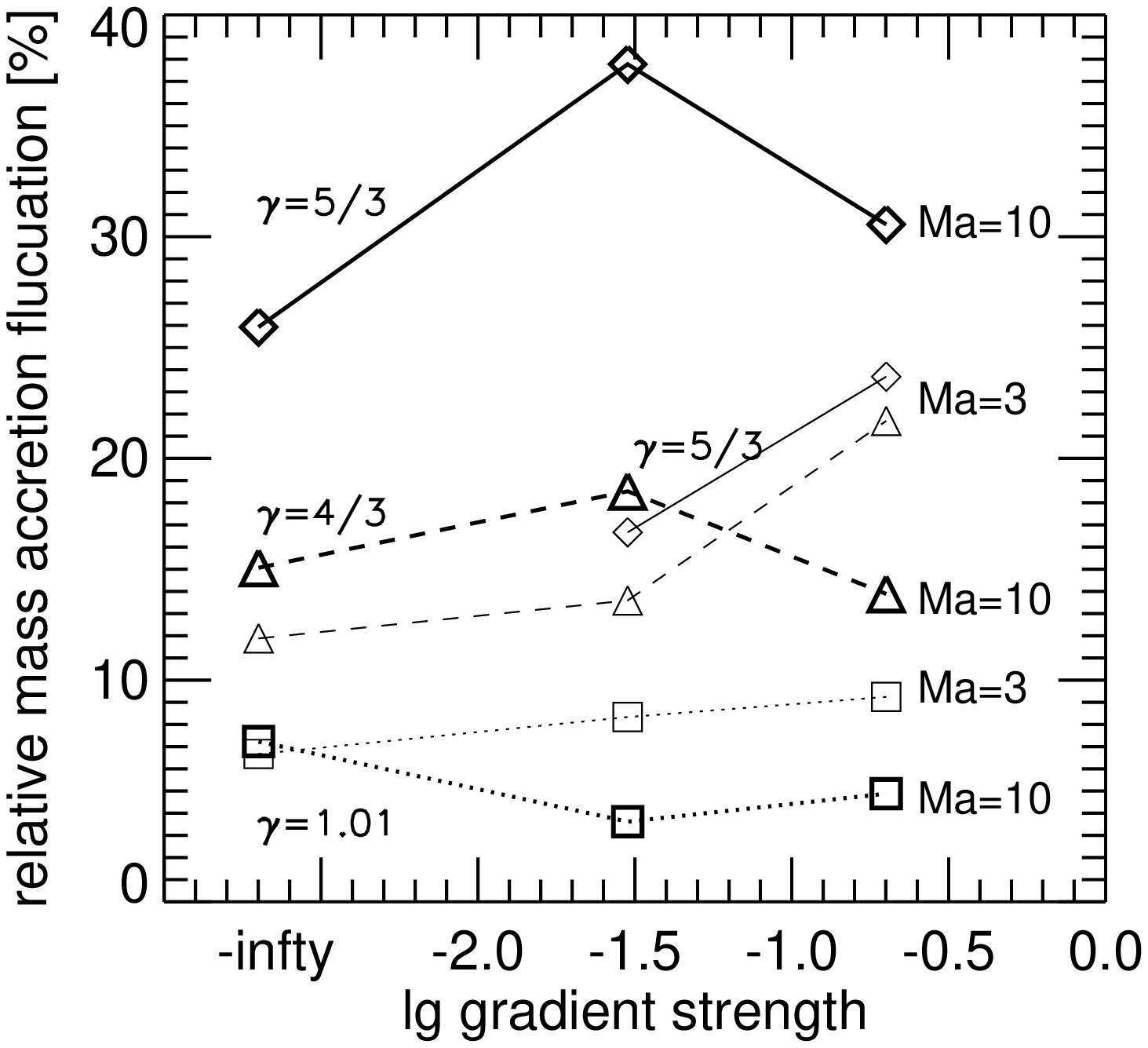}  \\[-7ex]
  \parbox[t]{8.5cm}{\caption[]{
  Mass accretion rates (units: $\dot{M}_{\rm BH}$) are shown
  as a function of the strength of the density gradient: the points to
  the right of lg gradient = -2
  are the results from this work, while the values for models without
  gradient (at the x-axis position ``-infty'') are taken from
  Ruffert~(1994) and Ruffert \& Arnett~(1994).
  Diamonds ($\Diamond$) denote models in which $\gamma=5/3$,
  triangles ($\triangle$) models with  $\gamma=4/3$,
  and squares ($\Box$) show $\gamma=1.01$
  The large bold symbols belong to models with a speed of 
  ${\cal M}_\infty=10$,
  while the smaller symbols belong to models with ${\cal M}_\infty=3$.
  The adiabatic index $\gamma$ and Mach number are also written near each set
  of points.
  The ``error bars'' extending from the symbols indicate one standard
  deviation from the mean ($S$ in Table~\ref{tab:models}).
  Some points were slightly shifted horizontally to be able to discern the 
  error bars.}
  \label{fig:massacc}} &
  \parbox[t]{8.5cm}{\caption[]{
  The relative mass fluctuations,
  i.e.~ the standard deviation $S$ divided by the average mass accretion
  rate $\overline{\dot{M}}$ (cf.~Table~\ref{tab:models}), is shown as
  a function of the strength of the density gradient: 20\% and 3\% are
  the results from this work, while the values for models without
  gradient (at the x-axis position ``-infty'') are taken from
  Ruffert~(1994) and Ruffert \& Arnett~(1994).
  Diamonds ($\Diamond$) denote models in which $\gamma=5/3$,
  triangles ($\triangle$) models with  $\gamma=4/3$,
  and squares ($\Box$) show $\gamma=1.01$
  The large bold symbols belong to models with a speed of 
  ${\cal M}_\infty=10$,
  while the smaller symbols belong to models with ${\cal M}_\infty=3$.
  } \label{fig:relfluct}} \\
  \epsfxsize=8.5cm  \epsfclipon \epsffile{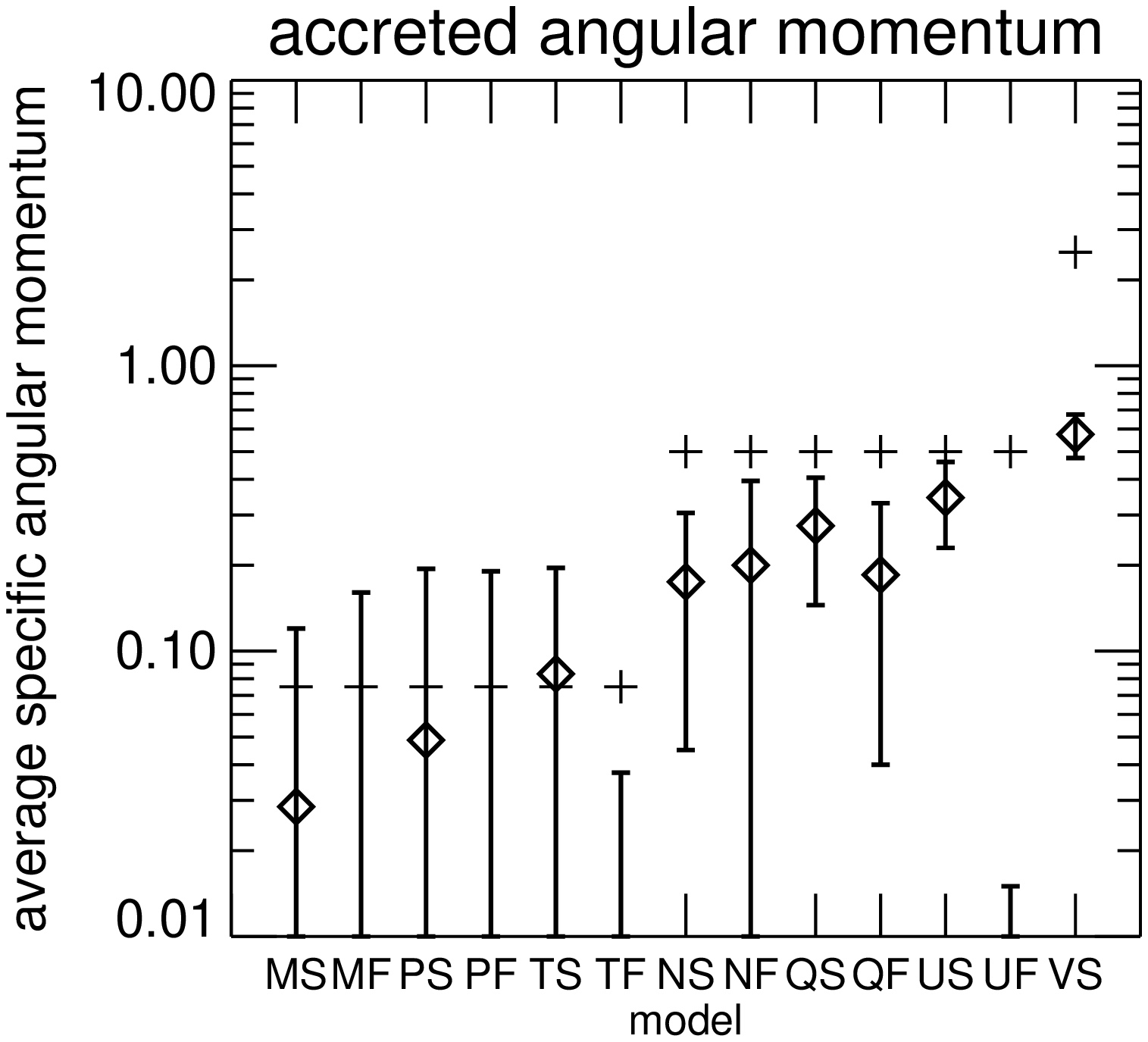} &
  \epsfxsize=8.5cm  \epsfclipon \epsffile{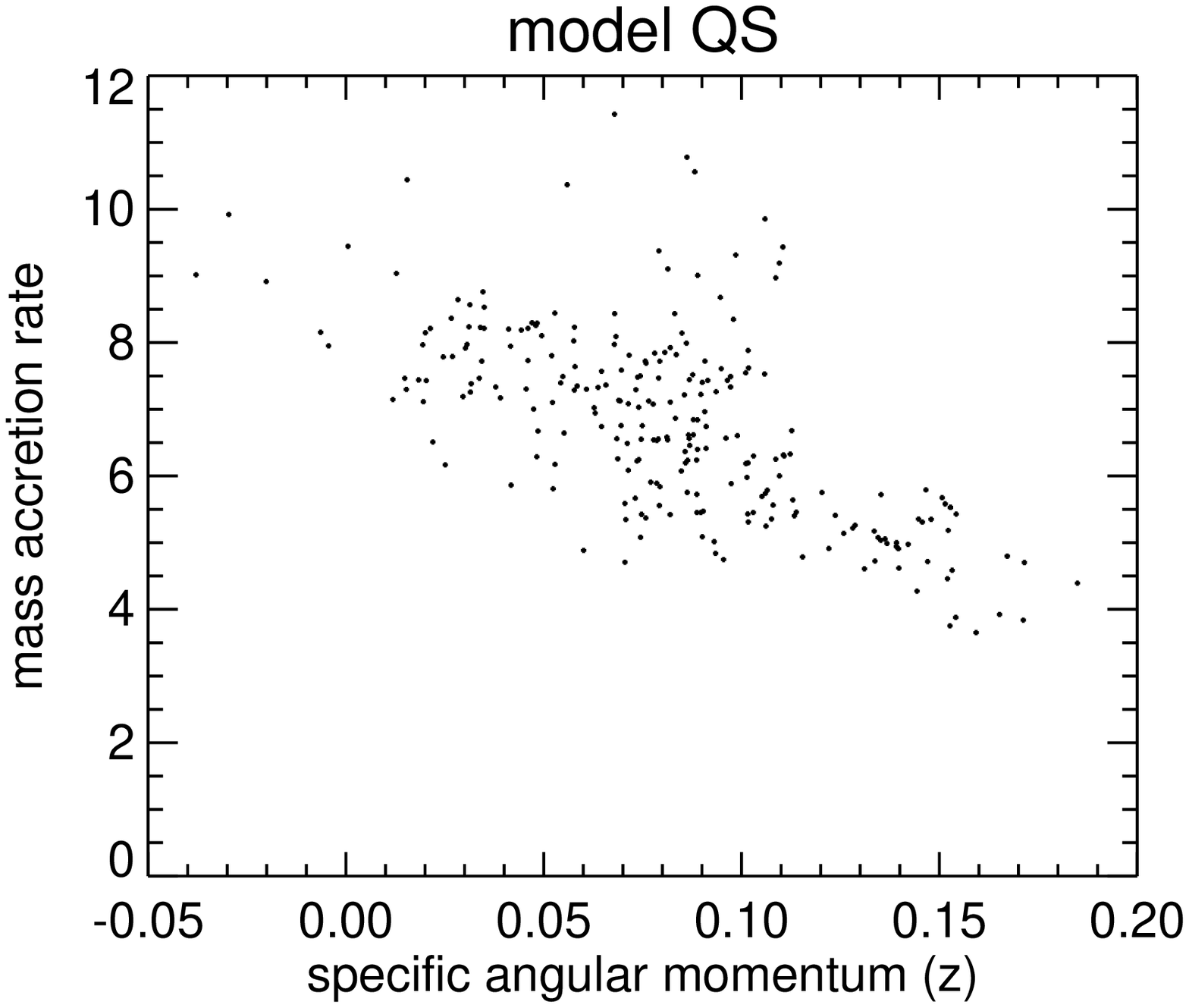} \\[-7ex]
  \parbox[t]{8.5cm}{\caption[]{
  The average accreted
  specific angular momentum (units: $l_{\rm s}$,
  Kepler velocity vortex at surface of accretor,  as given by
  Eq.~(10) in R1) is shown for most models by diamond
  symbols ($l_{\rm z}$ in Table~\ref{tab:models}).
  The ``error bars'' extending from the symbols indicate one standard
  deviation from the mean ($\sigma_{\rm z}$ in Table~\ref{tab:models}).
  The long error bars extending to the bottom axis are an indication
  that the fluctuations of the respective model are so large, that the
  specific angular momentum changes sign from time to time.
  The plus signs above the diamonds indicate the specific angular
  momentum $j_{\rm z}$ according to Eq.~(\ref{eq:specmomang}).
  } \label{fig:angacc}} &
  \parbox[t]{8.5cm}{\caption[]{
  The mass accretion rate is plotted versus the specific angular
  momentum of the accreted matter for model~QS.
  Each dot displays the two quantities at one moment in time.
  Only dots at time later than $t\ga0.8$ are plotted to avoid the
  initial transients. 
  } \label{fig:correl}} \\[-8ex]
\end{tabular}
\end{figure*}

\section{Dynamics and accretion rates\label{sec:descr1}}

\subsection{Results of models with $\gamma$=5/3 and 4/3}

I will describe the results of the models for which a ratio of
specific heats of $\gamma$=5/3 was chosen together with those models
of $\gamma$=4/3 because the evolution is very similar. 
The only exception is model~MV: its slow relative bulk velocity of
Mach~1.4 produces a significantly more stable flow.

Figs.~\ref{fig:Mdens} and~\ref{fig:Pdens} show snapshots of the flow
velocities and density distribution in the x-y-plane containing the
accretor.
The velocity pattern as well as the density contours within the shock
cone indicate a strongly unstable flow.
Also the shock cone itself has many bumps and kinks.

Note how the density contours strongly bend over upstream of the shock 
(at $x\approx-0.8$) for models~N and~Q indicating the large gradient
as compared to models~M and~P, where only the contour of $\rho=1$ is
seen to be detached from the shock.
The higher densities on the lower side of $y$ (where it is negative)
seem to produce an asymmetric shock cone: the lower part of the cone
subtends a smaller angle with the $y=0$-axis than the upper half.
However, the quickly varying accretion flow pattern tends to impinge
on the shock front and dislodge it.
This asymetry has been observed and commented on in 
Ishii et al.~(1993) and Soker \& Livio (1984).
In the models with smaller $\gamma$ (models~P and~Q) the shock cone
tends to be narrower and its minimum distance from the accretor
smaller than for the models~M and~N.

In contrast to what has been said, model~MV with a slow (but still
supersonic relative velocity), shows a very regular flow pattern: the
stagnation point is about 0.3~$R_{\rm A}$ downstream from the accretor.
Matter that comes within this point gets accreted while matter that
stays outside just passes the accretor.
The mass accretion rate of model~MV (cf.~Fig.~\ref{fig:valueMV}) rises
slowly and nearly monotonically to saturate close to the Bondi-Hoyle
formula value.
Only towards the end is there an indication that the flow might become
unstable for this model too. 
The component of interest ($z$) for the specific angular momentum
saturates at about 75\% of the value estimated analytically by
Eq.~(\ref{eq:specmomang}). 
This is probably due to the fact that for bulk velocities close to a
Mach number of unity the Hoyle-Lyttleton approximation brakes down
because a certain fraction of mass is accreted practically spherically
symmetrically, i.e.~as Bondi tried to describe it with the
approximative formula (Bondi, 1952).
The other two components ($x$ and $y$) fluctuate around zero (which is
given by the symmetry of the boundary conditions), but indicate at a very
early stage that the flow is mildly unstable.

The other 8 models (MS, MF, NS, NF, PS, PF, QS, QF) show very strong
fluctuations of the accretion rates of mass and all angular momentum
components (Figs.~\ref{fig:valueMS}, \ref{fig:valueMF},
\ref{fig:valuePS}, and \ref{fig:valuePF}).
A variation of factors of two is not uncommon, so the averages stated
in Table~\ref{tab:models} should be used with care and where possible
the standard deviations taken into account. 
I include the accretion rate plots for all models in order to
facilitate the judgement of how representative the average values are
for the whole temporal evolution.

A few trends can be discerned. All four models MS, MF, PS, PF,
i.e.~the ones with small $\varepsilon_\rho=0.03$ 
display a fairly quiet initial transient phase.
The~$x$ and~$y$-components of the angular momentum fluctuate mildly
around zero and the $z$-component reaches fairly precisely (to
within~10\%) the analytic estimate Eq.~(\ref{eq:specmomang}).
This shows the applicability of the analytic result only to the
initial quasi-stationary state.
The models with large $\varepsilon_\rho=0.2$ do not have this quiet
phase but become chaotic much quicker.
This decreased stability also produces lower accretion rates for
mass as well as lower specific angular momenta.
The initial quiet transient phase has already been seen in R1:
compare the present models to the ones from R1, IS, JS, KS, LS in
Figs.~4, 6, 8, and 9 in R1, respectively. 

Note that model PF is one for which the simulation was run fairly long
compared to the timescale of fluctuations. This increases the
confidence that the average mass accretion rate is a representative
value and not a random one of a transient state.
The angular momenta, however, still do not display the marginal
positive shift that the z-component should have compared to the other
two. 

No significant difference is observed when comparing two models that
only differ in Mach number.
However the mass accretion rates are significantly larger in the
models with smaller $\gamma$, again {\it ceteris paribus}.
The same applied to the velocity gradient models of R1, cf.~model~IS
and~SS in Figs.~4 and~11 respectively.
This dependence of the accretion rate on the adiabatic index $\gamma$
is well known for stationary flows from analytic calculations,
e.g.~Bondi (1952) for spherically symmetric flows and Sect.~5 and~6
and Figs.~9 and~10 in Foglizzo \& Ruffert (1997).

\subsection{Results of models with $\gamma$=1.01}

The first main obvious difference between the nearly isothermal models
and the more adiabatic ones is that the shock cone is attached to the
accretor, as can be seen in Fig.~\ref{fig:Tdens}.
The pressure around the accretor in this case is not sufficient to
push away and support the shock cone.
Not even the large density gradient is able to dislodge the shock from
touching the surface of the accretor.

Both models with high Mach number (TF and UF) hardly show any activity
of unstable flow within the shock cone, contrary to the moderately
supersonic cases (TS and US).
Again (as in model~MV) a clear and stable stagnation point is present
downstream of the accretor for these quiet models.
It is a common feature of practically all simulations (including the
ones by other authors) that when the shock cone is attached to the
accretor no (or hardly any) instability is observed. Whether or not
the shock cone is attached depends on many physical attributes of the
models (e.g.~stiffness of the equation of state, etc.) and numerical
parameters (not least resolution). However, when these conditions
collude to produce an attached shock, invariably the flow remains
stable. 

The very active flow of the slower models (TS and US) reflects itself
in a higher variability of the mass accretion rate as compared to
models TF and UF (Figs.~\ref{fig:valueTS} and~\ref{fig:valueTF}).
However the fluctuation of the mass accretion rates of all these
$\gamma=1.01$ models is much smaller than the fluctuations shown by
the more adiabatic models described further above.
The average mass accretion rates of the slow models slightly exceed the
values predicted by the Hoyle-Lyttleton theory, while the faster
models only reach 80\% of $\dot{M}_{\rm BH}$.
The $z$-component of the specific angular momentum in the models with
large gradient (US, UF) practically never reach the analytical estimate
Eq.~(\ref{eq:specmomang}) while the models with small gradient (TS, TF)
fluctuate strongly about this value, exceeding it by a factor of three
or even reversing sign.

These models can be compared to models without gradients 
(Ruffert,~1996) but with the same remaining parameters: models~TF
and~UF should be compared to model~HS in Fig.~7e, while models~TS
and~US can be compared to model~GS in Fig.~5e of Ruffert (1996).
Both the different behaviour of the mass accretion rates as well as
the amplitudes of fluctuation of the angular momentum accretion rates
are comparable between the models, indicating that the presence of
a density gradient does not significantly alter the accretion
properties of such a strongly unstable flow.
Of course, the mean of the $z$-component is not zero in flows with
gradients. 
Fig.~\ref{fig:valueTS} can also directly be compared with Fig.~11 from
Ishii et al.~(1993). Both the mass and angular momentum accretion
rates show very similar magnitudes and evolution.

\subsection{Results of model VS}

The density gradient in the previous set of models of
$\varepsilon_\rho=0.03$ and $\varepsilon_\rho=0.2$ was chosen in order
to facilitate the direct comparison between the results presented in
this paper and the ones with velocity gradients shown in R1.
However, Eq.~\ref{eq:specmomang} then predicts that the specific
angular momentum accreted will be six times smaller for the density
gradients as compared to accretion with velocity gradients.
That this is true becomes clear when comparing the plots for models 
MS, MF, PS, NS and NF with equivalent models from paper R1: IS, JS,
SS, KS and LS, respectively:
In the models presented in this paper, the z-component of the
angular momentum, which is the one influenced by the gradient, hardly
rises above the random fluctuations of the other two components.
In order to check the correct separation of the effect of the density
gradient from the unstable nature of the flow, a model with larger
gradient is helpful and will be presented in this subsection.

The particular value of $\varepsilon_\rho=1$ is suggested because for
this case, Taam \& Fryxell (1988) have found that a quasi-steady disk
forms which does not change its sense of direction.
Fig.~\ref{fig:valueVS} (left panels)
shows the temporal evolution of the mass and
angular momentum accretion for this large density model.
The mass accretion rate remains very low over the whole simulated time
on average being less than half the value of model QS (which is
similar to VS in all parameters except $\varepsilon_\rho$).
Note that these average values listed in Table~\ref{tab:models} are
normalised to Eq.~\ref{eq:specmomang} and not to Eq.~\ref{eq:coeffspec}.
Thus a big part of the reduction of specific angular momentum between
model~QS (0.55) and model~VS (0.23) is probably due to the
``tanh''-term represented by the factor $f_j$ (Fig.~\ref{fig:shali}):
$f_j\approx0.25$ for model~QS while $f_j\approx0.16$ for model~VS.

Also note that model~VS does not display as drastic a reduction in mass
accretion rate as shown in Fig.~22 of Taam \& Fryxell (1988)
[note the different choice of time units between this paper and the
one in Taam \& Fryxell, 1988].
The reason probably being that in their 2D models accretion gets
effectively shut off as soon as a stable disk structure forms, while
in my 3D simulations accretion can still proceed practically unimpeded
via the poles.

The lower left panel of Fig.~\ref{fig:valueVS} shows the angular momentum
accretion. In this large gradient model, the z-component clearly
dominates compared to the other two components, i.e.~the fluctuating
flow cannot compete with the angular momentum available in the bulk
flow gradient. Note also that the z-component never crosses the $y=0$
line which indicates the formation of a disk-like flow structure with
rotation in an unchanging sense. However, I do not expect the
specific angular momentum to reach a quasi-steady state because of the
three-dimensional nature of my simulations: angular momentum can
continue to be accreted from directions outside the plane of the disk
and thus probably disturbing the disk, too.

\section{Comparison of accretion rates\label{sec:descr2}}

\subsection{Mass accretion rates}

I collected all the means of the mass accretion rates and their
standard deviation into Fig.~\ref{fig:massacc} together with 
the appropriate values for models without gradients.
The standard deviation is shown as ``error bar'' in order to give an
indication on how precisely one should interpret the mean values.

(a) The mass accretion rates are fairly independent of the density
gradient strength for small $\gamma$, but a decrease of the rates
might be present when increasing the gradients:
a slight trend toward smaller rates for larger gradients seems 
possible.
(b)
A clear increase of the rates is visible when decreasing the index
$\gamma$. 
When going from 5/3 to 1.01 the accretion rate increases by
over a factor of two.
(c) Models with smaller Mach numbers have larger rates.
This trend also applies to models with a velocity gradient as can
be seen in Fig.~12 of R1.

Because the fluctuations of the mass accretion rate slightly decrease
with decreasing $\gamma$, while at the same time the rate itself
increases, it follows that the relative fluctuation decreases strongly
with decreasing $\gamma$ as can be seen in Fig.~\ref{fig:relfluct}.
No obvious trend seems visible on how the relative fluctuations change
with gradient strength; a slight increase might be present when
comparing the larger gradient fluctuation with the smaller gradient
ones.

The dependence of the relative fluctuations on velocity seems to
invert: while they are clearly larger for smaller Mach numbers for the
$\gamma=1.01$ models, they are smaller for smaller Mach
numbers in the $\gamma=5/3$ models.
And last, the simulations with $\gamma=4/3$ produce relative
fluctuations that vary more strongly with gradient strength than with
Mach number. 

\subsection{Specific angular momentum}

As has already been described in Sect.~\ref{sec:descr1} the specific
angular momentum reaches the analytic values given by
Eq.~(\ref{eq:specmomang}) to within 10\%, but only as long as the
accretion flow is roughly stable and only for the models with small
density gradients, $\varepsilon_\rho=0.03$.
In Fig.~\ref{fig:angacc} I show the specific angular momentum averaged
over the whole time the models were evolved, excluding the initial
transients. 
This gives a better picture of how the very unstable flow tends to
decrease the average and produce a large fluctuation of the specific
angular momentum.
So I included as ``error bars'' the standard deviation of the
fluctuations around the mean.

The unit I chose in Fig.~\ref{fig:angacc} is the specific angular
momentum that a 
vortex just at the surface of the accretor would have if it spun with
the local Kepler velocity. Thus unity in these units indicates a
Kepler orbit and matter that has more specific angular momentum than
this would be flung off. 
One can see in Fig.~\ref{fig:angacc} that the matter accreted is well
below this value.
For comparison, I also plot, using plus signs (+), the specific angular
momentum as given by Eq.~(\ref{eq:specmomang}).
These values tend to lie above the mean, especially for the models
with large gradients (N,Q,U), and are well within one standard
deviation from the mean for the small gradient models (M,P,T).
For the latter models the fluctuations are so large that occasionally
the sign of the accreted specific angular momentum reverses; this is
indicated by the ``error bars'' extending completely down to the $x$-axis.

If one is only interested in the average specific angular momentum
that is accreted, an inspection of Table~\ref{tab:models} yields that
the whole range of values between zero and about 70\% is attained
depending on the model parameters.
Livio et al.~(1986) reported values between 10\% and 20\% for their
parameters. 

If one reduces the size of the accretor, some point will be
reached when the maximum specific angular momentum that can be
accreted will become smaller than the amount present in the accretion
cylinder. 
Setting Eq.~(\ref{eq:specmomang}) equal to Eq.(10) in R1 a relation 
is obtained between the gradients in the flow ($\varepsilon$) and the
radius of the accretor $R_\star$:
\begin{equation}
  R_\star = 
    \frac{1}{32} (6\varepsilon_{\rm v}+\varepsilon_\rho)^2 R_{\rm A}  \quad.
\end{equation}
So for accretors smaller than this radius the angular momentum
accretion should no longer be dominated by what is given in the
accretion cylinder.
For $\varepsilon_{\rm v}$=0 and $\varepsilon_\rho$=0.03 and 0.2 we
obtain $R_\star\approx3\cdot10^{-5}R_{\rm A}$ and 
$R_\star\approx10^{-3}R_{\rm A}$.
Both these values are below what is currently possible to simulate
numerically, but could be important in astrophysical objects.
On the other hand for model~VS, which has $\varepsilon_{\rm v}$=0 
and $\varepsilon_\rho$=1.0, an accretor radius of 
$0.03R_{\rm A}$ results, which is larger than the
numerically used radius of $0.02R_{\rm A}$.
Thus in model~VS not all the specific angular momentum offered in the
incoming bulk flow can be accreted.
Vice-versa, the maximum gradient $\varepsilon$ that can be
accommodated by an accretor of given radius $R_\star$ is
\begin{equation}
  \varepsilon^{\rm max}_\rho = \sqrt{ 32 \frac{R_\star}{R_{\rm A}} }
  \quad \quad \quad \quad
  \varepsilon^{\rm max}_{\rm v}=\sqrt{\frac{16}{3}\frac{R_\star}{R_{\rm A}} }
\label{eq:limepsi}
\end{equation}

The right panel of Fig.~\ref{fig:valueVS} compares the specific
angular momentum accreted between the density gradient models
presented here and the velocity gradient models shown in R1.
The ratio (cf.~legend of the x-axis of Fig.~\ref{fig:valueVS})
is plotted of the specific angular momentum (values given
in Table~\ref{tab:models}) for these two sets of models.
If a pair of models experiences the same reduction in accretion of
specific angular momentum, the ratio plotted would be one. 
If the density-gradient model of the pair suffers a greater decrease
of accretion (due to e.g.~stronger relative fluctuations of the
unstable flow) this will be reflected by a ratio that is smaller than
unity. 

The pair MF/JS is less than zero, because the average angular momentum
accreted by model MF is actually retrograde to the bulk flow momentum.
For this pair, as well as for MS/IS, the very large 'error' bar
indicates that the fluctuations due to the unstable flow are very
large compared to the average and so the latter value does not permit
a strong statement. So although for four of the five models the ratios
are above unity, which would indicate that the unstable flow actually
increases the angular momentum accretion for the density gradient
models as compared to the velocity gradient models, this reasoning is
not a credible one.
Additionally one has to keep in mind that 
the specific angular momentum of the incoming bulk flow of models~KS
and~LS is actually larger than the maximum permitted by the
Eq.~\ref{eq:limepsi}. Thus the accreted momentum will be decreased due
to this effect, too (angular momentum barrier), which explains why
these model-pairs have the largest ratios.

\subsection{Correlations}

A correlation was found in R1 (Fig.~17) between the mass accretion
rate and the specific angular momentum: the rate decreases when the
magnitude of the momentum is largest.
In Fig.~\ref{fig:correl} I draw a similar plot as in R1, each dot
connecting the two quantities for every second timestep of the
numerical simulation.
A similar trend can be seen for model QS, which confirms that the
dynamics is similar: when the flow does not rotate (low specific
angular momentum) if falls down the potential to the surface of the
accretor and thus produces a higher mass accretion rate.
For all other models a correlation is not obvious.

\section{Summary\label{sec:conc}}

For the first time a comprehensive numerical {\it three}-dimensional
study is presented of wind-accretion with a {\it density} gradient
using a high resolution hydrodynamic code. 
I vary the following parameters: Mach number of the relative flow
(Mach ${\cal M}$=3 and~10), strength of the density gradient
perpendicular to this flow ($\varepsilon_\rho$=3\%, 20\% and 100\%
over one accretion radius), and adiabatic index 
($\gamma$=5/3, 4/3, and 1.01). 
The results are compared (a) among the models with different parameters,
(b) to some previously published simulations of models with density
and velocity gradients, and (c) also to the analytic estimates of the
specific angular momentum.

\begin{enumerate}
\item
All models exhibit active unstable phases, which are very
similar to the models without gradients.
Only the mildly supersonic case (${\cal M}$=1.4) displayed a
relatively steady flow (as compared to the faster flow models).
The accretion rates of mass, linear and angular momentum fluctuate
with time, although not as strongly as published previously for 2D
models.
\item
Depending on the model parameters, the average specific angular
momentum accreted is roughly between zero and 70\% of the analytical
estimate, which assumes that all angular momentum within the accretion
cylinder is actually accreted.
\item
The mass accretion rates of all models with density gradients are
equal, to within the fluctuation amplitudes, to the rates of the
models without gradients (published previously), although the
accretion rates might seem to decrease slightly when increasing the
density gradient. 
The fluctuations of the mass accretion rate in all models
hardly vary with gradient strength.
\item
The overall qualitative flow dynamics as well as the mass accretion
rates are very similar to what has been published on models with 
{\it velocity} gradients. 
Of course the accretion rate of angular momentum and its specific
values (i.e~per mass unit) are reduced if one compares equal gradients
for both velocity and density, well in accordance with the analytic
estimates.
This reduction means that in the density gradient models the
fluctuations due to the unstable accretion flow have a greater
influence on the angular momentum accretion than for the velocity
gradient models.
\item
The models with small gradients ($\varepsilon_\rho$=0.03) display
an initially quiet stable phase, in which the specific angular momentum of
the matter accreted is within 10\% of the analytic estimate.
Thus for the quiescent phases the analytic values are appropriate.
The average drops when the flow becomes unstable.
\item
The model with very large density gradient ($\varepsilon_\rho=100\%$
over one accretion radius) was the only one for which the accreted
angular momentum was always prograde with respect to the angular momentum
available in the incoming flow. Here the amplitude of the perturbation
due to the unstable flow is much smaller than the average angular
momentum accreted. However the specific angular momentum of the
incoming flow in this case is larger than the maximum given by the
Kepler velocity times the radius of the accretor surface.

\end{enumerate}

%Movies in mpeg format of the dynamical evolution of some  models are
%available in the WWW at
%{\tt http:\nix//www.mpa-garching.mpg.de\nix/\lower0.7ex\hbox{$\!$\~~$\!$}mor\nix/bhla.html}

\begin{acknowledgements}
I would like to thank the referee for
many constructive suggestions, not the least of which was model~VS;
Dr.~U.~Anzer and Dr.~T.~Foglizzo for carefully reading the manuscript
and suggesting improvements, and U.~Kolb for helpful discussions.
I gratefully acknowledge support by a PPARC Advanced Fellowship.
The calculations were done mostly at the Rechenzentrum Garching of the
Max Planck Gesellschaft.
\end{acknowledgements}


\begin{thebibliography}{}

\bibitem[]{an1} Anzer U., B\"orner G., Monaghan J.J., 1987, A\&A~176, 235

\bibitem[]{ben} Benensohn J.S., Lamb D.Q., Taam R.E., 1997, ApJ~478, 723

\bibitem[]{ber} Berger M.J., Colella P., 1989, JCP~82, 64

\bibitem[]{bof} Boffin H.M.J., 1991, IAU~Symp.~151

\bibitem[]{bon} Bondi H., 1952, MNRAS~112, 195

\bibitem[]{boh} Bondi H., Hoyle F., 1944, MNRAS~104, 273

\bibitem[]{col} Colella P., Woodward P.R., 1984, JCP~54, 174

\bibitem[]{dav} Davies R.E., Pringle J., 1980, MNRAS~191, 599

\bibitem[]{dod} Dodd K.N., McCrea W.H., 1952, MNRAS~112, 205

\bibitem[]{fog} Foglizzo T., Ruffert M., 1997, A\&A~320, 342

\bibitem[]{fry} Fryxell B.A., Taam R.E., 1988, ApJ~335, 862

\bibitem[]{ho1} Ho C., Taam R.E., Fryxell B.A., Matsuda T., Koide H.,
     Shima E., 1989, MNRAS~238, 1447

\bibitem[]{ho3} Hoyle F., Lyttleton R.A., 1939,
{\it Proc.\ Cam.\ Phil.\ Soc.}~35, 405

\bibitem[]{h4a} Hoyle F., Lyttleton R.A., 1940a,
{\it Proc.\ Cam.\ Phil.\ Soc.}~36, 323

\bibitem[]{h4b} Hoyle F., Lyttleton R.A., 1940b,
{\it Proc.\ Cam.\ Phil.\ Soc.}~36, 325

\bibitem[]{h4c} Hoyle F., Lyttleton R.A., 1940c,
{\it Proc.\ Cam.\ Phil.\ Soc.}~36, 424

\bibitem[]{ill} Illarionov A.F., Sunyaev R.A., 1975, A\&A~39, 185

\bibitem[]{ish} Ishii T., Matsuda T., Shima E., Livio M., Anzer U., 
                B\"orner G., 1993, ApJ~404, 706

\bibitem[]{li1} Livio M., Soker N., deKool M., Savonije G.J., 
                1986, MNRAS~222, 235

\bibitem[]{r94} Ruffert M., 1994, A\&AS~106, 505

\bibitem[]{r96} Ruffert M., 1996, A\&A~311, 817

\bibitem[]{r97} Ruffert M., 1997, A\&A~317, 793 (R1)

\bibitem[]{ruf:anz} Ruffert M., Anzer U., 1995, A\&A~295, 108

\bibitem[]{ruf:arn} Ruffert M., Arnett D., 1994, ApJ~427, 351

\bibitem[]{saw} Sawada K., Matsuda T., Anzer U., B\"orner G., 
         Livio M., 1989, A\&A~221, 263

\bibitem[]{sha} Shapiro S.L., Lightman A.P., 1976, ApJ~204, 555

\bibitem[]{sha} Shima E., Matsuda T., Anzer U., B\"orner G., 
          Boffin H.M.J, 1998, A\&A 337, 311 
\bibitem[]{taa} Soker N., Livio M., 1984, MNRAS~211, 927

\bibitem[]{taa} Taam R.E., Fryxell B.A., 1989, ApJ~339, 297

\bibitem[]{wan} Wang Y.-M., 1981, A\&A~102, 36

\end{thebibliography}
\end{document}